\documentclass[hidelinks,onefignum,onetabnum]{siamart250211}
\pdfoutput=1 %This is for the ArXiv submission only
\usepackage{graphicx}
\usepackage{amsfonts}
\usepackage{url}
\usepackage{epstopdf}
\usepackage{color}
\usepackage{bm}
\usepackage{comment}
\usepackage{rotating}
\usepackage{multirow}
\usepackage{cite}

\def\x{\bm{x}}
\def\X{\bm{X}}

\def\kmax{k_{\rm max}}

\def\ve{\varepsilon}
\def\w{{w}}

\def\pa{\partial\Omega}
\def\E{{\mathbb E}}
\def\I{{\mathbb I}}
\def\H{{\mathbb H}}
\def\P{{\mathbb P}}
\def\R{{\mathbb R}}

\def\G{{\mathcal G}}
\def\AA{{\bf A}}
\def\GG{{\bf G}}
\def\MM{{\bf M}}
\def\N{{\mathcal N}}

\def\Li{\mathrm{Li}}

\def\M{{\mathcal M}}
\def\ctanh{\mathrm{ctanh}}

\def\a{{\mathcal A}}

\def\VV{{\bf V}}

\headers{Steklov-Neumann spectral problem}{D. S. Grebenkov}

\title{Steklov-Neumann spectral problem: asymptotic analysis 
and applications to diffusion-controlled reactions\thanks{Submitted to the editors DATE.}}
\author{Denis S. Grebenkov\thanks{Laboratoire de Physique de la Mati\`{e}re Condens\'{e}e (UMR 7643), 
CNRS -- Ecole Polytechnique, Institut Polytechnique de Paris, 91120 Palaiseau, France; 
and CNRS - Universit\'e de Montr\'eal CRM - CNRS, 6128 succ Centre-Ville, Montr\'eal QC H3C 3J7, Canada
(\email{denis.grebenkov@polytechnique.edu})}}

\begin{document}

\maketitle

\date{\today}

\begin{abstract}
We consider the mixed Steklov-Neumann spectral problem for the
modified Helmholtz equation in a bounded domain when the Steklov
condition is imposed on a connected subset of the smooth boundary.  In
order to deduce the asymptotic behavior in the limit when the size of
the subset goes to zero, we reformulate the original problem in terms
of an integral operator whose kernel is the restriction of a suitable
Green's function (or pseudo-Green's function) to the subset.  Its
singular behavior on the boundary yields the asymptotic formulas for
the eigenvalues and eigenfunctions of the Steklov-Neumann problem.
While this analysis remains at a formal level, it is supported by
extensive numerical results for two basic examples: an arc on the
boundary of a disk and a spherical cap on the boundary of a ball.
Solving the original Steklov-Neumann problem numerically in these
domains, we validate the asymptotic formulas and reveal their high
accuracy, even when the subset is not small.  A straightforward
application of these spectral results to first-passage processes and
diffusion-controlled reactions is presented.  We revisit the
small-target limit of the mean first-reaction time on perfectly or
partially reactive targets.  The effect of multiple failed reaction
attempts is quantified by a universal function for the whole range of
reactivities.  Moreover, we extend these results to more sophisticated
surface reactions that go beyond the conventional narrow escape
problem.
\end{abstract}


\section{Introduction}
\label{sec:intro}

A macroscopic theory of diffusion-controlled reactions usually relies
on the diffusion equation with appropriate boundary conditions
\cite{North66,Wilemski73,Calef83,Berg85,Rice85,Grebenkov23c}.  It is
therefore common to employ the Laplacian eigenfunctions to get
spectral expansions for most quantities of interest, such as the
diffusion propagator, the survival probability, the moments and the
probability density of the first-reaction time on a target region, the
concentration of diffusing particles, or the diffusive flux
\cite{Redner,Schuss,Metzler,Benichou14,Holcman14,Masoliver,Lindenberg,Grebenkov,Dagdug}.
Despite its impressive progress, the conventional theory is limited to
basic surface reactions on perfectly or partially reactive regions of
the boundary that correspond respectively to Dirichlet and Robin
boundary conditions.  In turn, recent developments of the
encounter-based approach
\cite{Grebenkov19,Grebenkov19c,Grebenkov20,Grebenkov20b,Grebenkov20c,Grebenkov21a,Grebenkov22a,Grebenkov22b,Grebenkov22d}
employ the Steklov eigenfunctions to describe repetitive returns of a
diffusing particle to the reactive regions and thus to incorporate
more sophisticated reactions, such as non-Markovian binding,
activation/passivation of a catalytic germ, targets with
encounter-dependent reactivity, resetting mechanisms, permeation
across membranes, etc
\cite{Bressloff22d,Benkhadaj22,Bressloff22b,Bressloff22a,Bressloff22c,Grebenkov23a,Grebenkov23b,Bressloff23a,Bressloff23b}
(see also Sec. \ref{sec:application}).

For a bounded Euclidean domain $\Omega \subset \R^d$ with a smooth
boundary $\pa$, the conventional Steklov problem consists in finding
the eigenpairs $\{\mu, V\}$ satisfying
\begin{equation}
\Delta V = 0 \quad \textrm{in} ~\Omega,  \qquad \partial_n V = \mu V \quad \textrm{on}~\pa,
\end{equation}
where $\Delta$ is the Laplace operator and $\partial_n$ is the normal
derivative on the boundary $\pa$ oriented outwards the domain $\Omega$
\cite{Steklov1902,Kuznetsov14}.  The reminiscent feature of the
Steklov problem is that the spectral parameter $\mu$ stands in the
boundary condition.  For a broad class of domains, the spectrum is
known to be discrete, i.e., there is a countable sequence of
eigenpairs solving the Steklov problem \cite{Girouard17,Colbois24}.
This spectral problem plays an important role in spectral geometry
\cite{Levitin} and finds numerous applications in applied mathematics
and physics such as approximation of harmonic functions
\cite{Auchmuty04,Auchmuty13,Auchmuty14,Auchmuty15,Auchmuty18}, domain
decomposition \cite{Smith96,Levitin08,Delitsyn12,Delitsyn18}, electric
impedance tomography
\cite{Cheney99,Calderon80,Borcea02,Sylvester87,Curtis91}, and the
aforementioned encounter-based approach to diffusion-controlled
reactions.

\begin{figure}
\begin{center}
\includegraphics[width=100mm]{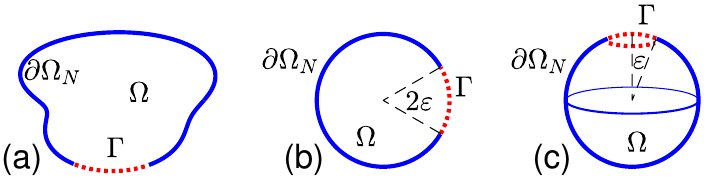} 
\end{center}
\caption{
Schematic illustration of the mixed Steklov-Neumann problem
(\ref{eq:SteklovN}) in a generic Euclidean domain {\bf (a)}, in a disk
{\bf (b)}, and in a ball {\bf (c)}.  The solid blue line indicates the
reflecting boundary $\pa_N$ with Neumann boundary condition, while the
dashed red line shows the subset $\Gamma$ with the Steklov
condition. }
\label{fig:scheme}
% A_DN_disk_mixed_scheme2();
\end{figure}

When surface reactions occur on an open subset $\Gamma$ of the
otherwise impenetrable reflecting boundary, one needs to treat
separately the reactive subset and the remaining passive surface of
the confinement, denoted as $\pa_N = \pa\backslash
\overline{\Gamma}$ (Fig. \ref{fig:scheme}a).  The subset
$\Gamma$ can represent a specific target, a catalytic germ, an ion
channel, an enzyme, an escape window, etc., depending on the physical,
chemical or biological context.
Accordingly, the above Steklov problem needs to be extended by
imposing the Steklov condition on $\Gamma$ and the Neumann condition
on $\pa_N$.  In this setting, one deals with the mixed Steklov-Neumann
problem and searches for the eigenpairs $\{\mu_k^{(p,\Gamma)},
V_k^{(p,\Gamma)}\}$, enumerated by the index $k = 0,1,2,\ldots$ and
satisfying
\begin{subequations}  \label{eq:SteklovN}
\begin{align}  \label{eq:Helm}
(p - D\Delta) V_k^{(p,\Gamma)} & = 0 \quad \textrm{in} ~\Omega, \\   \label{eq:Vk_SteklovBC}
\partial_n V_k^{(p,\Gamma)} & = \mu_k^{(p,\Gamma)} V_k^{(p,\Gamma)} \quad \textrm{on}~\Gamma, \\
\partial_n V_k^{(p,\Gamma)} & = 0 \quad \textrm{on}~ \pa_N, 
\end{align}
\end{subequations}
where $D > 0$ is a diffusion constant and $p \geq 0$ is a nonnegative
parameter, which is introduced to give a broader perspective of the
Steklov problem, as explained below.

The mixed Steklov-Neumann spectral problem for the Laplace equation
(i.e., at $p = 0$), also known as the sloshing problem (or ice-fishing
problem) in hydrodynamics
\cite{Henrici70,Troesch72,Miles72,Fox83,Kozlov04}, was discussed in
the mathematical literature (see \cite{Levitin22} and references
therein).  For instance, Ba\~nuelos {\it et al.} derived several
inequalities for the eigenvalues \cite{Banuelos10}, whereas Mayrand
{\it et al.}  studied the asymptotic behavior of the eigenvalues for
the three-dimensional sloshing problem on a triangular prism
\cite{Mayrand20}.
Ammari {\it et al.} analyzed the behavior of the Green's function in
planar domains and proposed a method for optimizing this function at a
given source-receiver pair \cite{Ammari20b}.  In particular, they
presented asymptotic expressions for the change in the Steklov-Neumann
eigenpairs when a small portion of the boundary is changed from
Steklov to Neumann condition.  Despite this progress, some practically
relevant questions remain unattended, e.g., the asymptotic behavior of
eigenvalues and eigenfunctions as the subset $\Gamma$ shrinks.  In the
context of first-passage phenomena
\cite{Redner,Schuss,Metzler,Benichou14,Holcman14,Masoliver,Lindenberg,Grebenkov,Dagdug},
this limit is known as the narrow escape problem; in particular, the
asymptotic behavior of the mean first-escape time and Laplacian
eigenvalues under singular perturbations of the boundary was
thoroughly investigated
\cite{Ward93,Grigoriev02,Holcman04,Kolokolnikov05,Singer06a,Singer06b,Singer06c,Schuss07,Benichou08,Singer08,Reingruber09,Pillay10,Cheviakov10,Cheviakov11,Cheviakov12,Caginalp12,Rojo12,Rupprecht15,Gomez15,Grebenkov16,Marshall16,Grebenkov17,Grebenkov19d,Kaye20,Guerin23}.
We aim to inspect the behavior of the Steklov-Neumann eigenvalues and
eigenfunctions in the small-target limit $\Gamma \to \emptyset$ and
then to relate the derived spectral results to the narrow escape
problem.

The paper is organized as follows.  In Sec. \ref{sec:theory}, we start
by recalling the basic properties of the Steklov-Neumann problem and
its spectral expansions.  In particular, we argue that the restriction
of the Green's function $\tilde{G}_0(\x,p|\x_0)$ of the modified
Helmholtz equation (\ref{eq:Helm}) to $\Gamma \times
\Gamma$ is the kernel of an integral operator that determines
the eigenpairs $\{\mu_k^{(p,\Gamma)}, V_k^{(p,\Gamma)}\}$ of the
spectral problem (\ref{eq:SteklovN}).  This property allows one to
construct the eigenpairs numerically for any subset $\Gamma$ and to
highlight the impact of its shape.  We also discuss the specific
features of this analysis in the limit $p = 0$ and the central role of
the pseudo-Green's function.
Section \ref{sec:limit} presents the main asymptotic results for
a bounded domain $\Omega \subset\R^d$ with a smooth boundary $\pa$.
We consider either a connected subset $\Gamma \subset \pa$ of
perimeter $2\epsilon$ for the case $d = 2$, or a (curved) disk-like
subset $\Gamma$ of radius $\epsilon$ for the case $d = 3$ (e.g., a cap
on a sphere).  In the limit $\epsilon\to 0$, we obtain the following
leading-order asymptotic behavior of the eigenvalues
$\mu_k^{(p,\Gamma)}$ and eigenfunctions $V_k^{(p,\Gamma)}$ restricted
to $\Gamma$:
\begin{equation}  \label{eq:main_asympt}
\mu_k^{(p,\Gamma)} \approx \frac{\hat{\mu}_k}{\epsilon} , \qquad V_k^{(p,\Gamma)}(\x) 
\approx \frac{\hat{V}_k(\x/\epsilon)}{\epsilon^{(d-1)/2}}  \quad (\x\in\Gamma),
\end{equation}
for any $0 \leq p \ll 1/(D\epsilon^2)$.  Here $\{\hat{\mu}_k,
\hat{V}_k\}$ are the eigenpairs of an auxiliary Steklov-Neumann
problem for the exterior of an interval in the half-plane ($d = 2$),
or for the exterior of a disk in the half-space ($d = 3$).  We discuss
different ways to construct these eigenpairs, in particular, by
reformulating the Steklov-Neumann problem as a spectral problem for an
integral operator whose kernel is explicitly derived.
Section \ref{sec:examples} provides numerical illustrations for two
relevant examples: an arc-shaped subset $\Gamma$ on the boundary of a
disk (Fig. \ref{fig:scheme}b) and a spherical cap $\Gamma$ on the
boundary of a ball (Fig. \ref{fig:scheme}c).  In these settings, the
Green's function and the pseudo-Green's function are known explicitly,
which allows us to construct efficiently the eigenpairs
$\{\mu_k^{(p,\Gamma)}, V_k^{(p,\Gamma)}\}$ for any size of the subset
$\Gamma$ and any positive $p$.  In this way, we illustrate the
accuracy and the validity range of the asymptotic relations
(\ref{eq:main_asympt}).
Section \ref{sec:application} presents a straightforward application
of these spectral results to first-passage problems.  In particular,
we express the mean first-reaction time (MFRT) in terms of the
eigenpairs $\{\mu_k^{(0,\Gamma)}, V_k^{(0,\Gamma)}\}$, discuss its
asymptotic behavior and re-analyze the validity of the constant-flux
approximation for the MFRT \cite{Grebenkov17}.  Most importantly, we
reveal the effect of partial reactivity over the whole range of
reactivities and extend the asymptotic analysis to more sophisticated
surface reactions that go beyond the conventional narrow escape
problem.  Section \ref{sec:conclusion} summarizes the main results and
concludes the paper.

We emphasize that the presented analysis remains at a formal level and
lacks rigorous proofs.  For instance, the convergence of spectral
expansions, integrations by parts, the use of Dirac distributions, and
probabilistic interpretations are not yet properly demonstrated.  For
this reason, the related statements are not formulated as theorems.
The omission of proofs is partly compensated by an extensive numerical
analysis that validates the asymptotic results and illustrates their
accuracy.  We expect that the presented results can stimulate a
broader interest among mathematicians to this topic to ensure its more
rigorous treatment in the future.

\section{General spectral properties}
\label{sec:theory}

This section prepares the theoretical ground to deal with the mixed
Steklov-Neumann problem (\ref{eq:SteklovN}).  In
Sec. \ref{sec:mixed_intro}, we recall the basic properties of the
eigenvalues and eigenfunctions and their relation to the Green's
function of the modified Helmholtz equation and to the Neumann
Laplacian.  Section \ref{sec:p0} focuses on the asymptotic behavior of
the Steklov eigenpairs $\{\mu_k^{(p,\Gamma)}, V_k^{(p,\Gamma)}\}$ in
the limit $p\to 0$.  In particular, we show how $\mu_k^{(0,\Gamma)}$
and $V_k^{(0,\Gamma)}$ can be obtained as eigenpairs of the integral
equation (\ref{eq:eigen_G}), whose kernel in Eq. (\ref{eq:G_def}) is
related to the pseudo-Green's function.  Section \ref{sec:role}
presents useful interpretations of the first-order corrections
$\a_\Gamma$ and $W_0^{(\Gamma)}$ to the principal eigenvalue
$\mu_0^{(p,\Gamma)}$ and eigenfunction $V_0^{(p,\Gamma)}$ in the limit
$p\to 0$.  For instance, we show how $\a_\Gamma$ determines the
variance of the boundary local time on $\Gamma$ and the leading-order
term of the mean first-passage time (MFPT) in the narrow escape
problem.

\subsection{Mixed Steklov-Neumann problem}
\label{sec:mixed_intro}

We consider a bounded Euclidean domain $\Omega \subset\R^d$ with a
smooth connected boundary $\pa$.  An open connected set $\Gamma
\subset \pa$ is called a ``reactive subset'' of the boundary $\pa$,
whereas its reflecting part is denoted as $\pa_N = \pa
\backslash \overline{\Gamma}$ (Fig. \ref{fig:scheme}a).  To avoid
possible technical issues on the ``junction'' between $\Gamma$ and
$\pa_N$, we assume that $\partial\Gamma$ is smooth.
In this setting, the mixed Steklov-Neumann problem (\ref{eq:SteklovN})
is known to have a discrete spectrum for any $p \geq 0$, i.e., there
is a countable sequence of eigenpairs $\{\mu_k^{(p,\Gamma)},
V_k^{(p,\Gamma)} \}$ satisfying (\ref{eq:SteklovN}); in particular,
the eigenvalues can be enumerated in increasing order:
\begin{equation}
0 \leq \mu_0^{(p,\Gamma)} < \mu_1^{(p,\Gamma)} \leq \mu_2^{(p,\Gamma)} \leq \ldots  \nearrow +\infty ,
\end{equation}
and the first eigenvalue $\mu_0^{(p,\Gamma)}$ is simple (see, e.g.,
\cite{Levitin,Banuelos10} for mathematical details in the case $p =
0$; an extension to $p > 0$ is straightforward).  The
eigenfunctions $V_k^{(p,\Gamma)}$ belong to the Sobolev space
$H^1(\Omega)$; moreover, since $p\geq 0$, the eigenfunctions
$V_k^{(p,\Gamma)}$ can be chosen to be real.
Their restrictions to $\Gamma$, denoted as $v_k^{(p,\Gamma)} =
V_k^{(p,\Gamma)}|_{\Gamma}$, form a complete orthonormal basis of the
space $L^2(\Gamma)$ of square-integrable functions on $\Gamma$, i.e.,
\begin{equation}  \label{eq:L2norm}
\int\limits_\Gamma  v_k^{(p,\Gamma)} \, v_{k'}^{(p,\Gamma)} = \delta_{k,k'},
\end{equation}
where $\delta_{k,k'} = 1$ for $k = k'$ and $0$ otherwise.  This
relation fixes the normalization of $v_k^{(p,\Gamma)}$ and thus of
$V_k^{(p,\Gamma)}$.  In turn, the $L^2(\Omega)$ norm of
$V_k^{(p,\Gamma)}$ can be found via the identity
\cite{Friedlander91,Grebenkov19,Chaigneau24}:
\begin{equation}  \label{eq:dmu_dp}
\int\limits_{\Omega} |V_k^{(p,\Gamma)}|^2 = D \partial_p \mu_k^{(p,\Gamma)} 
\end{equation}
(see Appendix \ref{sec:identity} for a formal derivation and
discussion).  Note that $\mu_k^{(p,\Gamma)}$ and $v_k^{(p,\Gamma)}$
turn out to be the eigenvalues and eigenfunctions of the
Dirichlet-to-Neumann operator $\M_p^{(\Gamma)}$:
\begin{equation}  \label{eq:Mp_eigenproblem}
\M_p^{(\Gamma)} v_k^{(p,\Gamma)} = \mu_k^{(p,\Gamma)} v_k^{(p,\Gamma)} .
\end{equation}
This operator acts from $H^{\frac12}(\Gamma)$ to
$H^{-\frac12}(\Gamma)$ and associates to a given function $f$ on
$\Gamma$ another function $g$ on $\Gamma$ such that $\M_p^{(\Gamma)} f
= g = (\partial_n u)|_{\Gamma}$, where $u$ is the unique solution of
the boundary value problem:
\begin{equation}
(p - D \Delta) u = 0 \quad \textrm{in} ~\Omega, \qquad u|_{\Gamma} = f,  \qquad
(\partial_n u)|_{\pa_N} = 0
\end{equation} 
(see \cite{Levitin,McLean} for mathematical details and the definition
of the Sobolev spaces $H^{\pm \frac12}(\Gamma)$).

We also introduce the Green's function $\tilde{G}_q(\x,p|\x_0)$ of the
modified Helmholtz equation, which satisfies for any $\x_0 \in \Omega$:
\begin{subequations}   \label{eq:Gq_problem}
\begin{align}  \label{eq:Gq_Helm}
(p - D \Delta) \tilde{G}_q(\x,p|\x_0) & = \delta(\x - \x_0) \quad (\x\in\Omega), \\  \label{eq:Gq_Robin}
\partial_n \tilde{G}_q(\x,p|\x_0) + q \tilde{G}_q(\x,p|\x_0) & = 0 \quad (\x\in \Gamma), \\  \label{eq:Gq_Neumann}
\partial_n \tilde{G}_q(\x,p|\x_0) & = 0 \quad (\x\in\pa_N),
\end{align}
\end{subequations}
where $\delta(\x-\x_0)$ is the Dirac distribution, and $0 \leq q
\leq +\infty$ is a fixed parameter; in the following, we mainly use
two limits: $q = 0$ (Neumann condition on $\Gamma$) and $q = \infty$
(Dirichlet condition on $\Gamma$).

Following \cite{Grebenkov20}, we get the spectral expansion of the
Green's function in terms of the Steklov-Neumann eigenpairs:
\begin{equation}  \label{eq:Gq_spectral}
\tilde{G}_q(\x,p|\x_0) = \tilde{G}_\infty(\x,p|\x_0) 
+ \frac{1}{D}\sum\limits_{k=0}^\infty \frac{V_k^{(p,\Gamma)}(\x) V_k^{(p,\Gamma)}(\x_0)}{q + \mu_k^{(p,\Gamma)}} 
\end{equation}
(see Appendix \ref{sec:Gq_spectral} for a formal derivation and
discussion on its convergence).
Setting $q = 0$ and restricting both $\x$ and $\x_0$ to $\Gamma$, we
have
\begin{equation}  \label{eq:Gq0_spectral}
D \tilde{G}_0(\x,p|\x_0) = \sum\limits_{k=0}^\infty \frac{v_k^{(p,\Gamma)}(\x) v_k^{(p,\Gamma)}(\x_0)}{\mu_k^{(p,\Gamma)}}  \quad (\x,\x_0\in\Gamma).
\end{equation}
One sees that the restriction of $D \tilde{G}_0(\x,p|\x_0)$ to
$\Gamma\times\Gamma$ is the kernel of an integral operator, which is
the inverse of the Dirichlet-to-Neumann operator $\M_p^{(\Gamma)}$ on
$\Gamma$.  As a consequence, one can search for its eigenvalues
$\mu_k^{(p,\Gamma)}$ and eigenfunctions $v_k^{(p,\Gamma)}$ as the
eigenpairs of this integral operator:
\begin{equation}  \label{eq:Gp_problem}
\int\limits_\Gamma d\x \, D \tilde{G}_0(\x,p|\x_0) \, v_k^{(p,\Gamma)}(\x) 
= \frac{v_k^{(p,\Gamma)}(\x_0)}{\mu_k^{(p,\Gamma)}} 
\qquad (\x_0 \in \Gamma, ~ k \geq 0).
\end{equation}
Once $v_k^{(p,\Gamma)}$ is obtained on $\Gamma$, its extension
$V_k^{(p,\Gamma)}(\x_0)$ to the whole domain $\x_0\in \Omega$ follows
as
\begin{equation}  \label{eq:Vkp_vkp}
V_k^{(p,\Gamma)}(\x_0) = \mu_k^{(p,\Gamma)} \int\limits_{\Gamma} d\x \, D\tilde{G}_0(\x,p|\x_0)\, v_k^{(p,\Gamma)}(\x)
\qquad (\x_0 \in \overline{\Omega}).
\end{equation}
This relation is obtained by multiplying Eq. (\ref{eq:Gq_Helm}) with
$q = 0$ by $V_k^{(p,\Gamma)}(\x)$, multiplying Eq. (\ref{eq:Helm}) by
$\tilde{G}_0(\x,p|\x_0)$, subtracting these equations, integrating
over $\x\in\Omega$, and using the Green's formula and the boundary
conditions (\ref{eq:Vk_SteklovBC}, \ref{eq:Gq_Robin}).

Importantly, the Green's function $\tilde{G}_0(\x,p|\x_0)$ does not
depend on $\Gamma$.  Indeed, at $q = 0$, the Robin boundary condition
(\ref{eq:Gq_Robin}) is reduced to the Neumann one so that one deals
with the Neumann boundary condition on the whole boundary $\pa$.  In
other words, if the Green's function $\tilde{G}_0(\x,p|\x_0)$ is known
for a given domain $\Omega$, its restriction to $\Gamma \times
\Gamma$ determines the integral operator, whose eigenmodes solve the
mixed Steklov-Neumann problem.  For some simple domains, the Green's
function is known explicitly (e.g., see Appendices
\ref{sec:Adisk} and \ref{sec:Aball} for a disk and a ball).  In
general, it admits the standard spectral expansion
\begin{equation}  \label{eq:G0_Laplacian}
\tilde{G}_0(\x,p|\x_0) = \sum\limits_{k=0}^\infty \frac{u_k^N(\x) \, u_k^N(\x_0)}{\lambda_k^N + p/D}   \qquad (\x,\x_0 \in \overline{\Omega})
\end{equation} 
over the $L^2(\Omega)$-normalized eigenfunctions $u_k^N$ and
eigenvalues $\lambda_k^N$ of the Laplace operator with Neumann
boundary condition:
\begin{equation}
-\Delta u_k^N = \lambda_k^N u_k^N \quad \textrm{in}~\Omega, \qquad \partial_n u_k^N = 0 \quad \textrm{on}~\pa .
\end{equation}
The expansion (\ref{eq:G0_Laplacian}) can be verified by applying the
operator $(p-D\Delta)$ to the right-hand side and using the
completeness relation
\begin{equation}  \label{eq:ukN_completeness}
\sum\limits_{k=0}^\infty u_k^N(\x) \, u_k^N(\x_0) = \delta(\x-\x_0)  \qquad (\x,\x_0 \in \Omega),
\end{equation}
which reflects the completeness of the basis of the Neumann Laplacian
eigenfunctions $\{u_k^N\}$ in $L^2(\Omega)$ (this relation can also be
formally understood as the expansion of the Dirac distribution onto
the basis $\{u_k^N\}$).  Alternatively, one can use the expansion
(\ref{eq:Gq0_spectral}) with $\Gamma = \pa$:
\begin{equation}  \label{eq:G0_whole}
\tilde{G}_0(\x,p|\x_0) = \sum\limits_{k=0}^\infty \frac{v_k^{(p,\pa)}(\x) \, v_k^{(p,\pa)}(\x_0)}{D \mu_k^{(p,\pa)}}  \qquad (\x,\x_0\in \pa),
\end{equation} 
which is based on the Steklov eigenfunctions $v_k^{(p,\pa)}$ on the
whole boundary $\pa$ and is thus independent of the subset $\Gamma$.

\subsection{Limit $p = 0$}
\label{sec:p0}

At $p = 0$, the smallest Steklov eigenvalue is zero, while the
corresponding eigenfunction is constant:
\begin{equation}
\mu_0^{(0,\Gamma)} = 0,  \qquad V_0^{(0,\Gamma)} = \frac{1}{\sqrt{|\Gamma|}} \,,
\end{equation}
where $|\Gamma|$ denotes the area of $\Gamma$, and we used the
normalization (\ref{eq:L2norm}) of the Steklov eigenfunctions.  As a
consequence, its contribution to the spectral expansion
(\ref{eq:Gq0_spectral}) diverges, which reflects the well-known fact
that the conventional Green's function for the Laplace equation does
not exist for the interior Neumann problem.  This divergence can be
easily amended by subtracting the divergent term and thus dealing with
pseudo-Green's function (also known as Neumann Green's function
\cite{Kolokolnikov05}):
\begin{equation}  \label{eq:G0_limit}
\G_0(\x|\x_0) = \lim\limits_{p\to 0} \biggl(D \tilde{G}_0(\x,p|\x_0) - \frac{1}{|\Omega|(p/D)}\biggr) ,
\end{equation}
where $|\Omega|$ denotes the volume of $\Omega$.  Since $\lambda_0^N =
0$ and $u_0^N = 1/\sqrt{|\Omega|}$, the subtracted term is precisely
the contribution of the principal eigenmode ($k = 0$) to the spectral
expansion (\ref{eq:G0_Laplacian}).  As a consequence, the
pseudo-Green's function admits the same spectral expansion without the
principal eigenmode:
\begin{equation}  \label{eq:G0_spectral}
\G_0(\x|\x_0) = \sum\limits_{k=1}^\infty \frac{u_k^N(\x) \, u_k^N(\x_0)}{\lambda_k^N}   \qquad (\x,\x_0 \in \overline{\Omega}).
\end{equation}

Applying the Laplace operator to the spectral expansion
(\ref{eq:G0_spectral}), one can check that the pseudo-Green's function
satisfies for any $\x_0 \in \Omega$:
\begin{subequations}  \label{eq:G0_problem}
\begin{align}  \label{eq:G0_problem_eq}
\Delta \G_0(\x|\x_0) & = \frac{1}{|\Omega|} - \delta(\x-\x_0) \quad (\x\in\Omega), \\
\partial_n \G_0(\x|\x_0) & = 0 \quad (\x\in \pa),
\end{align}
\end{subequations}
where we used the completeness relation
(\ref{eq:ukN_completeness}).
While the limit in Eq. (\ref{eq:G0_limit}) and the spectral expansion
(\ref{eq:G0_spectral}) determine $\G_0(\x|\x_0)$ uniquely, the
boundary value problem (\ref{eq:G0_problem}) defines the
pseudo-Green's function up to an additive constant, which can be
fixed by the additional condition:
\begin{equation}  \label{eq:G0_constant} 
\int\limits_{\Omega} d\x \, \G_0(\x|\x_0) = 0 .
\end{equation}
In fact, this condition is consistent with the fact that the integral
of the spectral expansion (\ref{eq:G0_spectral}) over $\Omega$ is zero
because all eigenfunctions $u_k^N$ with $k\geq 1$ are orthogonal to
the constant function $u_0^N$.  Note that the symmetry of the
pseudo-Green's function, $\G_0(\x|\x_0) = \G_0(\x_0|\x)$, implies that
\begin{equation}  \label{eq:G0_constant2} 
\int\limits_{\Omega} d\x_0 \, \G_0(\x|\x_0) = 0.
\end{equation}
We stress that the unusual constant $1/|\Omega|$ in
Eq. (\ref{eq:G0_problem_eq}) is necessary to satisfy the convergence
theorem; indeed, the integral of Eq. (\ref{eq:G0_problem_eq}) over
$\x\in \Omega$ reads
\begin{align*}
0 = \int\limits_{\Omega} d\x \biggl(\frac{1}{|\Omega|} - \delta(\x-\x_0)\biggr) & = \int\limits_{\Omega} d\x \, \Delta \G_0(\x|\x_0) 
 = \int\limits_{\pa} d\x \, \partial_n \G_0(\x|\x_0) = 0,
\end{align*}
and this identity could not be satisfied without the constant
$1/|\Omega|$.  As no such constant is present in
Eq. (\ref{eq:Gq_Helm}) for other Green's functions, $\G_0(\x|\x_0)$ is
called {\it pseudo}-Green's function.

The last step consists in expressing $\G_0(\x|\x_0)$ in terms of the
Steklov eigenfunctions $V_k^{(0,\Gamma)}$, in analogy to
Eq. (\ref{eq:G0_spectral}).  For this purpose, we need to compute the
limit in Eq. (\ref{eq:G0_limit}).  This computation relies on
the analyticity of Dirichlet-to-Neumann maps, which was established in
\cite{Behrndt15} (Corollary 4.7).  When $p \geq 0$, the
Dirichlet-to-Neumann operators $\M_p^{(\Gamma)}$ form a self-adjoint
holomorphic family of type (A), in the terminology used in Kato's book
\cite{Kato}.  A general analytic perturbation theory for linear elliptic
operators \cite{Kato} (Chapter 7) allows therefore to establish the
analyticity of eigenvalues and eigenprojectors (under certain
conditions).
Alternatively, the analyticity of the eigenvalues of the Steklov
problem (i.e., with $\Gamma = \pa$) follows from a corresponding
result for Robin Laplacian eigenvalues \cite{Bucur17} (p. 82) and the
duality between Robin and Steklov spectral problems \cite{Levitin}.
An extension of this result to the Steklov-Neumann problem is
straightforward.  Indeed, the eigenvalues of the Laplace operator with
mixed Robin-Neumann boundary conditions remain analytic in $p$ near
$0$ because the domain $H^1(\Omega)$ of the Laplace operator does not
change, while the associated form is still analytic in $p$.

The analyticity near $0$ justifies a Taylor expansion of the principal
eigenmode of the Steklov-Neumann problem in powers of $p$ as $p\to 0$:
\begin{subequations}  \label{eq:mu0v0_p0}
\begin{align}  \label{eq:mu0_p0}
\mu_0^{(p,\Gamma)} & = \frac{p |\Omega|}{D |\Gamma|}\biggl(1 - \a_{\Gamma} \frac{p}{D}|\Omega| + O(p^2) \biggr)  ,\\    \label{eq:V0_p0}
V_0^{(p,\Gamma)}(\x) & = \frac{1}{\sqrt{|\Gamma|}} \biggl(1 + W_0^{(\Gamma)}(\x) \frac{p}{D} |\Omega| + O(p^2)\biggr),
\end{align}
\end{subequations}
where $\a_\Gamma$ and $W_0^{(\Gamma)}(\x)$ are the first-order
corrections to the principal eigenvalue and eigenfunction,
respectively:  
\begin{subequations}
\begin{align}  \label{eq:a_def}
\a_\Gamma & = - \frac{D^2 |\Gamma|}{2|\Omega|^2}  \lim\limits_{p\to 0} \partial_p^2 \mu_0^{(p,\Gamma)}  , \\  \label{eq:W0_def}
W_0^{(\Gamma)}(\x) & = \frac{D \sqrt{|\Gamma|}}{|\Omega|} \lim\limits_{p\to 0} \partial_p V_0^{(p,\Gamma)}(\x).
\end{align}
\end{subequations}
According to Eq. (\ref{eq:dmu_dp}), the leading term in
Eq. (\ref{eq:mu0_p0}) is determined by the first derivative of
$\mu_0^{(p,\Gamma)}$, evaluated at $p = 0$:
\begin{equation}
\lim\limits_{p\to 0} \partial_p \mu_0^{(p,\Gamma)} = \frac{1}{D} \int\limits_{\Omega} d\x \, 
[\underbrace{V_0^{(0)}}_{=|\Gamma|^{-1/2}}]^2 = \frac{|\Omega|}{D |\Gamma|}  \,.
\end{equation} 
Since the first eigenvalue is simple, there is no ambiguity in
choosing the analytic branch.
We stress that the above arguments on the analyticity of eigenvalues
and eigenfunctions of the mixed Steklov-Neumann problem
(\ref{eq:SteklovN}) lack technical details and rigorous proofs.  For
this reason, the Taylor expansion (\ref{eq:V0_p0}) is treated at a
formal level, and its rigorous proof remains an open problem.

Substituting Eqs. (\ref{eq:mu0v0_p0}) into Eq. (\ref{eq:G0_limit}), we
evaluate the pseudo-Green's function as
\begin{align}  \nonumber
\G_0(\x|\x_0) & = D\tilde{G}_\infty(\x,0|\x_0) + \a_\Gamma + W_0^{(\Gamma)}(\x) + W_0^{(\Gamma)}(\x_0) \\ 
&  + \sum\limits_{k = 1}^\infty \frac{V_k^{(0,\Gamma)}(\x) V_k^{(0,\Gamma)}(\x_0)}{\mu_k^{(0,\Gamma)}}   \qquad (\x,\x_0\in\Omega).
\end{align}
In turn, its restriction to $\Gamma \times \Gamma$ eliminates the
first term and gives
\begin{align}  \label{eq:G0}
\G_0(\x|\x_0) & = \a_\Gamma + w_0^{(\Gamma)}(\x) + w_0^{(\Gamma)}(\x_0)   
 + \sum\limits_{k = 1}^\infty \frac{v_k^{(0,\Gamma)}(\x) v_k^{(0,\Gamma)}(\x_0)}{\mu_k^{(0,\Gamma)}}   \qquad (\x,\x_0\in\Gamma),
\end{align}
where $w_0^{(\Gamma)}$ is the restriction of $W_0^{(\Gamma)}$ to
$\Gamma$.  Integrating this expression over $\x\in \Gamma$ yields
\begin{equation}  \label{eq:auxil5}
\int\limits_{\Gamma} d\x \, \G_0(\x|\x_0)  = |\Gamma| \bigl(w_0^{(\Gamma)}(\x_0) + \a_\Gamma\bigr)  \qquad (\x_0\in\Gamma),
\end{equation}
where we used the orthogonality of $w_0^{(\Gamma)}$ to a constant,
\begin{equation}  \label{eq:w0_int}
\int\limits_{\Gamma} w_0^{(\Gamma)} = 0,
\end{equation}
which follows from the normalization of the eigenfunction
$v_0^{(p,\Gamma)}$:
\begin{equation}   
1 = \int\limits_{\Gamma} |v_0^{(p,\Gamma)}|^2 
= \underbrace{\int\limits_{\Gamma} |v_0^{(0,\Gamma)}|^2}_{=1} +
p \,\frac{2|\Omega|}{D|\Gamma|} \underbrace{\int\limits_{\Gamma} w_0^{(\Gamma)} }_{=0}  + O(p^2).
\end{equation}
The integral of Eq. (\ref{eq:auxil5}) over $\x_0\in\Gamma$ determines
the constant $\a_\Gamma$,
\begin{equation}   \label{eq:a_eps}
\a_\Gamma = \frac{1}{|\Gamma|^2} \int\limits_{\Gamma} d\x_0 \int\limits_{\Gamma} d\x \, \G_0(\x|\x_0)  ,
\end{equation}
from which the function $w_0^{(\Gamma)}(\x_0)$ follows as
\begin{equation}   \label{eq:omega_eps}
w_0^{(\Gamma)}(\x_0) = - \a_\Gamma + \frac{1}{|\Gamma|} \int\limits_{\Gamma} d\x \, \G_0(\x|\x_0)    \qquad (\x_0\in\Gamma).
\end{equation}
Combining these results, we define the integral kernel:
\begin{equation}  \label{eq:G_def}
\G(\x|\x_0) = \G_0(\x|\x_0) - \bigl(w_0^{(\Gamma)}(\x) + w_0^{(\Gamma)}(\x_0) + \a_\Gamma\bigr), \\
\end{equation}
which admits the following expansion according to (\ref{eq:G0}):
\begin{equation}    \label{eq:G}
\G(\x|\x_0) = \sum\limits_{k = 1}^\infty \frac{v_k^{(0,\Gamma)}(\x) v_k^{(0,\Gamma)}(\x_0)}{\mu_k^{(0,\Gamma)}}   \qquad (\x,\x_0\in\Gamma).
\end{equation}
Once the kernel $\G(\x|\x_0)$ is found, one can search for the
eigenvalues and eigenfunctions of the associated integral operator:
\begin{equation}   \label{eq:eigen_G}
\int\limits_{\Gamma} d\x \, \G(\x|\x_0) v_k^{(0,\Gamma)}(\x) = \frac{1}{\mu_k^{(0,\Gamma)}} v_k^{(0,\Gamma)}(\x_0)  
\qquad (\x_0 \in \Gamma, ~ k \geq 1),
\end{equation}
and thus determine the spectral properties of the Dirichlet-to-Neumann
operator $\M_0^{(\Gamma)}$, cf. Eq. (\ref{eq:Mp_eigenproblem}).  As
the constant eigenfunction $v_0^{(0,\Gamma)}$ is excluded from the sum
in Eq. (\ref{eq:G}), one has
\begin{equation}  \label{eq:Gintegral0}
\int\limits_{\Gamma} d\x \, \G(\x|\x_0) = 0  \qquad (\x_0\in \Omega).
\end{equation}
This property ensures that the eigenfunctions of the above integral
operator are orthogonal to a constant, as it should be.

As previously, once the eigenfunction $v_k^{(0,\Gamma)}$ is found on
$\Gamma$, it can be extended into $\Omega$.  As $V_0^{(0,\Gamma)} =
v_0^{(0,\Gamma)} = 1/\sqrt{|\Gamma|}$ is known, we focus on $k > 0$.
For this extension, one can multiply Eq. (\ref{eq:Helm}) with $p
= 0$ by $\G_0(\x|\x_0)$, multiply Eq. (\ref{eq:G0_problem_eq}) by
$V_k^{(0,\Gamma)}(\x)$, subtract them, integrate over $\x\in\Omega$,
and use the Green's formula and the boundary conditions to get
\begin{align} \label{eq:Vk_p0}
V_k^{(0,\Gamma)}(\x_0) = b_k^{(\Gamma)} + \mu_k^{(0,\Gamma)} \int\limits_{\Gamma} d\x \, \G_0(\x|\x_0)\, v_k^{(0,\Gamma)}(\x) 
 \qquad (\x_0\in \Omega),
\end{align}
where
\begin{equation}  \label{eq:Vk0_int}
b_k^{(\Gamma)} = \frac{1}{|\Omega|}\int\limits_{\Omega} d\x \, V_k^{(0,\Gamma)}(\x)  \qquad (k = 1,2,\ldots).
\end{equation}
Expectedly, the pseudo-Green's function $\G_0(\x|\x_0)$ determines
$V_k^{(0,\Gamma)}$ up to an additive constant $b_k^{(\Gamma)}$, which
can be fixed from the values of $v_k^{(0,\Gamma)}$ on $\Gamma$.  For
this purpose, we first substitute the expansion (\ref{eq:V0_p0}) into
the boundary value problem (\ref{eq:SteklovN}) to show that the
first-order correction $W_0^{(\Gamma)}(\x)$ of the principal
eigenfunction satisfies
\begin{subequations}  \label{eq:W0_problem}
\begin{align}  \label{eq:W0_problem_1}
\Delta W_0^{(\Gamma)}(\x) & = \frac{1}{|\Omega|}  \quad (\x\in\Omega), \\
\partial_n W_0^{(\Gamma)} & = \frac{{\mathbb I}_{\Gamma}(\x)}{|\Gamma|} \quad (\x\in\pa),
\end{align}
\end{subequations}
where ${\mathbb I}_{\Gamma}(\x)$ is the indicator function of
$\Gamma$, which is equal to $1$ for $\x\in\Gamma$ and $0$ otherwise.
The solution of this problem is determined up to an additive constant,
which can be fixed by using Eq. (\ref{eq:w0_int}).  Equivalently,
substituting Eqs. (\ref{eq:mu0v0_p0}) into both sides of the identity
(\ref{eq:dmu_dp}) at $k = 0$, we get
\begin{align} \nonumber
\int\limits_{\Omega} |V_0^{(p,\Gamma)}|^2 & = \frac{|\Omega|}{|\Gamma|} 
+ p \frac{2|\Omega|}{D|\Gamma|} \int\limits_{\Omega} W_0^{(\Gamma)} + O(p^2) \\   \label{eq:auxil11}
& = \frac{|\Omega|}{|\Gamma|} - 2 \a_\Gamma \frac{|\Omega|^2}{D |\Gamma|} p + O(p^2) = D \partial_p \mu_0^{(p,\Gamma)} ,
\end{align}
from which
\begin{equation}  \label{eq:W0_int}
\frac{1}{|\Omega|} \int\limits_{\Omega} W_0^{(\Gamma)} = - \a_\Gamma \,,
\end{equation}
that fixes the additive constant.

In order to fix $b_k^{(\Gamma)}$, one can multiply Eq. (\ref{eq:Helm})
with $p = 0$ by $W_0^{(\Gamma)}(\x)$, multiply
Eq. (\ref{eq:W0_problem_1}) by $V_k^{(0,\Gamma)}(\x)$, subtract them,
integrate over $\x\in\Omega$, and use the Green's formula and the
boundary conditions:
\begin{equation}  \label{eq:bk_w0}
b_k^{(\Gamma)} = - \frac{1}{\mu_k^{(0,\Gamma)}}\int\limits_{\Gamma} d\x \, w_0^{(\Gamma)}(\x) \, v_k^{(0,\Gamma)}(\x) \qquad (k = 1,2,\ldots).
\end{equation}
Note also that the integral of Eq. (\ref{eq:Helm}) with $p = 0$ over
$\x\in \Omega$ yields the identity
\begin{equation}
\frac{1}{|\Omega|} \int\limits_{\Omega} d\x \, V_k^{(p,\Gamma)} 
= \frac{D \mu_k^{(p,\Gamma)}}{p |\Omega|} \int\limits_{\Gamma} d\x \, v_k^{(p,\Gamma)}   \qquad (p > 0, ~ k = 0,1,\ldots).
\end{equation}
In the limit $p\to 0$, the left-hand side approaches $b_k^{(\Gamma)}$,
yielding an alternative interpretation of $b_k^{(\Gamma)}$ as
\begin{equation}  \label{eq:wk_def}
b_k^{(\Gamma)} = \frac{D}{|\Omega|} \lim\limits_{p\to 0} \frac{\mu_k^{(p,\Gamma)}}{p} \int\limits_{\Gamma}  v_k^{(p,\Gamma)}   \qquad (k = 1,2,\ldots)
\end{equation}
(when the associated eigenvalue $\mu_k^{(0,\Gamma)}$ is degenerate,
one needs to keep following the $k$-th analytic branch).

\subsection{Role of the first-order corrections}
\label{sec:role}

According to Eq. (\ref{eq:G_def}), the kernel $\G(\x|\x_0)$ is
determined by the pseudo-Green's function $\G_0(\x|\x_0)$, which was
thoroughly investigated in the mathematical and physical literature
(see
\cite{Kolokolnikov05,Condamin07,Cheviakov11,Giuggioli20} and
references therein).  However, Eq. (\ref{eq:G_def}) includes two other
contributions, namely, the first-order corrections $\a_\Gamma$ and
$\w_0^{(\Gamma)}$ to the principal eigenvalue and eigenfunction of
$\M_p^{(\Gamma)}$ as $p\to 0$.  In other words, our construction of
the eigenmodes of the operator $\M_0^{(\Gamma)}$ is not limited to the
case $p = 0$ but also requires the knowledge of spectral properties of
$\M_p^{(\Gamma)}$ in the vicinity of $p \approx 0$.  As most former
works dealt directly with the conventional problem at $p = 0$, the
role of the first-order corrections $\a_\Gamma$ and $w_0^{(\Gamma)}$
seems to be overlooked.  In this section, we briefly discuss their
properties.

The correction $w_0^{(\Gamma)}$ was formally introduced in
Eq. (\ref{eq:W0_def}) via the derivative $\partial_p v_0^{(p,\Gamma)}$
at $p=0$ but it could also be obtained by solving the boundary value
problem (\ref{eq:W0_problem}), with the condition (\ref{eq:w0_int}).
Curiously, the function $W_0^{(\Gamma)}$ turns out to be related to
the constant-flux approximation $T_q^{(\rm app)}(\x_0)$ of the MFRT
$T_q(\x_0)$
\cite{Grebenkov17,Grebenkov19d,Grebenkov17g,Grebenkov18a,Grebenkov21g},
which can be written as
\begin{equation}  \label{eq:Tapp_W0}
T_q^{(\rm app)}(\x_0) = \frac{|\Omega|}{D} \biggl(\frac{1}{q|\Gamma|}
- W_0^{(\Gamma)}(\x_0)\biggr)  \qquad (q > 0, ~ \x_0\in \overline{\Omega}).
\end{equation}
One can easily check that this function satisfies for any $q > 0$:
\begin{subequations}
\begin{align}  \label{eq:Tapp_Poisson}
D \Delta T_q^{(\rm app)}(\x_0) & = - 1 \quad (\x_0\in\Omega), \\  \label{eq:Tapp_BC}
-\partial_n T_q^{(\rm app)} & = \frac{|\Omega|}{D|\Gamma|} \, \I_{\Gamma}(\x_0)  \quad (\x_0 \in \pa), \\ \label{eq:Tapp_closure}
qD\int\limits_\Gamma d\x_0 \, T_q^{(\rm app)} & = |\Omega| .
\end{align}
\end{subequations}
In turn, the MFRT $T_q(\x_0)$ satisfies (see
Sec. \ref{sec:application} for more details)
\begin{subequations}  \label{eq:Tq_problem}
\begin{align}   \label{eq:Tq_problem_1}
D \Delta T_q(\x_0) & = - 1 \quad (\x_0\in\Omega), \\  \label{eq:Tq_problem_2}
- \partial_n T_q & = q \, T_q(\x_0)\,  \I_{\Gamma}(\x_0)  \quad (\x_0 \in \pa).
\end{align}
\end{subequations}
One sees that the constant-flux approximation consists in replacing
the Robin boundary condition on the subset $\Gamma$, $-\partial_n T_q
= q \,T_q$, by the constant-flux condition $-\partial_n T_q^{\rm
(app)} = |\Omega|/(D|\Gamma|)$.  In addition, the self-consistent
closure relation (\ref{eq:Tapp_closure}) is imposed to fix the
constant and thus to get the unique solution.  We return to the
approximation (\ref{eq:Tapp_W0}) in Sec. \ref{sec:application}.

We also highlight the versatile roles of the constant $\a_\Gamma$
emerging in different contexts.  This constant was introduced in
Eq. (\ref{eq:a_def}) via the second derivative of the principal
eigenvalue $\mu_0^{(p,\Gamma)}$, evaluated at $p = 0$.  In turn,
Eq. (\ref{eq:a_eps}) gives its representation as the double integral
of the pseudo-Green's function $\G_0(\x|\x_0)$ over the subset
$\Gamma$.  For the conventional Steklov problem (with $\Gamma =
\pa$), this relation was earlier recognized in \cite{Grebenkov19c}.
In this reference, the constant $\a_\Gamma$ was shown to determine the
long-time behavior of the variance of the boundary local time, which
is the proxy of the number of encounters between a particle and the
boundary.  In Appendix \ref{sec:variance}, we extend these former
results to our setting of a given subset $\Gamma$ and get the
long-time asymptotic behavior of the variance of the boundary local
time $\ell_t$ on $\Gamma$:
\begin{equation}  \label{eq:variance}
\mathrm{Var}\{ \ell_t \} \approx  \a_\Gamma \frac{2|\Gamma|^2}{|\Omega|} \, D t + O(1)  \qquad (t\to \infty).
\end{equation}
Finally, we will show in Sec. \ref{sec:application} that $\a_\Gamma$
also controls the asymptotic behavior of the MFPT to $\Gamma$ in the
small-target limit.
In this limit, the correction $\a_\Gamma$ admits a universal
asymptotic behavior (see Sec. \ref{sec:limit}).

We conclude this section by noting that Dittmar used the Neumann
function $\N(\x|\x_0)$ to analyze the Steklov eigenfunctions for the
conventional case $p = 0$ and $\Gamma = \pa$ in planar domains
\cite{Dittmar04}.  For any $\x_0\in \Omega$, this function is defined
as the unique solution of the boundary value problem:
\begin{subequations}
\begin{align}  
- \Delta \N(\x|\x_0) & = \delta(\x-\x_0)  \quad (\x\in\Omega), \\ \label{eq:N_BC}
\partial_n \N & = -\frac{1}{|\pa|} \quad (\x\in \pa), \\
\int\limits_{\pa} d\x \, \N(\x|\x_0) & = 0 ,
\end{align}
\end{subequations}
and thus differs from the pseudo-Green's function $\G_0(\x|\x_0)$
satisfying Eqs. (\ref{eq:G0_problem}, \ref{eq:G0_constant}).  Dittmar
showed that the restriction of $\N(\x|\x_0)$ to $\pa \times \pa$ reads
\begin{equation}  \label{eq:Neumann}
\N(\x|\x_0) = \sum\limits_{k = 1}^\infty \frac{v_k^{(0,\pa)}(\x) \, v_k^{(0,\pa)}(\x_0)}{\mu_k^{(0,\pa)}}  \quad (\x,\x_0\in\pa),
\end{equation}
and therefore allows one to compute the eigenvalues and eigenfunctions
of the Dirichlet-to-Neumann operator $\M_0^{(\pa)}$ on the boundary
$\pa$.  In this setting, the restriction of $\N(\x|\x_0)$ to the
boundary $\pa$ is precisely our kernel $\G(\x|\x_0)$ defined in
Eq. (\ref{eq:G}).  An explicit form of $\N(\x|\x_0)$ for the interior
and exterior of a sphere is well known \cite{Hitotumatu54} (see also
\cite{McCann01} for rectangular domains), whereas some general
properties of the Neumann function and its applications in electrical
impedance imaging were discussed in \cite{Friedman89}.

The Dittmar's approach can be extended to deal with the mixed
Steklov-Neumann problem by modifying the definition of the Neumann
function, namely, by replacing $1/|\pa|$ by ${\mathbb
I}_\Gamma(\x)/|\Gamma|$ in the Neumann boundary condition
(\ref{eq:N_BC}).  One can easily get the spectral expansion of the
modified function, denoted as $\N^{(\Gamma)}(\x|\x_0)$:
\begin{align} \label{eq:N_whole}
\N^{(\Gamma)}(\x|\x_0) & = D\tilde{G}_\infty(\x,0|\x_0) 
 + \sum\limits_{k = 1}^\infty \frac{V_k^{(0,\Gamma)}(\x) V_k^{(0,\Gamma)}(\x_0)}{\mu_k^{(0,\Gamma)}}   \qquad (\x,\x_0\in\Omega),
\end{align}
so that its restriction to $\Gamma \times \Gamma$ becomes again
identical to the kernel $\G(\x|\x_0)$ and thus allows the construction
of the eigenmodes of the Dirichlet-to-Neumann operator
$\M_0^{(\Gamma)}$.  However, the major drawback of this construction
is that the modified Neumann function $\N^{(\Gamma)}(\x|\x_0)$ depends
on $\Gamma$, whereas the pseudo-Green's function $\G_0(\x|\x_0)$ that
we used, does not.  In summary, both the modified Neumann function and
the pseudo-Green's function can be employed but the latter is
independent of $\Gamma$ and thus simpler.

\section{Asymptotic results}
\label{sec:limit}

In this section, we present our main asymptotic results in the
small-target limit $\Gamma \to \emptyset$.  In Sec. \ref{sec:generic},
we employ a dilation argument to reduce the original Steklov-Neumann
problem (\ref{eq:SteklovN}) for a small subset $\Gamma$ to an
auxiliary Steklov-Neumann problem, either for an interval $(-1,1)$ in
the half-plane ($d = 2$), or for the unit disk in the half-space ($d =
3$).  We recall the spectral properties of these two auxiliary
problems.  In Sec. \ref{sec:small_asympt}, we reformulate these two
problems in terms of integral operators and compute explicitly their
kernels.  We also describe the asymptotic behavior of the corrections
$\a_\Gamma$ and $w_0^{(\Gamma)}(\x)$ and of the coefficients
$b_k^{(\Gamma)}$.  The role of boundary connectivity is discussed in
Sec. \ref{sec:connectivity}.

\subsection{Dilation argument and two auxiliary Steklov-Neumann problems}
\label{sec:generic}

\begin{figure}
\begin{center}
\includegraphics[width=85mm]{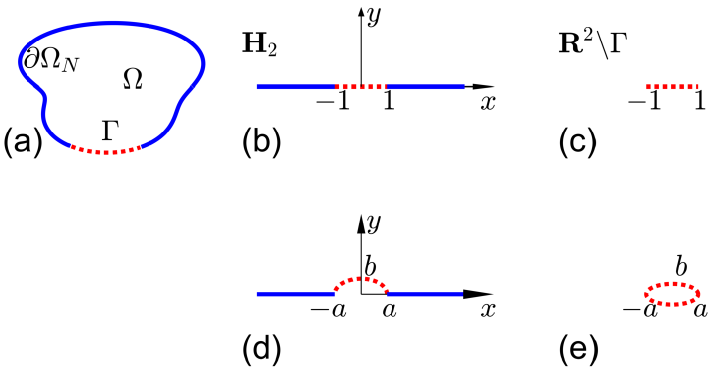} 
\end{center}
\caption{
In the small-target limit, zooming in near the subset $\Gamma$
transforms the original Steklov-Neumann problem in a bounded domain
{\bf (a)} into the auxiliary spectral problem
(\ref{eq:Steklov_interval}) in the upper half-plane $\H_2 = \R \times
\R_+$ {\bf (b)}, with Steklov condition on the interval $(-1,1)$
(dashed red line) and Neumann condition on the remaining horizontal
axis (solid blue line).  The latter is equivalent to the Steklov
problem in the exterior of the interval $(-1,1)$ in the plane $\R^2$
{\bf (c)}.  The last two problems can be seen as the limiting cases
($b = 0$) of two equivalent problems for the exterior of an ellipse
with semi-axes $a = 1$ and $b$ in the plane {\bf (e)} and for the
exterior of a half-ellipse in the upper half-plane {\bf (d)}. }
\label{fig:scheme_ellipse}
% A_DN_disk_mixed_scheme_ellipse3();
\end{figure}

In two dimensions, we consider $\Gamma \subset \pa$ to be a
connected subset of perimeter $2\epsilon$ (i.e., a smooth curve of
arbitrary shape).  In three dimensions, we focus on the particular
case when the subset $\Gamma$ is a (curved) disk of radius $\epsilon$
(an extension to other shapes of $\Gamma$ will be mentioned below).
We are interested in the limit $\epsilon \to 0$; in practice, the
derived results are applicable when the size $\epsilon$ of the subset
$\Gamma$ is the smallest length of the problem.  In particular, we
require that
\begin{equation}  \label{eq:curvature}
\epsilon \ll \frac{1}{\max\limits_{\x\in\pa} \{ H(\x)\}} \,, 
\end{equation}
where $H(\x)$ is the mean curvature of the boundary at the point $\x
\in\pa$.  Moreover, the reflecting part of the boundary, $\pa_N$,
should not appear close to the center of the subset $\Gamma$ (more
formally, we require that the radius of the largest ball that is
inscribed into $\Omega$ and touches any point of $\Gamma$, is much
larger than $\epsilon$).

Denoting by $\x_\Gamma \in \pa$ the center of the subset $\Gamma$, we
rescale the coordinates as $\x \to \hat{\x} = (\x -
\x_\Gamma)/\epsilon$ that dilates $\Omega$, $\pa$, and $\Gamma$ into
$\hat{\Omega}$, $\partial\hat{\Omega}$, and $\hat{\Gamma}$, and
transforms the original Steklov-Neumann problem (\ref{eq:SteklovN}) to
\begin{subequations}  \label{eq:SteklovNbis}
\begin{align} 
(\epsilon^2 p/D - \Delta) V_k^{(p,\Gamma)}(\hat{\x}) & = 0 \quad (\hat{\x} \in \hat{\Omega}), \\ 
\partial_n V_k^{(p,\Gamma)}(\hat{\x}) & = \epsilon \mu_k^{(p,\Gamma)} V_k^{(p,\Gamma)}(\hat{\x}) \quad (\hat{\x} \in \hat{\Gamma}), \\
\partial_n V_k^{(p,\Gamma)}(\hat{\x}) & = 0 \quad (\hat{\x} \in \partial\hat{\Omega}\backslash \overline{\hat{\Gamma}}). 
\end{align}
\end{subequations}
In the limit $\epsilon \to 0$, the domain $\hat{\Omega}$ blows up and
can be substituted by the half-plane or the half-space, whereas
$\partial \hat{\Omega}$ is substituted by the horizontal axis or the
horizontal plane (an additional rotation may be needed to choose a
suitable orientation of the coordinates).  In turn, the size of the
rescaled subset $\hat{\Gamma}$ remains unchanged, but it becomes
flatter as $\epsilon\to 0$, and can thus be substituted either by the
interval $(-1,1)$ lying on the horizontal axis
(Fig. \ref{fig:scheme_ellipse}b), or by the unit disk lying on the
horizontal plane.  The condition (\ref{eq:curvature}) ensures that the
curvature of $\Gamma$ can be neglected in the leading order.
Moreover, for a fixed $p \geq 0$, the term $\epsilon^2 p/D$ vanishes
in the limit $\epsilon\to 0$.
As a consequence, the rescaled problem (\ref{eq:SteklovNbis}) can be
formally substituted by an auxiliary Steklov-Neumann problem:

(i) in two dimensions, one searches for the eigenpairs $\{\hat{\mu}_k,
\hat{V}_k(x,y)\}$, satisfying in the half-plane $\H_2 = \R\times
\R_+$:
\begin{subequations}  \label{eq:Steklov_interval}
\begin{align}   \label{eq:Steklov_interval_1}
\Delta \hat{V}_k &= 0 \qquad (y > 0), \\
- (\partial_y \hat{V}_k)_{y=0} & = \begin{cases} 0 \hskip 17mm (|x| \geq 1),\cr
\hat{\mu}_k \hat{V}_k(x,0) \quad (|x| < 1), \end{cases} \\  \label{eq:Vk2d_inf}
|\hat{\x}| \, |\nabla \hat{V}_k| & \to 0 \qquad (|\hat{\x}|\to \infty),
\end{align}
\end{subequations}
with $\hat{\x} = (x,y)$, and the normalization
\begin{equation}  \label{eq:Vk_2d_norm}
\int\limits_{-1}^1 dx \, |\hat{v}_k(x)|^2 = 1,
\end{equation}
where $\hat{v}_k(x) = \hat{V}_k(x,0)$ is the restriction of
$\hat{V}_k(\hat{\x})$ to the interval $(-1,1)$.

(ii) in three dimensions, one searches for the eigenpairs
$\{\hat{\mu}_k, \hat{V}_k(x,y,z)\}$, satisfying in the half-space
$\H_3 = \R^2\times \R_+$:
\begin{subequations}  \label{eq:Steklov_disk}
\begin{align}   \label{eq:Steklov_disk_1}
\Delta \hat{V}_k & = 0 \qquad (z > 0), \\
- (\partial_z \hat{V}_k)_{z=0} & = \begin{cases} 0 \hskip 20.5mm (x^2+y^2 \geq 1),\cr
\hat{\mu}_k \hat{V}_k(x,y,0) \quad (x^2+y^2 < 1), \end{cases} \\ \label{eq:Vk3d_inf} 
|\hat{\x}|^2 \, |\nabla \hat{V}_k| & \to 0 \qquad (|\hat{\x}|\to \infty),
\end{align}
\end{subequations}
with $\hat{\x} = (x,y,z)$, and the normalization
\begin{equation}  \label{eq:Vk_3d_norm}
\int\limits_{\Gamma_1} d\hat{\x} \, |\hat{v}_k(\hat{\x})|^2 = 1,
\end{equation}
where $\hat{v}_k(\hat{\x})$ is the restriction of
$\hat{V}_k(\hat{\x})$ to the unit disk $\Gamma_1$ on the horizontal
plane $z = 0$.

Even though the domains $\H_2$ and $\H_3$ are unbounded, the subset
with the Steklov condition (interval or disk) is bounded, so that the
spectrum of each auxiliary Steklov-Neumann problem is discrete (see
\cite{Henrici70,Troesch72,Miles72,Fox83,Kozlov04} for details).  In
particular, the eigenvalues $\hat{\mu}_k$ can be ordered as:
\begin{equation}
0 = \hat{\mu}_0 < \hat{\mu}_1 \leq \hat{\mu}_2 \leq \cdots \nearrow +\infty ,
\end{equation}
whereas the eigenfunctions $\{\hat{v}_k\}$ form an orthonormal
complete basis of the $L^2$ space on the interval $(-1,1)$ ($d = 2$)
or the unit disk ($d = 3$).  Note that the condition
(\ref{eq:Vk3d_inf}) was formulated in \cite{Henrici70} to ensure that
the eigenfunctions $\{\hat{v}_k\}$ are orthogonal to a constant
$\hat{v}_0 \propto 1$.  We emphasize that the conditions
(\ref{eq:Vk2d_inf}, \ref{eq:Vk3d_inf}) allow for $\hat{V}_k(\hat{\x})$
to have a constant nonzero limit $\hat{V}_k(\infty)$ as $|\hat{\x}|\to
\infty$.  These limiting values will play an important role in the
following asymptotic analysis.

Comparing the rescaled problem (\ref{eq:SteklovNbis}) to the auxiliary
problems (\ref{eq:Steklov_interval}, \ref{eq:Steklov_disk}), one gets
immediately the asymptotic relations (\ref{eq:main_asympt}).  Note
that the factor $\epsilon^{-(d-1)/2}$ provides the correct rescaling
of the normalization conditions (\ref{eq:Vk_2d_norm},
\ref{eq:Vk_3d_norm}) to ensure the $L^2(\Gamma)$ normalization
(\ref{eq:L2norm}) of the eigenfunctions $v_k^{(p,\Gamma)}$.  We stress
that the above dilation argument provides an asymptotic approximation
of the eigenvalues $\mu_k^{(p,\Gamma)}$ and of the eigenfunctions
$V_k^{(p,\Gamma)}(\x)$ {\it on or near} the subset $\Gamma$.  In turn,
the behavior of $V_k^{(p,\Gamma)}(\x)$ far from $\Gamma$ (i.e., when
the Euclidean distance $|\x - \Gamma|$ is large) is not well captured
by $\hat{V}_k(\hat{\x})$.
We also emphasize that the dilation argument is not a rigorous proof;
in particular, we do not discuss the convergence of eigenfunctions,
nor a more subtle situation of degenerate eigenvalues, in which case
eigenprojectors are needed.  A rigorous reformulation and validation
of this formal analysis presents an interesting perspective of this
work.

Quite remarkably, the asymptotic relations (\ref{eq:main_asympt})
imply that the eigenpairs $\{\mu_k^{(p,\Gamma)}, v_k^{(p,\Gamma)}\}$
do not depend on a fixed parameter $p$ in the leading order as
$\epsilon\to 0$.  In practice, if $\epsilon$ is fixed, the above
dilation argument is valid when $\epsilon^2 p/D \ll 1$.  In other
words, the asymptotic relations (\ref{eq:main_asympt}) provide an
approximation when $0 \leq p \ll D/\epsilon^2$.  Since $\sqrt{D/p}$ is
a lengthscale, this condition is consistent with our earlier claim
that $\epsilon$ should be the smallest length of the problem.

In three dimensions, the axial symmetry of the Steklov-Neumann
problem (\ref{eq:Steklov_disk}) implies that its eigenfunctions
$\hat{V}_k$ can be searched in the cylindrical coordinates
$(r,z,\varphi)$ as $[c_{m,n}^1 \cos(m\varphi) + c_{m,n}^2
\sin(m\varphi)] \hat{V}_{m,n}(r,z)$, for any $m = 0,1,2,\ldots$, with
arbitrary constants $c_{m,n}^{1}$ and $c_{m,n}^{2}$.  It is therefore
convenient to employ the double index $(m,n)$ to account for the
$2\pi$-periodicity of an eigenfunction on the angle $\varphi$, and the
index $n = 0,1,2,\ldots$ to distinguish different eigenfunctions
within the $m$-th family.
Accordingly, the associated eigenvalues are denoted as
$\hat{\mu}_{m,n}$.  An extensive numerical analysis from
\cite{Grebenkov24} for spheroids suggested that the eigenvalues
$\hat{\mu}_{0,n}$ are simple, whereas $\hat{\mu}_{m,n}$ are twice
degenerate for $m > 0$, in agreement with the above linear combination
of sine and cosine functions.
Re-ordering the ensemble of all eigenvalues $\{\hat{\mu}_{m,n}\}$ into
an increasing sequence, one can recover the former enumeration by a
single index $k$; in other words, for each $k = 0,1,2,\ldots$, there
exist indices $m_k$ and $n_k$ such that $\hat{\mu}_k =
\hat{\mu}_{m_k,n_k}$ and
\begin{equation}  \label{eq:double_index}
\hat{V}_k \propto \bigl[c_{m_k,n_k}^1 \cos(m_k\varphi) + c_{m_k,n_k}^2
\sin(m_k\varphi) \bigr] \hat{V}_{m_k,n_k}(r,z).
\end{equation}
In the following, we will use interchangeably both the double index
$(m,n)$ and the single index $k$.

There are different ways to construct numerically the eigenpairs
$\{\hat{\mu}_k, \hat{V}_k\}$.  For instance, the interval $(-1,1)$ in
the plane $\R^2$ can be seen as the limit $b\to 0$ of ellipses with
semi-axes $1$ and $b$ (Fig. \ref{fig:scheme_ellipse}e), so that
solutions of the spectral problem (\ref{eq:Steklov_interval}) can be
obtained by inspecting the Steklov problem for the exterior of an
ellipse (see Appendix \ref{sec:ellipse}).  Similarly, the unit disk in
$\R^3$ can be seen as the limit of oblate spheroids, allowing one to
construct solutions of the spectral problem (\ref{eq:Steklov_disk}) by
looking at the Steklov problem for the exterior of spheroids (see
Appendix \ref{sec:spheroid} and \cite{Grebenkov24}).  In turn, the
next section presents an alternative way to compute the eigenpairs
$\{\hat{\mu}_k, \hat{V}_k\}$, which is based on the asymptotic
behavior of the kernel $\G(\x|\x_0)$ and the limit of the integral
equation (\ref{eq:eigen_G}) (a similar approach was used in
\cite{Henrici70}).  This alternative computation re-enforces the
dilation argument described above.

We conclude that the asymptotic relations (\ref{eq:main_asympt})
do not depend on the shape of the domain $\Omega$ (see also
\cite{Bonnetier22} for a proper justification of this statement in the
related context of the narrow-escape problem).
In contrast, the shape of the subset $\Gamma$ is relevant, as it
defines the eigenpairs of the Steklov-Neumann problem
(\ref{eq:Steklov_disk}) in the half-space.  For instance, if the
subset $\Gamma$ had an elliptic shape, the dilation argument would
still work but the asymptotic behavior would be determined by the
eigenpairs of an auxiliary Steklov-Neumann problem for an ellipse in
the half-space.

\begin{table} 
\begin{center}
\begin{tabular}{|c|c|c|c|c|c|c|} \hline
\multirow{6}{3mm}{\begin{turn}{90}\centering 2D \end{turn}}
   & $k$                      & 1      & 2       & 3      & 4       & 5       \\ 
   & $\hat{\mu}_k$            & 2.0061 & 3.4533  & 5.1253 & 6.6286  & 8.2600  \\ 
   & $[\hat{V}_k(\infty)]^2$  & 0      & 0.0664  & 0      & 0.0391  & 0       \\ \cline{2-7}
   & $k$                      & 6       & 7      &  8      & 9      & 10      \\ 
   & $\hat{\mu}_k$            & 9.7839  & 11.398 & 12.933  & 14.538 & 16.079 \\  
   & $[\hat{V}_k(\infty)]^2$  & 0.0279  & 0      & 0.0217  &  0     & 0.0178 \\ \hline \hline
%%%%%%%%%  3D  %%%%%%%%%%
\multirow{6}{3mm}{\begin{turn}{90}\centering 3D \end{turn}}
   & $k$                      & 1      & 2       & 3      & 4       & 5       \\ 
   & $\hat{\mu}_{0,k}$        & 4.1213 & 7.3421 & 10.517 & 13.677 & 16.831   \\  
   & $[\hat{V}_{0,k}(\infty)]^2$& 0.0380 & 0.0249 & 0.0187 & 0.0150 & 0.0125 \\  \cline{2-7}
   & $k$                      & 6       & 7      &  8      & 9      & 10      \\ 
   & $\hat{\mu}_{0,k}$        &  19.981 & 23.128 & 26.275 & 29.420 & 32.565  \\  
   & $[\hat{V}_{0,k}(\infty)]^2$ & 0.0108 & 0.0095 & 0.0084 & 0.0076 & 0.0069  \\ \hline
\end{tabular}
\end{center}
\caption{
{\bf (Top six rows)} First ten eigenvalues $\hat{\mu}_k$ of the
Steklov-Neumann problem (\ref{eq:Steklov_interval}), apart from
$\hat{\mu}_0 = 0$, as well as the squared coefficients
$[\hat{V}_k(\infty)]^2$ determining the asymptotic behavior of
$b_k^{(\Gamma)}$ and the function $\Psi_{\rm 2d}(z)$ from
Eq. (\ref{eq:Tq_Tinf_2d}).  Note that $1/\hat{\mu}_k$ are the
eigenvalues of the kernel $\hat{\G}(x|x_0)$ from
Eq. (\ref{eq:Glim_disk}).  They were obtained by a numerical
diagonalization of the matrix $\MM(0)$ from Eq. (\ref{eq:M_ellipse})
of size $200\times 200$, see Appendix \ref{sec:ellipse}.  {\bf (Bottom
six rows)} First ten eigenvalues $\hat{\mu}_{0,k}$ corresponding to
axially symmetric eigenfunctions of the Steklov-Neumann problem
(\ref{eq:Steklov_disk}), apart from $\hat{\mu}_{0,0} = 0$, as well as
the squared coefficients $[\hat{V}_{0,k}(\infty)]^2$ determining the
asymptotic behavior of $b_{0,k}^{(\Gamma)}$ and the function
$\Psi_{\rm 3d}(z)$ from Eq. (\ref{eq:Tq_Tinf_3d}).  They were obtained
by a numerical diagonalization of the matrix $\GG$ from
Eq. (\ref{eq:GG_disk_upper}) of size $100\times 100$; note that only
even eigenmodes were kept, see Appendix \ref{sec:spheroid}.  For all
data, the shown digits do not change as the truncation size increases.
These eigenvalues are in excellent agreement with those reported in
\cite{Henrici70}, see Tables 6.1 and 6.2. }
\label{tab:eta_2D}
%   [G,mu,V] = A_DN_disk_mixed_G_Cheb2(30);   mu(1:10)'
% [bk, vk, y, muint] = A_DN_ellipse_int_bk;
%%%  A_DN_disk_ext_bk;
\end{table}

\subsection{Asymptotic behavior of the pseudo-Green's function and related quantities}
\label{sec:small_asympt}

As in Sec. \ref{sec:theory}, the auxiliary Steklov-Neumann
problems (\ref{eq:Steklov_interval}, \ref{eq:Steklov_disk}) can be
reformulated as spectral problems for suitable integral operators,
like Eq. (\ref{eq:eigen_G}).  For this purpose, one needs to determine
the asymptotic form of the kernel $\G(\x|\x_0)$ from
Eq. (\ref{eq:G_def}).  Given that both $\x$ and $\x_0$ belong to a
small subset $\Gamma$ and thus are close to each other, one can use
the well-known asymptotic behavior of the pseudo-Green's function
$\G_0(\x|\x_0)$ as $\x\to \x_0$.  Re-delegating technical computations
to Appendix \ref{sec:small}, we summarize the main results.

\subsubsection{Two dimensions}

When $\x_0\in\pa$, the pseudo-Green's function in the limit
$\x\to\x_0$ behaves as (see \cite{Kolokolnikov05} and references
therein)
\begin{equation}  \label{eq:Gpseudo_2d_asympt}
\G_0(\x|\x_0) = -\frac{1}{\pi} \ln |\x-\x_0| + R_0(\x_0) + o(1) \qquad (\x\to\x_0 \in \pa),
\end{equation}
where the regular part $R_0(\x_0)$ accounts for the contribution of
order $O(1)$, while $o(1)$ denotes higher-order corrections vanishing
as $\x\to \x_0$.  We outline the minor abuse of notation when the
dimensional distance $|\x-\x_0|$ stands in the argument of the
logarithmic function; a more accurate expression would be
$\G_0(\x|\x_0) = -\frac{1}{\pi} \ln (|\x-\x_0|/\ell) +
\bar{R}_0(\x_0) + o(1)$ with a suitable length $\ell$, but we keep
using (\ref{eq:Gpseudo_2d_asympt}) in the following.
Substituting this expression into Eqs. (\ref{eq:a_eps},
\ref{eq:omega_eps}) yields (see details in Appendix \ref{sec:small}):
\begin{align}  \label{eq:aGamma_asympt_2d}
\a_\Gamma & \approx - \frac{\ln (2\epsilon)}{\pi} + \frac{3}{2\pi} + R_0(\x_\Gamma) + o(1), \\  \label{eq:hatw0_asympt_2d}
w_0^{(\Gamma)}(\x_0) & \approx \hat{w}_0(s_0/\epsilon) \qquad (\x_0\in \Gamma),
\end{align}
where $\x_\Gamma \in \pa$ is the center of the subset $\Gamma$, $s_0$
is the curvilinear coordinate of $\x_0 \in \Gamma$ (ranging from
$-\epsilon$ to $\epsilon$), and
\begin{equation}  \label{eq:hatw0_2d}
\hat{w}_0(x) = - \frac{(1+x)\ln(1+x) + (1-x)\ln(1-x) + 1 - 2\ln 2}{2\pi} \,.
\end{equation}
Moreover, the coefficients $b_k^{(\Gamma)}$ from
Eq. (\ref{eq:Vk0_int}) behave in the leading order as
\begin{equation}  \label{eq:bk_2d}
b_k^{(\Gamma)} \approx \epsilon^{-1/2}\, \hat{V}_k(\infty)  ,
\end{equation}
where the factor $\epsilon^{-1/2}$ agrees with the rescaling of
eigenfunctions in Eq. (\ref{eq:main_asympt}), and $\hat{V}_k(\infty)$
is the limit of $\hat{V}_k(\hat{\x})$ as $|\hat{\x}|\to \infty$.  The
first ten eigenvalues $\hat{\mu}_k$ and the associated limits
$\hat{V}_k(\infty)$ are reported in Table
\ref{tab:eta_2D}.

Substituting Eqs. (\ref{eq:aGamma_asympt_2d},
\ref{eq:hatw0_asympt_2d}) into Eq. (\ref{eq:G_def}), we get
\begin{equation}
\G(\x|\x_0) \approx \hat{\G}(s/\epsilon|s_0/\epsilon) + o(1),
\end{equation}
where $s$ and $s_0$ are the curvilinear coordinates of $\x$ and $\x_0$
on the subset $\Gamma$, and 
\begin{equation}  \label{eq:Glim_disk}
\hat{\G}(x|x_0) = -\frac{1}{\pi} \ln |x-x_0| + \frac{1}{2\pi} \hat{\G}_{\rm sym}(x|x_0),
\end{equation}
with
\begin{align} \nonumber
\hat{\G}_{\rm sym}(x|x_0) & = (1+x_0)\ln(1+x_0) + (1-x_0)\ln(1-x_0) \\
& + (1+x)\ln(1+x) + (1-x)\ln(1-x) - 1 - 2\ln(2) .
\end{align} % Checked, see A_DN_disk_mixed_Gkernel_lim_check();
This explicit kernel defines an integral operator on $L^2((-1,1))$
that determines the eigenpairs of the Steklov-Neumann problem
(\ref{eq:Steklov_interval}):
\begin{equation}  \label{eq:eigen_disk_eps0}
\int\limits_{-1}^{1} dx \, \hat{v}_k(x)\, \hat{\G}(x|x_0) = \frac{1}{\hat{\mu}_k} \hat{v}_k(x_0) 
 \qquad (-1 < x_0 < 1, ~ k=1,2,\ldots)
\end{equation}
(note that the trivial eigenpair $\hat{\mu}_0 = 0$ and $\hat{v}_0 =
1/\sqrt{2}$ is not captured by this equation and should be added
separately).  If the eigenvalue $\hat{\mu}_k$ is simple, the symmetry
of the kernel, $\hat{\G}(-x|x_0) = \hat{\G}(x|-x_0)$, implies that the
associated eigenfunction $\hat{v}_k(x)$ is either symmetric, or
antisymmetric.  If an eigenfunction $\hat{v}_k(x)$ is antisymmetric,
its integral with the symmetric part $\hat{\G}_{\rm sym}(x|x_0)$ of
the kernel $\hat{\G}(x|x_0)$ vanishes.  In other words, the
antisymmetric eigenfunctions can be obtained by solving the simpler
spectral problem for the integral operator with the logarithmic
kernel:
\begin{equation}  \label{eq:eigen_disk_eps00}
\int\limits_{-1}^{1} dx \, \hat{v}_{2k-1}(x)\, \frac{1}{\pi} \ln \biggl(\frac{1}{|x-x_0|}\biggr) 
= \frac{\hat{v}_{2k-1}(x_0)}{\hat{\mu}_{2k-1}}    \qquad (-1 < x_0 < 1),
\end{equation}
where we used odd indices $2k-1$ to enumerate antisymmetric
eigenfunctions.
Such spectral problems were thoroughly studied in the past (see
\cite{Widom64,Simic06,Chen09,Mandal21,Polosin22} and references
therein).  In particular, the large-$k$ asymptotic behavior of the
eigenvalues $\hat{\mu}_{2k-1}$ of the antisymmetric eigenfunctions of
Eq. (\ref{eq:eigen_disk_eps00}) was given in \cite{Polosin22}.  In the
leading order, one has $\hat{\mu}_{2k-1} \approx \pi k + O(1)$.  While
all eigenvalues that we computed numerically were simple (see Table
\ref{tab:eta_2D}), a rigorous proof of this conjecture for all
eigenvalues remains an open question.  We also checked numerically
that the eigenvalues $\hat{\mu}_{2k}$ corresponding to the symmetric
eigenfunctions of the kernel $\hat{\G}(x|x_0)$ obey the same
asymptotic relation in the leading order.  We conclude that
\begin{equation}  \label{eq:etak_asympt_2d}
\hat{\mu}_k \approx \frac{\pi}{2} k + O(1)   \qquad (k\to\infty),
\end{equation}
which agrees with lower and upper bounds discussed in Appendix
\ref{sec:bounds}.

\subsubsection{Three dimensions}

When $\x_0\in\pa$, the pseudo-Green's function in the limit
$\x\to\x_0$ behaves as
\begin{equation}  \label{eq:G0_asympt_3d}
\G_0(\x|\x_0) = \frac{1}{2\pi |\x-\x_0|} - \frac{H(\x_0)}{4\pi} \ln |\x-\x_0| + R_0(\x_0) + o(1) 
\qquad (\x\to \x_0\in \pa),
\end{equation}
where $R_0(\x_0)$ is the regular part, and $H(\x_0)$ is the mean
curvature of the boundary at the point $\x_0 \in\pa$
\cite{Popov92,Silbergleit03,Singer08}.
Substituting this expression into Eqs. (\ref{eq:a_eps},
\ref{eq:omega_eps}) yields (see details in Appendix \ref{sec:small}):
\begin{align}  \label{eq:aGamma_asympt_3d}
\a_\Gamma & \approx \frac{8}{3\pi^2 \epsilon} - \frac{H(\x_\Gamma)}{4\pi} \biggl(\ln \epsilon - \frac14\biggr) + R_0(\x_\Gamma) + o(1) , \\
\label{eq:w0_asympt_3d}
w_0^{(\Gamma)}(\x_0) & \approx \frac{1}{2\pi \epsilon} \hat{w}_0\bigl( r_0/\epsilon \bigr) + O(1) \qquad (\x_0\in\Gamma), 
\end{align}
where $r_0 = |\x_0 - \x_\Gamma|$, 
\begin{equation}  \label{eq:hatw0_3d}
\hat{w}_0(\hat{r}) = \frac{4}{\pi} \biggl(E(\hat{r}) - \frac43\biggr),
\end{equation}  % [w0,w0lim,y] = A_DN_sphere_mixed_w0_fig(w0lim);
and $E(k)$ is the complete elliptic integral of the second kind:
\begin{equation}  \label{eq:EllipticE}
E(k) = \int\limits_0^{\pi/2} dz \, \sqrt{1 - k^2 \sin^2 z} \,.
\end{equation}
As a consequence, the coefficients $b_k^{(\Gamma)}$ from
Eq. (\ref{eq:Vk0_int}) behave as
\begin{equation} \label{eq:bk_3d}
b_k^{(\Gamma)} = \epsilon^{-1} \hat{V}_k(\infty) + O(1)  \qquad (\epsilon\to 0),
\end{equation}
where the factor $\epsilon^{-1}$ agrees with the rescaling of
eigenfunctions in Eq. (\ref{eq:main_asympt}).  If $\hat{V}_k$ is a
periodically oscillating (along the angle $\varphi$),
non-axially-symmetric eigenfunction, it vanishes at infinity so that
the corresponding $b_k^{(\Gamma)}$ behaves as $O(1)$.  The first ten
eigenvalues $\hat{\mu}_{0,n}$ and the limiting values
$\hat{V}_{0,n}(\infty)$ of the associated axially symmetric
eigenfunctions are reported in Table \ref{tab:eta_2D}.

Let us now deduce an alternative spectral problem for determining the
eigenpairs $\{\hat{\mu}_k, \hat{v}_k\}$.  Substituting
Eqs. (\ref{eq:aGamma_asympt_3d}, \ref{eq:w0_asympt_3d}) into
Eq. (\ref{eq:G_def}), we determine the asymptotic behavior of the
kernel $\G(\x|\x_0)$:
\begin{align}  \label{eq:G_3d_auxil}
\G(\x|\x_0) & = \frac{1}{2\pi|\x-\x_0|} - \frac{H(\x_\Gamma)}{4\pi} \ln|\x-\x_0|   \\  \nonumber  
& - \frac{\hat{w}_0(r_0/\epsilon)}{2\pi \epsilon} - \frac{\hat{w}_0(r/\epsilon)}{2\pi \epsilon}  
 - \frac{8}{3\pi^2 \epsilon} + \frac{H(\x_\Gamma)}{4\pi} \ln \epsilon  + O(1),
\end{align}
where we replaced $H(\x_0)$ by $H(\x_\Gamma)$ in
Eq. (\ref{eq:G0_asympt_3d}), up to a negligible error $O(\epsilon)$.
One sees that the leading-order terms of the kernel $\G(\x|\x_0)$
respect the rotational symmetry of the disk and can thus be decomposed
in the cylindrical coordinates $\x = (r,0,\varphi)$ and $\x_0 =
(r_0,0,\varphi_0)$ as
\begin{equation}  \label{eq:G_Fourier_3d}
\G(\x|\x_0) = \frac{1}{2\pi \epsilon} \sum\limits_{m=-\infty}^\infty e^{-im(\varphi-\varphi_0)} \hat{\G}^{(m)}(r/\epsilon|r_0/\epsilon)
+ O(1),
\end{equation}
where the Fourier coefficients determine the kernels
\begin{align}
\hat{\G}^{(m)}(\hat{r}|\hat{r}_0) & = \lim\limits_{\epsilon \to 0} 
\epsilon \int\limits_0^{2\pi} d\varphi \, e^{im(\varphi-\varphi_0)} \G(\epsilon \hat{\x}|\epsilon \hat{\x}_0), 
\end{align}
with rescaled positions $\hat{\x} = \x/\epsilon = (\hat{r},0,\varphi)$
and $\hat{\x}_0 = \x_0/\epsilon = (\hat{r}_0,0,\varphi_0)$ in the unit
disk.  For any integer $m \ne 0$, we get
\begin{equation}  \label{eq:Ghatm_3d}
\hat{\G}^{(m)}(\hat{r}|\hat{r}_0) = \int\limits_0^{2\pi} d\varphi \, 
\frac{\cos(m\varphi)}{2\pi \sqrt{\hat{r}^2 + \hat{r}_0^2 - 2\hat{r} \hat{r}_0 \cos \varphi}} \,,
\end{equation}
which comes from the first term in Eq. (\ref{eq:G_3d_auxil}); note the
contribution of the second term was $O(\epsilon)$ and thus vanished.

For $m = 0$, we use Eqs. (\ref{eq:dist_average},
\ref{eq:logdist_average}) to get
\begin{align*}
\int\limits_0^{2\pi} d\varphi\, \G(\x|\x_0) & = \frac{2 K(r_</r_>)}{\pi r_>}  
 - \frac{H(\x_\Gamma)}{4} \ln \biggl(\frac{r^2+r_0^2+|r^2-r_0^2|}{2}\biggr)  \\
& - \frac{\hat{w}_0(r_0/\epsilon) + \hat{w}_0(r/\epsilon) + 16\pi/3}{\epsilon}
+ \frac{H(\x_\Gamma)}{2} \ln \epsilon + O(1) ,
\end{align*}
where $r_< = \min\{r,r_0\}$, $r_> = \max\{r,r_0\}$, and $K(k)$ is the
complete elliptic integral of the first kind:
\begin{equation}
K(k) = \int\limits_0^{\pi/2} \frac{dz}{\sqrt{1-k^2 \sin^2 z}} \,.
\end{equation}
Substituting $r = \epsilon \hat{r}$ and $r_0 = \epsilon \hat{r}_0$, we
see that the second term is subleading and thus vanishes in the limit
$\epsilon\to 0$, yielding
\begin{equation}   \label{eq:Ghat_3d}
\hat{\G}^{(0)}(\hat{r}|\hat{r}_0) = \frac{4}{\pi}  \biggl[\frac{K(\hat{r}_</\hat{r}_>)}{2\hat{r}_>} - E(\hat{r}) - E(\hat{r}_0) + \frac43\biggr].
\end{equation}

The Fourier expansion (\ref{eq:G_Fourier_3d}) of the kernel
$\G(\x|\x_0)$ suggests to search the leading-order term of an
eigenfunction $v_k^{(0,\Gamma)}$ of the Dirichlet-to-Neumann operator
$\M_0^{(\Gamma)}$ in the form $e^{im\varphi}
\hat{v}_{m,n}(r/\epsilon)$, with suitable indices $m$ and $n$.
Substitution of this form and Eq. (\ref{eq:G_Fourier_3d}) into
Eq. (\ref{eq:eigen_G}) yields
\begin{align*}
\frac{e^{im\varphi_0} \hat{v}_{m,n}(r_0/\epsilon)}{\mu_{m,n}} &= 
\int\limits_0^{\epsilon} dr \, r \int\limits_0^{2\pi} d\varphi \, e^{im\varphi} \hat{v}_{m,n}(r/\epsilon) 
\frac{1}{2\pi \epsilon} \sum\limits_{m'=-\infty}^\infty e^{-im(\varphi-\varphi_0)} \hat{\G}^{(m)}(r/\epsilon|r_0/\epsilon) \\
& = e^{im\varphi_0} \epsilon \int\limits_0^1 d\hat{r} \, \hat{r}\, \hat{v}_{m,n}(\hat{r}) 
\hat{\G}^{(m)}(\hat{r}|\hat{r}_0), 
\end{align*}
i.e.,
\begin{equation}  \label{eq:Geigen_3d}
\int\limits_0^1 d\hat{r}  \, \hat{r} \, \hat{v}_{m,n}(\hat{r})\, \hat{\G}^{(m)}(\hat{r}|\hat{r}_0)
= \frac{\hat{v}_{m,n}(\hat{r}_0)}{\hat{\mu}_{m,n}} 
 \qquad (0 < \hat{r}_0 < 1, ~ n=1,2,\ldots).
\end{equation}
One sees that for each integer $m$, the kernel
$\hat{\G}^{(m)}(\hat{r}|\hat{r}_0)$ defines an integral operator in
$L^2((0,1))$ (with the scalar product including the weight $\hat{r}$)
that determines the eigenvalues $1/\hat{\mu}_{m,n}$ and eigenfunctions
$\hat{v}_{m,n}(\hat{r})$.  As previously, we impose the normalization
\begin{equation}
2\pi \int\limits_0^1 d\hat{r} \, \hat{r} \, |\hat{v}_{m,n}(\hat{r})|^2 = 1 .
\end{equation}
As in the planar case, the principal eigenpair $\hat{\mu}_0 = 0$ and
$\hat{v}_0 = 1/\sqrt{\pi}$ should be added separately.  Note that we
used here the complex-valued form $e^{im\varphi}
\hat{v}_{m,n}(\hat{r})$ instead of the real-valued form in
Eq. (\ref{eq:double_index}).  Given that
$\hat{\G}^{(-m)}(\hat{r}|\hat{r}_0) =
\hat{\G}^{(m)}(\hat{r}|\hat{r}_0)$, one can easily translate one form
to the other.

Since the kernels $\hat{\G}^{(m)}(\hat{r}|\hat{r}_0)$ were obtained as
the leading-order contributions to the kernel $\G(\x|\x_0)$ in the
limit $\epsilon\to 0$, the eigenvalues $\hat{\mu}_{m,n}$ and
eigenfunctions $e^{im\varphi} \hat{v}_{m,n}(\hat{r})$ are suitable
candidates to represent the leading-order behavior of the eigenvalues
and eigenfunctions of the Dirichlet-to-Neumann operator
$\M_0^{(\Gamma)}$.  In other words, the spectral problem
(\ref{eq:Geigen_3d}) provides an alternative formulation of the
Steklov-Neumann problem (\ref{eq:Steklov_disk}).  We emphasize that
the equivalence between these two formulations requires a mathematical
proof (see also \cite{Henrici70}).  Note that the curvature of the
boundary, $H(\x_\Gamma)$, that appeared in the kernel $\G(\x|\x_0)$ in
the subleading term in Eq. (\ref{eq:G_3d_auxil}), does not affect the
leading-order contributions to the eigenvalues and eigenfunctions.
This analysis justifies the possibility of replacing a curved subset
$\Gamma$ by a disk in the limit $\epsilon\to 0$.

\subsection{Role of the boundary connectivity}
\label{sec:connectivity}

It is also instructive to outline the role of the assumption that the
smooth boundary $\pa$ is connected, i.e., the distance between the
subset $\Gamma$ and the reflecting part $\pa_N$ is zero.  This is a
typical setting for diffusive applications when a particle has to
react on a reactive patch $\Gamma$ on an inert boundary or to escape
from a domain through a ``hole'' $\Gamma$ on the confining wall.  In
other applications, however, reactive patches $\Gamma$ are dispersed
in the volume, which is enclosed by a reflecting boundary $\pa_N$ so
that the boundary $\pa$ (still formed by $\Gamma$ and $\pa_N$) is not
connected.  This setting is often referred to as the narrow capture
problem to distinguish it from the narrow escape problem.  In both
cases, the subset $\Gamma$ is referred to as a target to be reached or
searched.  If the target size were to be much smaller than the
distance between $\Gamma$ and $\pa_N$, the dilation of the domain
would push the boundary $\pa_N$ to infinity, and the asymptotic
behavior would be determined from the analysis of the exterior Steklov
problem for the rescaled target $\hat{\Gamma} = \Gamma/\epsilon$.  In
particular, even though the $1/\epsilon$ scaling of the eigenvalues
$\mu_k^{(0,\Gamma)}$ would remain valid, the coefficients
$\hat{\mu}_k$ would in general be different.

To illustrate this point, let us consider the annular domain $\Omega =
\{ \x\in\R^d ~:~ \ve R < |\x| < R\}$ between two concentric spheres of
radii $\epsilon = \ve R$ and $R$, with the Steklov condition on the
inner sphere $\Gamma$ of radius $\epsilon$ and the Neumann condition
on the outer sphere $\pa_N$ of radius $R$.  The rotational invariance
of this problem allows one to get the eigenvalues $\mu_k^{(0,\Gamma)}$
explicitly (see, e.g., \cite{Grebenkov20c,Colbois21}):
\begin{equation}
\mu_k^{(0,\Gamma)} = \frac{k}{\ve R} \, \frac{(k + d-2)(1 - \ve^{2k+d-2})}{k  + (k + d-2)\ve^{2k+d-2}} \,.
\end{equation}
When $\ve \ll 1$ and $d = 2$, one has $\mu_k^{(0,\Gamma)}
\approx k/\epsilon$.  In turn, the asymptotic relation
(\ref{eq:main_asympt}) for the planar case with a connected boundary
yields $\mu_k^{(0,\Gamma)} \approx \pi k/(2\epsilon)$ at large $k$,
where we used Eq. (\ref{eq:etak_asympt_2d}).  One sees that the
scaling of $\mu_k^{(0,\Gamma)}$ as $1/\epsilon$ is the same for both
settings, but the prefactor differs by $\pi/2$.

\section{Two examples}
\label{sec:examples}

In this section, we illustrate the above general results with two
examples: an arc-shaped subset $\Gamma$ on the boundary of a disk
(Sec. \ref{sec:disk}) and a spherical cap on the boundary of a ball
(Sec. \ref{sec:ball}).  In both cases, the explicit formulas for the
Green's function $\tilde{G}_0(\x,p|\x_0)$ and the pseudo-Green's
function $\G_0(\x|\x_0)$ allow us to compute accurately the
eigenvalues and eigenfunctions of the Steklov-Neumann problem for
$\Gamma$ of any size and for any $p \geq 0$.  In this way, we can
access the accuracy and the validity range of the asymptotic results
deduced in Sec. \ref{sec:limit}.

\subsection{Disk}
\label{sec:disk}

As the first example, we consider the domain $\Omega = \{ \x
\in\R^2~:~ |\x| < R\}$ to be a disk of radius $R$, and set $\Gamma =
\{ (R,\theta) \in \pa ~:~ |\theta| < \ve\}$ to be an arc of angle
$2\ve$ on the circular boundary $\pa$ (Fig. \ref{fig:scheme}b), where
we used the polar coordinates $\x = (r,\theta)$, and $\epsilon =
\ve R$.
In Appendix \ref{sec:disk_general}, we recall the exact explicit
expressions for the Green's function $\tilde{G}_0(\x,p|\x_0)$ and the
pseudo-Green's function $\G_0(\x|\x_0)$, and derive the corrections
$w_0^{(\Gamma)}(\x)$ and $\a_\Gamma$ for any $0 < \ve \leq \pi$.
These expressions allow us to determine the eigenpairs
$\{\mu_k^{(p,\Gamma)}, V_k^{(p,\Gamma)}\}$ numerically for any arc
$\Gamma$.  A numerical solution of the spectral problems
(\ref{eq:eigen_G}, \ref{eq:eigen_disk_eps0}) is described in Appendix
\ref{sec:numerics_2d}.

Figure \ref{fig:eigenvalues_disk} illustrates the dependence of the
first four eigenvalues $\mu_k^{(0,\Gamma)}$ on the arc half-angle
$\ve$.  At $\ve = \pi$, one retrieves the conventional Steklov problem
for the disk (i.e., $\Gamma = \pa$), for which $\mu_k^{(0,\pa)} =
k/R$.  The eigenvalue $\mu_0^{(0,\Gamma)} = 0$ (not shown) is simple,
whereas the other eigenvalues for $k > 0$ are twice degenerate.  In
turn, as soon as $\ve < \pi$, the degeneracy is removed, and the
considered first eigenvalues for the Steklov-Neumann problem turn out
to be simple.  Moreover, they rapidly become close to their asymptotic
form (\ref{eq:main_asympt}), with $\hat{\mu}_k$ reported in Table
\ref{tab:eta_2D}.
The bottom panel of Fig. \ref{fig:eigenvalues_disk} presents the ratio
between the eigenvalue and its asymptotic form (\ref{eq:main_asympt})
to highlight its approach to $1$ as $\ve$ decreases.  The asymptotic
form gets more and more accurate as $k$ increases, even for large
$\ve$.  We conclude that the asymptotic relation
(\ref{eq:main_asympt}) provides an accurate approximation for
$\mu_k^{(0,\Gamma)}$ even for moderately large $\ve$.

\begin{figure}
\begin{center}
\includegraphics[width=0.49\textwidth]{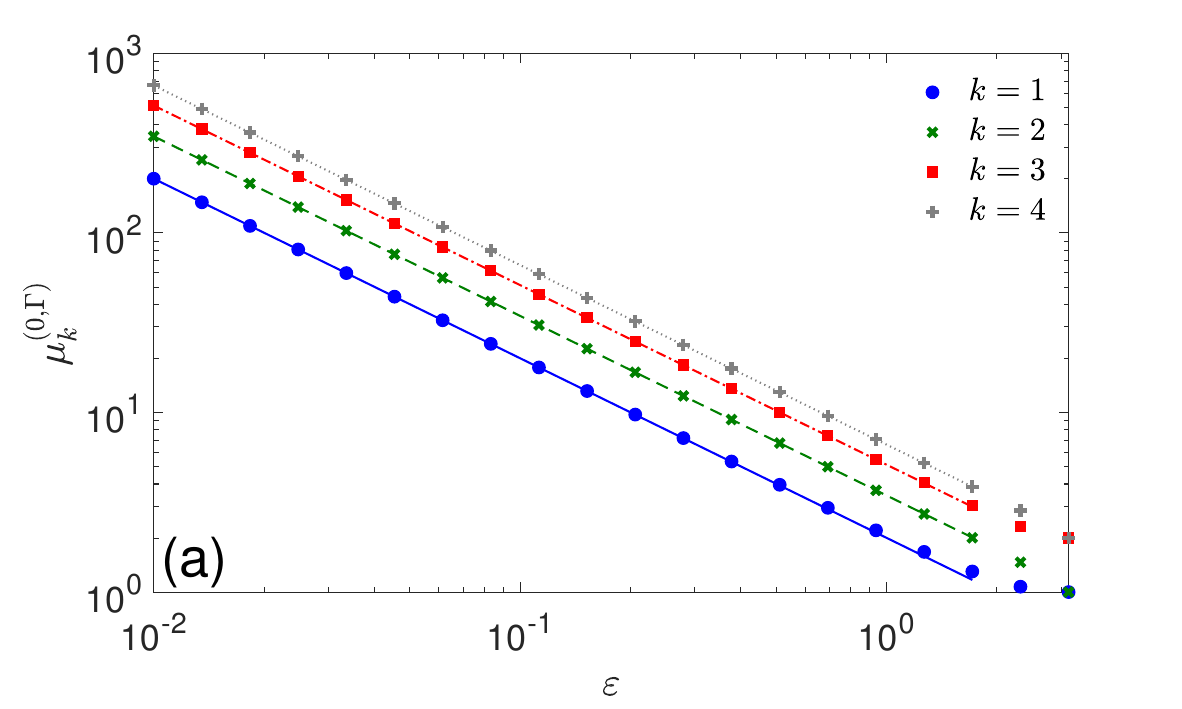} 
\includegraphics[width=0.49\textwidth]{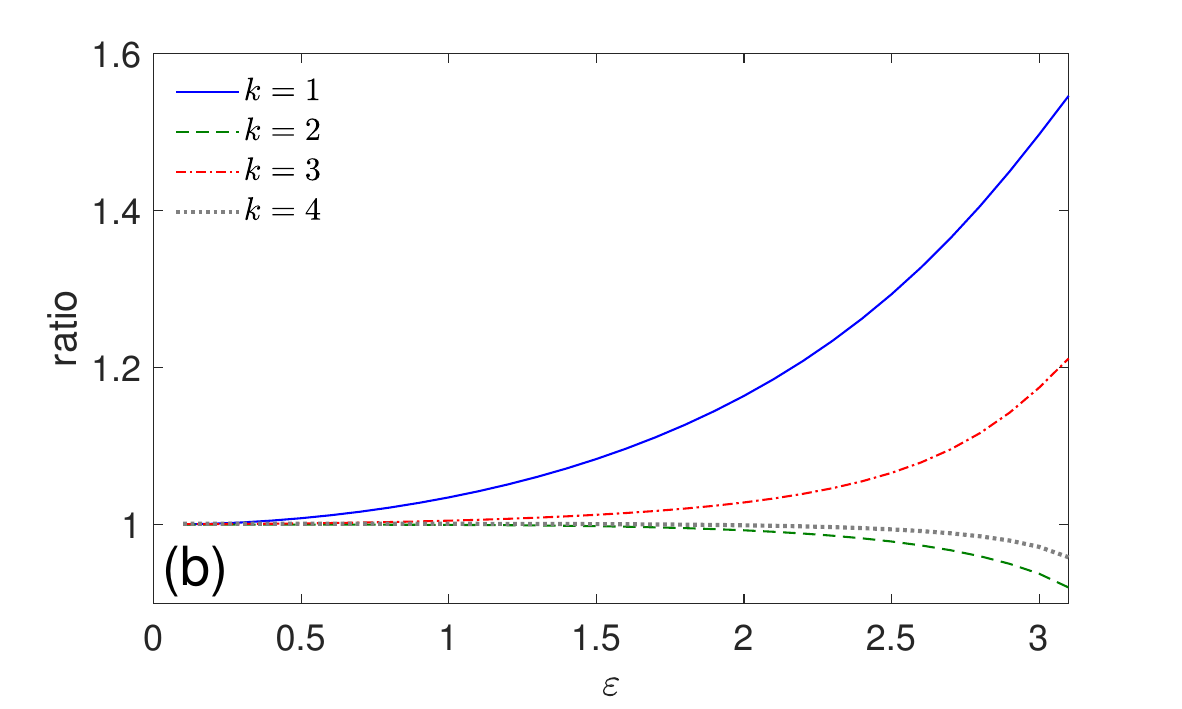} 
\end{center}
\caption{
{\bf (a)} First four eigenvalues $\mu_k^{(0,\Gamma)}$ of the
Dirichlet-to-Neumann operator $\M_0^{(\Gamma)}$ for the arc-shaped
subset $\Gamma$ on the boundary of the unit disk ($R = 1$) as
functions of the arc half-angle $\ve = \epsilon/R$.  Symbols present
the results of a numerical diagonalization of the matrix $\GG^{(\ve)}$
from Eq. (\ref{eq:Gmatrix_ve}), with $N = 30$ and $\kmax = 20000$.
Lines show the asymptotic relation (\ref{eq:main_asympt}), in which
$\hat{\mu}_k$ are given in Table \ref{tab:eta_2D}.  {\bf (b)} Ratio
$\mu_k^{(0,\Gamma)}/(\hat{\mu}_k/\epsilon)$ between the eigenvalue
$\mu_k^{(0,\Gamma)}$ and its asymptotic form.}
\label{fig:eigenvalues_disk}
% load('disk_fig1a.mat', 'muj', 'muj0', 'eps');
% [muj,muj0,eps] = A_DN_disk_mixed_eigenvalues_fig(muj,muj0);
% load('disk_fig1b.mat', 'muj', 'muj0', 'eps');
% [muj,muj0,eps] = A_DN_disk_mixed_eigenvalues_fig2(muj,muj0);
\end{figure}

\begin{figure}
\begin{center}
\includegraphics[width=0.49\columnwidth]{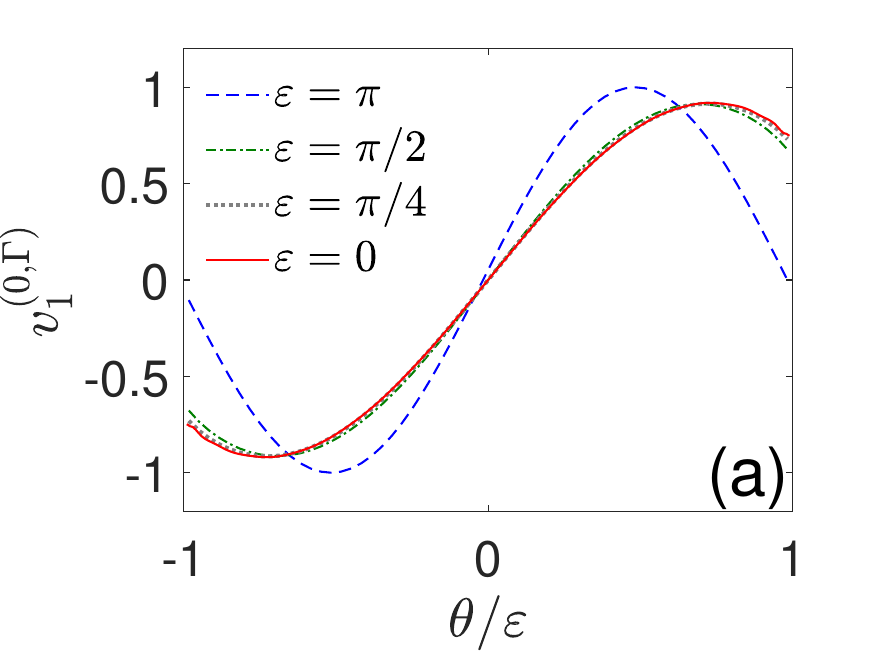} 
\includegraphics[width=0.49\columnwidth]{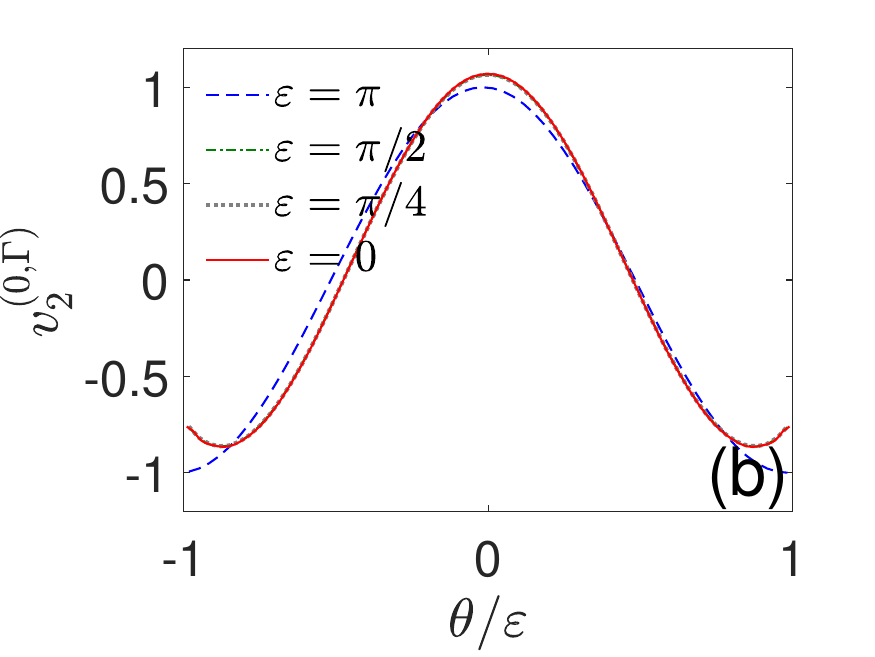} 
\includegraphics[width=0.49\columnwidth]{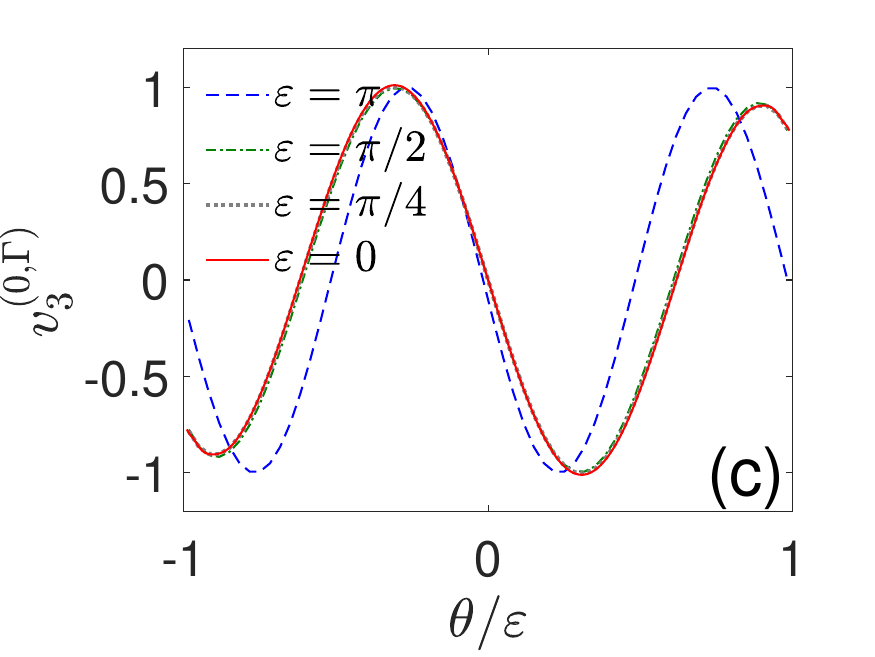} 
\includegraphics[width=0.49\columnwidth]{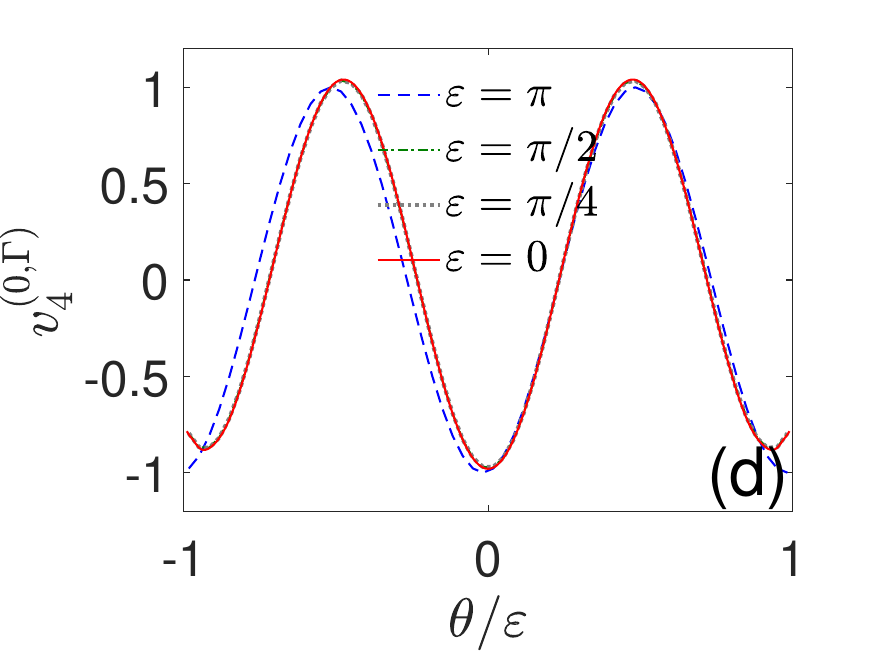} 
\end{center}
\caption{
First four eigenfunctions $\sqrt{\ve}\, v_k^{(0,\Gamma)}$ of the
Dirichlet-to-Neumann operator $\M_0^{(\Gamma)}$ for the arc-shaped
subset $\Gamma$ on the boundary of the unit disk ($R = 1$), for three
values of the arc half-angle $\ve$ as indicated in the legend.  They
were obtained by diagonalizing the matrix $\GG^{(\ve)}$ from
Eq. (\ref{eq:Gmatrix_ve}), with $N = 30$ and $\kmax = 20000$.  Note
that the factor $\sqrt{\ve}$ ensures the correct normalization and
comparable amplitudes of eigenfunctions for different $\ve > 0$.  For
$\ve = 0$, we plotted the eigenfunctions $\hat{v}_k(x)$ that were
obtained by diagonalizing the matrix $\GG$ from
Eq. (\ref{eq:Glim_matrix}) of size $30 \times 30$.  }
\label{fig:disk_vk}
% A_DN_disk_mixed_eigenvalues_fig3b(1);
\end{figure}

Figure \ref{fig:disk_vk} illustrates the behavior of the four
eigenfunctions $v_k^{(0,\Gamma)}$ for several values of $\ve$.  When
$\ve = \pi$ (i.e., $\Gamma = \pa$), the eigenfunctions are known
explicitly as $\cos(k\theta)$ and $\sin(k\theta)$.  These
eigenfunctions are shown by dashed line.  As $\ve$ decreases, the
eigenfunctions $v_k^{(0,\Gamma)}$ preserve their oscillating character
but deviate from sine and cosine functions.  Note that even for $\ve =
\pi/2$, the eigenfunctions are very close to the limiting ones,
$\hat{v}_k$, obtained by diagonalizing the matrix $\GG$ from
Eq. (\ref{eq:Glim_matrix}) as the eigenfunctions of the kernel
$\hat{\G}(x|x_0)$ from Eq. (\ref{eq:Glim_disk}).  As $k$ increases,
the eigenfunctions $\hat{v}_k$ become closer to sine and cosine
functions, i.e., to the case $\ve = \pi$.  In other words,
higher-order eigenfunctions do not almost depend on $\ve$ (up to
rescaling).

\subsection{Ball}
\label{sec:ball}

We replicate the above illustration for a ball of radius $R$, $\Omega
= \{\x\in \R^3 ~:~ |\x| < R\}$, with $\Gamma$ being the spherical cap
of angle $\ve$ around the North pole: $\Gamma = \{
\x\in\pa ~:~ 0 \leq \theta < \ve\}$ (Fig. \ref{fig:scheme}c), where we
used the spherical coordinates $(r,\theta,\varphi)$, and $\epsilon =
\ve R$.  The axial symmetry of the domain allows us to search the
eigenfunctions in the form $e^{im\varphi}
V_{m,n}^{(p,\Gamma)}(r,\theta)$, with an integer $m$, enumerated by $n
= 1,2,\cdots$.
In Appendix \ref{sec:ball_general}, we recall the exact explicit
expressions for the Green's function $\tilde{G}_0(\x,p|\x_0)$ and the
pseudo-Green's function $\G_0(\x|\x_0)$, and derive the corrections
$w_0^{(\Gamma)}(\x)$ and $\a_\Gamma$.
These expressions allow one to determine the eigenpairs
$\{\mu_{m,n}^{(p,\Gamma)}, V_{m,n}^{(p,\Gamma)}\}$ numerically for any
$0 < \ve \leq \pi$.  Throughout this section, we focus on axially
symmetric eigenpairs with $m = 0$, for which the asymptotic relations
(\ref{eq:main_asympt}) read as
\begin{equation}  \label{eq:main_asympt_3d}
\mu_{0,n}^{(0,\Gamma)} \approx \frac{\hat{\mu}_{0,n}}{\epsilon}, \qquad
V_{0,n}^{(0,\Gamma)}(R,\theta,\varphi) \approx \frac{\hat{v}_{0,n}(\theta/\ve)}{\epsilon} \,.
\end{equation}
A numerical solution of the spectral problems (\ref{eq:eigen_G},
\ref{eq:Geigen_3d}) is described in Appendix \ref{sec:numerics_3d}.

Figure \ref{fig:eigenvalues_sphere}(a) shows the first four
eigenvalues $\mu_{0,n}^{(0,\Gamma)}$ as functions of the angle $\ve$.
Expectedly, one retrieves the well-known eigenvalues $n/R$ in the
classical case $\ve = \pi$ when the subset $\Gamma$ covers the whole
boundary (i.e., $\Gamma = \pa$).  As $\ve$ decreases, the asymptotic
relation (\ref{eq:main_asympt_3d}) rapidly becomes an accurate
approximation of $\mu_{0,n}^{(0,\Gamma)}$.  The accuracy of this
approximation is illustrated on Fig. \ref{fig:eigenvalues_sphere}(b).
Note that the maximal relative deviation between
$\mu_{0,n}^{(0,\Gamma)}$ and its asymptotic form
$\hat{\mu}_{0,n}/\epsilon$ is below $27\%$ (see the minimum of the
blue curve corresponding to $n = 1$).  The first ten eigenvalues
$\hat{\mu}_{0,n}$ are listed in Table
\ref{tab:eta_2D}.

\begin{figure}
\begin{center}
\includegraphics[width=0.49\textwidth]{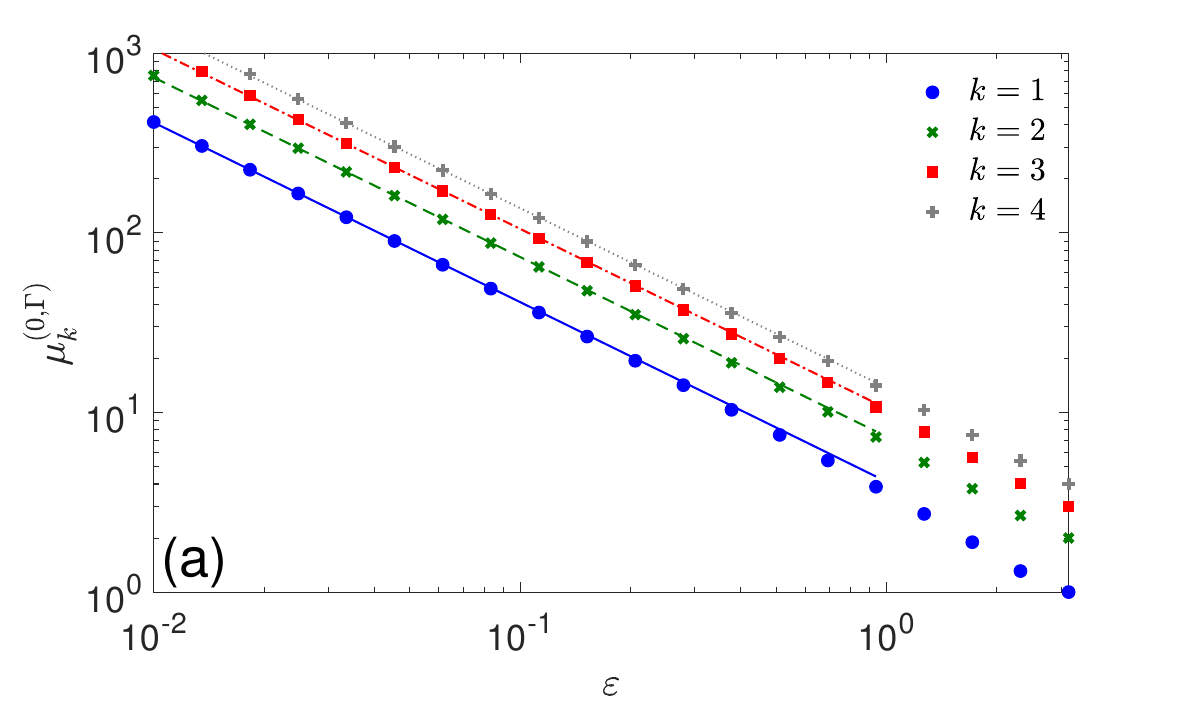} 
\includegraphics[width=0.49\textwidth]{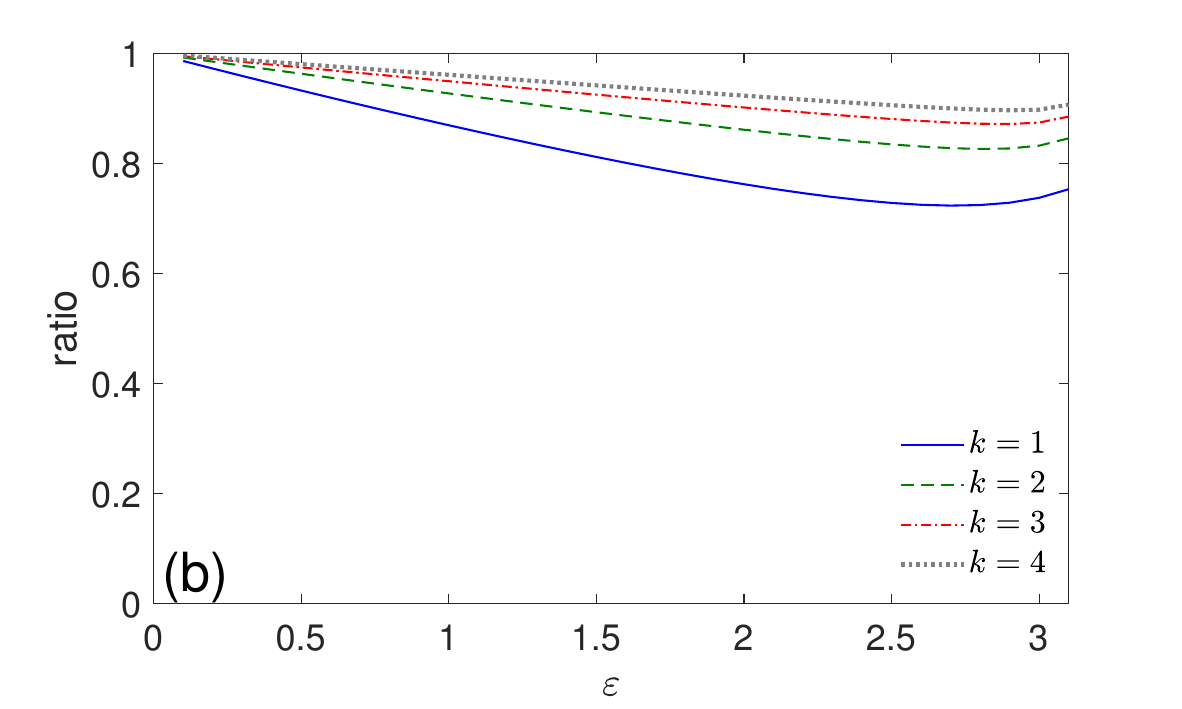} 
\end{center}
\caption{
{\bf (a)} First four eigenvalues $\mu_{0,n}^{(0,\Gamma)}$ of the
Dirichlet-to-Neumann operator $\M_0^{(\Gamma)}$ for the spherical cap
$\Gamma$ of angle $\ve$ on the unit sphere ($R = 1$) as functions of
the angle $\ve$.  Symbols present the results of a numerical
diagonalization of the matrix $\GG^{(\ve)}$ from
Eq. (\ref{eq:G_matrix_sphere}), with $N = 100$ and $\kmax = 1000$.
Lines show the asymptotic relations (\ref{eq:main_asympt_3d}), in
which $\hat{\mu}_{0,n}$ are given in Table \ref{tab:eta_2D}.  {\bf
(b)} Ratio $\mu_{0,n}^{(0,\Gamma)}/(\hat{\mu}_{0,n}/\epsilon)$ between
the eigenvalue $\mu_{0,n}^{(0,\Gamma)}$ and its asymptotic form.}
\label{fig:eigenvalues_sphere}
% load('sphere_fig1a.mat', 'muj', 'muj0', 'eps');
% [muj,muj0,eps] = A_DN_sphere_mixed_eigenvalues_fig;
% load('sphere_fig1b.mat', 'muj', 'muj0', 'eps');
% [muj,muj0,eps] = A_DN_sphere_mixed_eigenvalues_fig2(muj,muj0);
%% [G, Ed,V] = A_DN_sphere_mixed_Glim4;   % we used N = 150
\end{figure}

\begin{figure}
\begin{center}
\includegraphics[width=0.49\columnwidth]{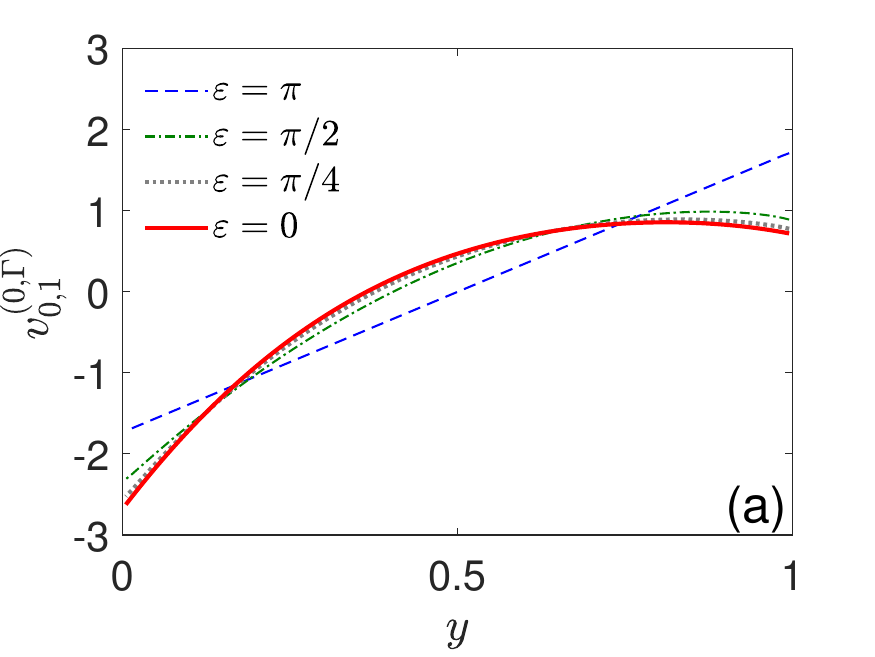} 
\includegraphics[width=0.49\columnwidth]{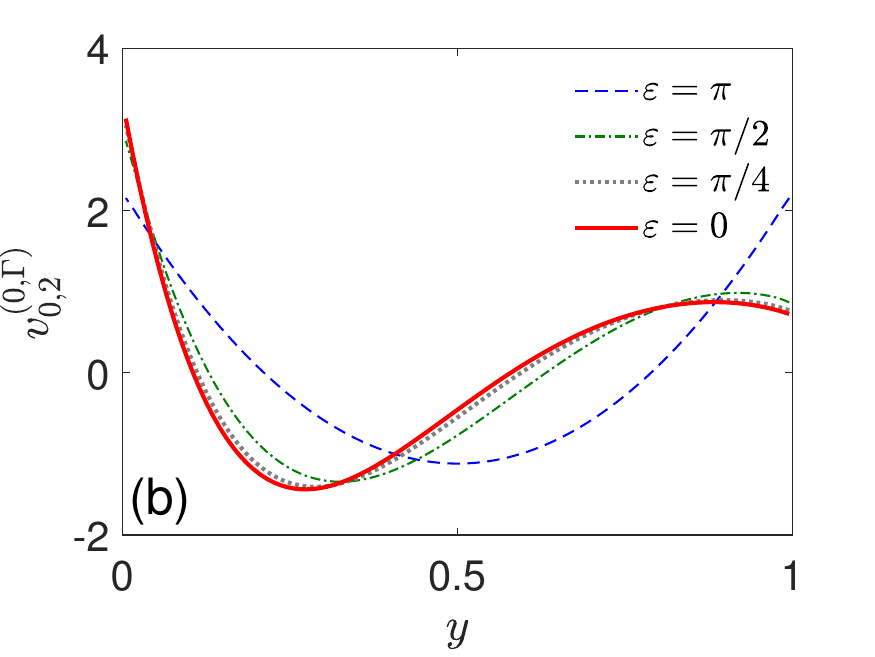} 
\includegraphics[width=0.49\columnwidth]{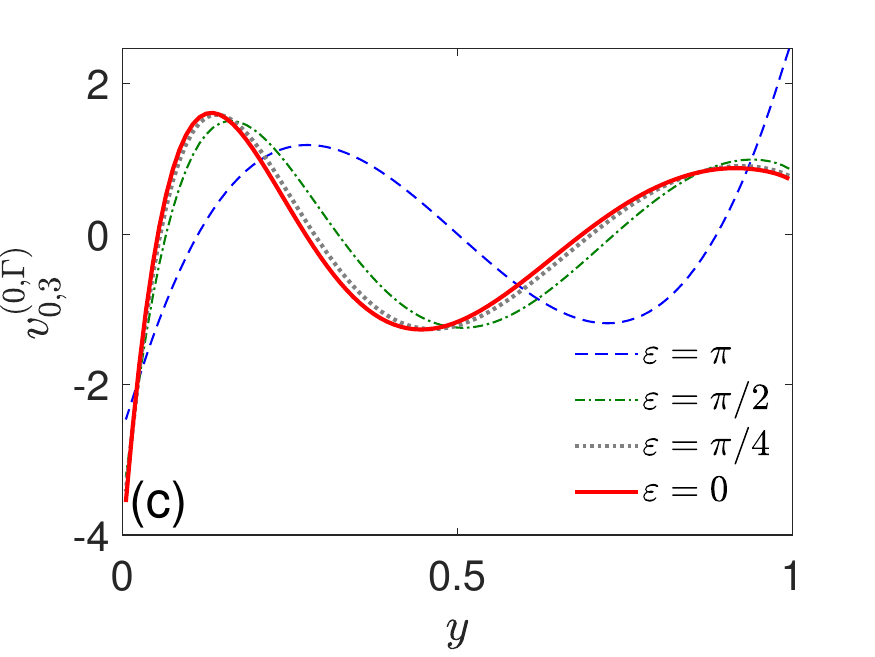} 
\includegraphics[width=0.49\columnwidth]{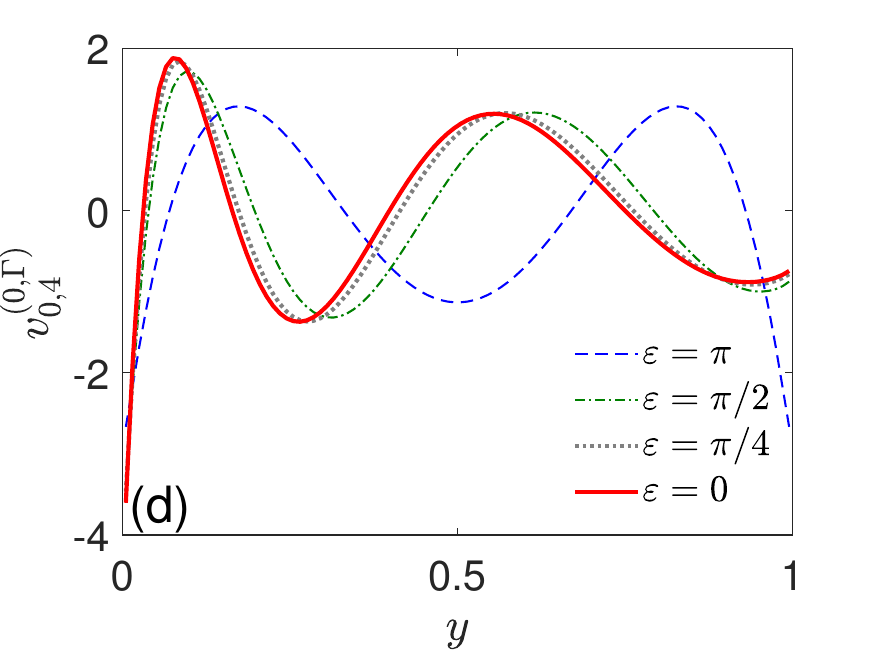} 
\end{center}
\caption{  
First four axially symmetric eigenfunctions $\sqrt{1-\cos\ve}\,
V_{0,n}^{(0,\Gamma)}(R,\theta,\varphi)$ of the Dirichlet-to-Neumann
operator $\M_0^{(\Gamma)}$ on the spherical cap $\Gamma$ of angle
$\ve$ on the unit sphere ($R = 1$), for three values of the angle
$\ve$ as indicated in the legend, with $y =
(1-\cos\theta)/(1-\cos\ve)$.  They were obtained by diagonalizing the
matrix $\GG^{(\ve)}$ from Eq. (\ref{eq:G_matrix_sphere}), with $N =
100$ and $\kmax = 1000$.  Note that the factor $\sqrt{1-\cos\ve}$
ensures the correct normalization and comparable amplitudes of
eigenfunctions for different $\ve >0$.  For $\ve = 0$, we plotted the
eigenfunctions $\hat{v}_{0,n}(\sqrt{y})$, that were obtained by
diagonalizing the matrix $\GG$ from Eq. (\ref{eq:GG_disk_upper}) of
size $30 \times 30$.  }
\label{fig:sphere_vk}
% load('scheme_vk.mat');
% [V,V0,y] = A_DN_sphere_mixed_eigenfunctions_fig2(V,V0);
\end{figure}

Figure \ref{fig:sphere_vk} presents the related axially symmetric
eigenfunctions $V_{0,n}^{(0,\Gamma)}(R,\theta,\varphi)$, restricted to
$\Gamma$, for three values of $\ve$.  When $\ve = \pi$, these
eigenfunctions are the (rescaled) Legendre polynomials,
$\sqrt{n+1/2}\, P_n(\cos\theta)$, as shown by blue dashed line for $n
= 1,2,3,4$ on four panels.  As $\ve$ decreases, the rescaled
eigenfunctions rapidly approach $\hat{v}_{0,n}$, highlighting the
accuracy of the asymptotic relation (\ref{eq:main_asympt_3d}), even
for $\ve = \pi/2$ when the subset $\Gamma$ covers half of the sphere.

\section{Mean first-reaction time}
\label{sec:application}

The eigenmodes of the Dirichlet-to-Neumann operator determine many
quantities characterizing first-passage processes and related
diffusion-controlled reactions \cite{Grebenkov20}.  In this section,
we illustrate a straightforward application of the asymptotic results
by considering restricted diffusion in a bounded domain $\Omega$ and
focusing on the first-reaction time $\tau$ on a reactive target
$\Gamma \subset \pa$
\cite{Redner,Schuss,Metzler,Benichou14,Holcman14,Masoliver,Lindenberg,Grebenkov,Dagdug}.
The moments and the distribution of the random variable $\tau$ were
thoroughly investigated in the past, especially in the narrow escape
limit
\cite{Ward93,Grigoriev02,Holcman04,Kolokolnikov05,Singer06a,Singer06b,Singer06c,Schuss07,Benichou08,Singer08,Reingruber09,Pillay10,Cheviakov10,Cheviakov11,Cheviakov12,Caginalp12,Rojo12,Rupprecht15,Gomez15,Grebenkov16,Marshall16,Grebenkov17,Grebenkov19d,Kaye20,Guerin23}.
Some of these former results will be used for comparison with our
predictions.  Most importantly, the knowledge of the eigenpairs of the
Dirichlet-to-Neumann operator allows us to go beyond the classical
setting and to investigate first-reaction times for more sophisticated
surface reaction mechanisms on $\Gamma$, including activation or
passivation of the target, encounter-dependent reactivity,
etc. \cite{Grebenkov20}.  We start with the classical setting and then
briefly describe its extensions.

\subsection{Constant reactivity}
\label{sec:constant}

For a diffusing particle started from a point $\x_0 \in \Omega$ at
time $0$, we are interested in the first-reaction time $\tau$ on a
target $\Gamma \subset \pa$ with a constant reactivity parameter $q >
0$.  The diffusing particle executes a partially reflected Brownian
motion inside $\Omega$, with normal reflections on $\pa_N$ and
eventual reactions on $\Gamma$ \cite{Grebenkov06,Grebenkov20}.  The
distribution of the random variable $\tau$ is characterized by the
survival probability $S_q(t|\x_0) = \P_{\x_0}\{ \tau > t\}$, i.e., the
probability of no reaction on $\Gamma$ up to time $t$.  The survival
probability $S_q(t|\x_0)$ satisfies the diffusion equation with mixed
Robin-Neumann boundary conditions:
\begin{subequations}
\begin{align}
\partial_t S_q(t|\x_0) & = D \Delta S_q(t|\x_0) \qquad (\x_0\in\Omega), \\  \label{eq:Sq_Robin}
- \partial_n S_q(t|\x_0) & = q \, S_q(t|\x_0)\, \I_\Gamma(\x_0)  \qquad (\x_0\in \pa), 
\end{align}
\end{subequations}
subject to the initial condition $S_q(0|\x_0) = 1$ (we recall that
$\I_\Gamma(\x_0)$ is the indicator function of $\Gamma$).  The notion
of partial reactivity was introduced by Collins and Kimball
\cite{Collins49} and later investigated in various contexts
(see
\cite{Grebenkov23c,Sapoval94,Grebenkov06,Singer08b,Grebenkov19b,Piazza22}
and references therein).  When the particle arrives onto a partially
reactive target, it can either react or be reflected to resume its
diffusion, until the next arrival on $\Gamma$, and so on.  Depending
on the considered application, such reflections can represent the
arrival at a passive boundary point (due to either microscopic
heterogeneity of reactive sites on $\Gamma$ or their temporal
activation/passivation dynamics), a failure to overcome an energy
activation barrier in a chemical reaction, a failure to squeeze
through a narrow channel to escape, etc.  In this way, the parameter
$q$ is related to the probability of the reaction event and thus
characterizes the reactivity of the target $\Gamma$, by ranging from
$0$ for an inert, passive target (without any reaction), to $+\infty$
for a perfectly reactive target, on which the particle reacts
instantly upon the first arrival \cite{Grebenkov23c}.

The Laplace transform of the survival probability,
\begin{equation}
\tilde{S}_q(p|\x_0) = \int\limits_0^\infty dt\, e^{-pt} \, S_q(t|\x_0)  \qquad (p \geq 0),
\end{equation}
satisfies the modified Helmholtz equation with mixed Robin-Neumann
boundary conditions, 
\begin{subequations}  \label{eq:Stilde_helm}
\begin{align}
(p - D\Delta) \tilde{S}_q(p|\x_0) & = 1  \quad (\x_0\in \Omega), \\
\partial_n \tilde{S}_q(p|\x_0) + q \tilde{S}_q(p|\x_0) \I_{\Gamma}(\x_0) & = 0 \quad (\x_0\in \pa).
\end{align}
\end{subequations}
As this problem can be obtained by integrating
Eq. (\ref{eq:Gq_problem}) over $\x\in \Omega$, one has
\begin{equation}
\tilde{S}_q(p|\x_0)  = \int\limits_\Omega d\x \, \tilde{G}_q(\x,p|\x_0),
\end{equation}
so that the spectral expansion (\ref{eq:Gq_spectral}) for the Green's
function yields after some simplifications (see
\cite{Grebenkov19c,Grebenkov20} for more details):
\begin{equation} \label{eq:tildeSq}
\tilde{S}_q(p|\x_0) = \frac{1}{p} - \frac{1}{p} 
\sum\limits_{k=0}^\infty \frac{V_k^{(p,\Gamma)}(\x_0)}{1 + \mu_k^{(p,\Gamma)}/q} \int\limits_\Gamma  v_k^{(p,\Gamma)}
\qquad (\x_0\in\overline{\Omega}, ~ q > 0).
\end{equation}
Since the negative time derivative of the survival probability
determines the probability density of the first-reaction time,
$H_q(t|\x_0) = -\partial_t S_q(t|\x_0)$, its Laplace transform reads
\begin{align} \label{eq:tildeHq}
\tilde{H}_q(p|\x_0) & = 1 - p \tilde{S}_q(p|\x_0)     
= \sum\limits_{k=0}^\infty \frac{V_k^{(p,\Gamma)}(\x_0)}{1 + \mu_k^{(p,\Gamma)}/q} \int\limits_\Gamma v_k^{(p,\Gamma)}
\qquad (\x_0\in\overline{\Omega},~ q > 0).
\end{align}
By definition, this is the moment-generating function of the random
time $\tau$: 
\begin{equation}
\tilde{H}_q(p|\x_0) = \int\limits_0^\infty dt \, e^{-pt} \, H_q(t|\x_0) = \E_{\x_0}\{ e^{-p\tau}\}.  
\end{equation}

In the following, we focus on the MFRT:
\begin{equation}
T_q(\x_0) = \E_{\x_0}\{ \tau \}  = \int\limits_0^\infty dt \, t  \underbrace{H_q(t|\x_0)}_{=-\partial_t S_q(t|\x_0)}
= \int\limits_0^\infty dt \, S_q(t|\x_0) = \tilde{S}_q(0|\x_0),
\end{equation}
which satisfies the boundary value problem (\ref{eq:Tq_problem}).  In
the limit $q\to \infty$, the Robin boundary condition on $\Gamma$
turns into the Dirichlet one, and $T_\infty(\x_0)$ becomes the mean
first-passage time to $\Gamma$.

To compute the MFRT, we evaluate a series expansion of
$\tilde{H}_q(p|\x_0)$ in powers of $p$ up to the first order (or,
equivalently, the series expansion of $\tilde{S}_q(p|\x_0)$ up to the
zeroth order).  Since $v_k^{(0,\Gamma)}$ for any $k > 0$ is orthogonal
to the constant function $v_0^{(0,\Gamma)} = 1/\sqrt{|\Gamma|}$, the
integral of $v_k^{(p,\Gamma)}$ over $\Gamma$ vanishes as $p\to 0$.  In
turn, we use the series expansions (\ref{eq:mu0v0_p0}) for the
eigenmode with $k = 0$ to get as $p\to 0$:
\begin{align*}
\tilde{H}_q(p|\x_0) & = \frac{1 + W_0^{(\Gamma)}(\x_0) \frac{p|\Omega|}{D}}{1 + \frac{p|\Omega|}{qD|\Gamma|}} 
\biggl(1 + \frac{p|\Omega|}{D|\Gamma|} \int\limits_\Gamma d\x \, w_0^{(\Gamma)}(\x)\biggr)  \\
& + \sum\limits_{k=1}^\infty \frac{V_k^{(0,\Gamma)}(\x_0)}{1 + \mu_k^{(0,\Gamma)}/q} 
\underbrace{\int\limits_\Gamma d\x \, v_k^{(p,\Gamma)}(\x)}_{= p b_k^{(\Gamma)} |\Omega|/(D\mu_k^{(0,\Gamma)}) + O(p^2)} + O(p^2) ,
\end{align*}
from which
\begin{equation} \label{eq:Tq_spectral} 
T_q(\x_0) = \frac{|\Omega|}{D} \biggl(\frac{1}{q|\Gamma|} - W_0^{(\Gamma)}(\x_0)
 -  \sum\limits_{k=1}^\infty \frac{b_k^{(\Gamma)} \, V_k^{(0,\Gamma)}(\x_0)}{\mu_k^{(0,\Gamma)}(1 + \mu_k^{(0,\Gamma)}/q)}  \biggr)
\qquad (\x_0\in\overline{\Omega}),
\end{equation}
where we used Eq. (\ref{eq:wk_def}), whereas the integral of
$w_0^{(\Gamma)}$ vanished according to Eq. (\ref{eq:w0_int}).  Solving
the boundary value problem (\ref{eq:W0_problem}), one can find
$W_0^{(\Gamma)}(\x_0)$ that determines the coefficients
$b_k^{(\Gamma)}$ via Eq. (\ref{eq:bk_w0}), while
$V_k^{(0,\Gamma)}(\x_0)$ are given by Eq. (\ref{eq:Vk_p0}).  By
recalling the relation (\ref{eq:Tapp_W0}), one can also represent the
above expression as
\begin{equation} \label{eq:Tqx0}
T_q(\x_0) = T_q^{\rm (app)}(\x_0) - \frac{|\Omega|}{D} 
\sum\limits_{k=1}^\infty \frac{b_k^{(\Gamma)} \, V_k^{(0,\Gamma)}(\x_0)}{\mu_k^{(0,\Gamma)}(1 + \mu_k^{(0,\Gamma)}/q)} \,,
\end{equation}
where $T_q^{\rm (app)}(\x_0)$ is the constant-flux approximation to
the MFRT \cite{Grebenkov17}.  In other words, the spectral expansion
in the second term provides the correction to this approximation.  The
exact expression (\ref{eq:Tqx0}) determines the dependence of the MFRT
on the reactivity parameter $q$ over the entire range of $0 < q \leq
+\infty$.

It is also instructive to consider the volume-averaged MFRT,
\begin{equation}
\overline{T_q} = \frac{1}{|\Omega|} \int\limits_{\Omega} d\x_0 \, T_q(\x_0),
\end{equation}
as if the starting point $\x_0$ was uniformly distributed in $\Omega$.
Using Eqs. (\ref{eq:Vk0_int}, \ref{eq:W0_int}), we get then
\begin{equation}  \label{eq:Tq_volume}  
\overline{T_q} = \frac{|\Omega|}{D}\biggl(\frac{1}{q|\Gamma|} + \a_\Gamma  
- \sum\limits_{k=1}^\infty \frac{[b_k^{(\Gamma)}]^2}{\mu_k^{(0,\Gamma)} (1 + \mu_k^{(0,\Gamma)}/q)} \biggr).
\end{equation}
In the limit $q\to \infty$, we find the volume-averaged MFPT to the
target $\Gamma$:
\begin{equation}  \label{eq:Tinf_volume}
\overline{T_\infty} = \frac{|\Omega|}{D} \biggl(\a_\Gamma  
- \sum\limits_{k=1}^\infty \frac{[b_k^{(\Gamma)}]^2}{\mu_k^{(0,\Gamma)}} \biggr).
\end{equation}
While these exact representations are valid for arbitrary bounded
domains with smooth boundaries, they require the knowledge of the
Steklov eigenfunctions and eigenvalues.  In the next section, we
discuss their consequences in the small-target limit.

\subsection{Small-target limit}

As stated earlier, the small-target limit of the MFPT was thoroughly
investigated
\cite{Ward93,Grigoriev02,Holcman04,Kolokolnikov05,Singer06a,Singer06b,Singer06c,Schuss07,Benichou08,Singer08,Reingruber09,Pillay10,Cheviakov10,Cheviakov11,Cheviakov12,Caginalp12,Rojo12,Rupprecht15,Gomez15,Grebenkov16,Marshall16,Grebenkov17,Grebenkov19d,Kaye20,Guerin23}.
As the detailed discussion of this topic can be found elsewhere (see,
e.g., the review \cite{Holcman14}), we just outline some novel
insights onto the volume-averaged quantities $\overline{T_\infty}$ and
$\overline{T_q}$ from our asymptotic results.

\subsubsection*{Two dimensions}

For planar bounded domains, we substitute the derived asymptotic
relations (\ref{eq:main_asympt}, \ref{eq:aGamma_asympt_2d},
\ref{eq:bk_2d}) into Eq. (\ref{eq:Tq_volume}) to get in the
small-target limit $\epsilon \to 0$:
\begin{equation}  \label{eq:Tq_volume_asympt}  
\overline{T_q} \approx \frac{|\Omega|}{D}\biggl(\frac{1}{2q\epsilon} - \frac{\ln (2\epsilon)}{\pi} + \frac{3}{2\pi} + R_0(\x_\Gamma)  
- \sum\limits_{k=1}^\infty \frac{[\hat{V}_k(\infty)]^2/\hat{\mu}_k}{1 + \hat{\mu}_k/(q\epsilon)} \biggr).
\end{equation}
Let us briefly comment on this asymptotic relation.

(i) for a perfectly reactive target ($q = \infty$), this relation is
reduced to
\begin{equation}  \label{eq:Tinf_volume_asympt}  
\overline{T_\infty} \approx \frac{|\Omega|}{\pi D}\biggl(\ln(2/\epsilon) + \pi R_0(\x_\Gamma) + o(1)\biggr),
\end{equation}
where we used Eq. (\ref{eq:sum_mu_bk2_2d}) to evaluate the sum.  The
logarithmic leading-order term, which was derived by Singer {\it et
al.} \cite{Singer06a}, is universal (see further discussions on a more
general form of the small parameter in \cite{Grebenkov16}).  In turn,
the next-order term $O(1)$ depends on the confining domain and the
location $\x_\Gamma$ of the target through the regular part
$R_0(\x_\Gamma)$ of the pseudo-Green's function.  For instance, for
the disk of radius $R$ with the arc-shaped target $\Gamma$ of angle
$2\ve$, one has $R_0(\x_\Gamma) = 1/(8\pi)$ and thus retrieves the
asymptotic result from \cite{Singer06b}.

(ii) For a partially reactive target ($0 < q < \infty$), the
situation is drastically different.  In fact, the dominant
contribution to Eq. (\ref{eq:Tq_volume_asympt}) comes from the first
term $|\Omega|/(2qD\epsilon)$, which diverges as $1/\epsilon$.  The
crucial role of partial reactivity in the narrow escape problem was
thoroughly discussed in \cite{Grebenkov17}.  When $q$ is fixed, the
sum in Eq. (\ref{eq:Tq_volume_asympt}) vanishes as $O(\epsilon)$ and
can thus be neglected.  However, if $q\epsilon$ is fixed (i.e., if $q$
grows as $\epsilon\to 0$), one gets a different scaling.  Moreover,
one can fix $\epsilon$ small enough and study the dependence of the
MFRT on the reactivity parameter $q$, in which case
Eq. (\ref{eq:Tq_volume_asympt}) provides a universal dependence of
$(\overline{T_q} - \overline{T_\infty} )/|\Omega|$ on $q\epsilon$,
regardless of the domain:
\begin{equation}  \label{eq:Tq_Tinf_2d}  
\overline{T_q} - \overline{T_\infty} \approx \frac{|\Omega|}{D} \Psi_2(q\epsilon),
\qquad \Psi_2(z) = \frac{1}{2z} + \sum\limits_{k=1}^\infty \frac{[\hat{V}_k(\infty)]^2}{\hat{\mu}_k + z} \,.
\end{equation}
This unique function monotonously decreases from $+\infty$ at $z = 0$
to $0$ as $z \to \infty$, and thus characterizes the impact of partial
reactivity onto the MFRT on a small target.  At small $z$, the
dominant contribution comes from the first term, $1/(2z)$, whereas the
sum gives higher-order corrections in $z = q\epsilon$:
\begin{equation}
z \Psi_2(z) \approx \frac12 + \sum\limits_{n=1}^\infty C_n z^n, 
\qquad C_n = \sum\limits_{k=1}^\infty \frac{[\hat{V}_k(\infty)]^2}{[\hat{\mu}_k]^n} \,.
\end{equation}
The first two coefficients $C_1$ and $C_2$ are given by
Eqs. (\ref{eq:sum_mu_bk2_2d}) and (\ref{eq:sum_bk2_2d}), respectively.
Comparing Eq. (\ref{eq:Tq_Tinf_2d}) with Eqs. [2,3] from
\cite{Guerin23}, we get the asymptotic behavior of the function
$\Psi_2(z)$ at large $z$:
\begin{equation}  \label{eq:Psi_2d_asympt}
z \Psi_2(z) \approx \frac{1}{\pi^2} \biggl(\ln(8z) + \gamma + 1\biggr)  \qquad (z \to\infty),
\end{equation}
where $\gamma \approx 0.5772$ is the Euler constant.  Figure
\ref{fig:psi}(a) presents the function $\Psi_2(z)$ and its asymptotic
behavior (\ref{eq:Psi_2d_asympt}), which are in excellent agreement.

\subsubsection*{Three dimensions}

In three dimensions, we substitute the asymptotic relations
(\ref{eq:main_asympt}, \ref{eq:aGamma_asympt_3d}, \ref{eq:bk_3d}) into
Eq. (\ref{eq:Tq_volume}) to get in the small-target limit:
\begin{align}    \label{eq:Tq_3d_auxil}
\overline{T_q} & \approx \frac{|\Omega|}{\pi D}\biggl(\frac{1}{q \epsilon^2} + \frac{8}{3\pi \epsilon}
+ \frac{H(\x_\Gamma)}{4} \ln (1/\epsilon)   
 - \frac{\pi}{\epsilon} \sum\limits_{k=1}^\infty \frac{[\hat{V}_k(\infty)]^2}{\hat{\mu}_k(1 + \hat{\mu}_k/(q\epsilon))} + O(1) \biggr).
\end{align}
As previously, we distinguish two cases:

(i) for a perfectly reactive target ($q = \infty$), one gets
\begin{equation}   \label{eq:Tinf_3d_auxil2}
\overline{T_\infty} \approx \frac{|\Omega|}{\pi D}\biggl(\frac{8}{3\pi \epsilon}
+ \frac{H(\x_\Gamma)}{4} \ln (1/\epsilon) 
 - \frac{\pi}{\epsilon} \sum\limits_{k=1}^\infty \frac{[\hat{V}_k(\infty)]^2}{\hat{\mu}_k} + O(1) \biggr).
\end{equation}
While the term $\a_\Gamma$ provided the dominant contribution to the
volume-averaged MFPT in the planar case, this is not true anymore in
three dimensions; indeed, the leading-order scaling $1/\epsilon$ comes
from both $\a_\Gamma$ and the spectral expansion.  This distinction
explains why the constant-flux approximation from
\cite{Grebenkov17} that ignores the spectral expansion, was more
accurate in two dimensions than in three dimensions.  Using
Eq. (\ref{eq:sum_mubk2_3d}) to evaluate the sum in
Eq. (\ref{eq:Tinf_3d_auxil2}), we get
\begin{equation}
\overline{T_\infty} \approx \frac{|\Omega|}{4D}\biggl(\frac{1}{\epsilon}
+ \frac{H(\x_\Gamma)}{\pi} \ln(1/\epsilon) + O(1) \biggr).
\end{equation}
The leading term of this expression goes back to Lord Rayleigh
\cite{Rayleigh} (see \cite{Singer06a} for more discussions).
As expected, the curvature of the boundary appears in the subleading,
logarithmic term.

\begin{figure}
\begin{center}
\includegraphics[width=0.49\textwidth]{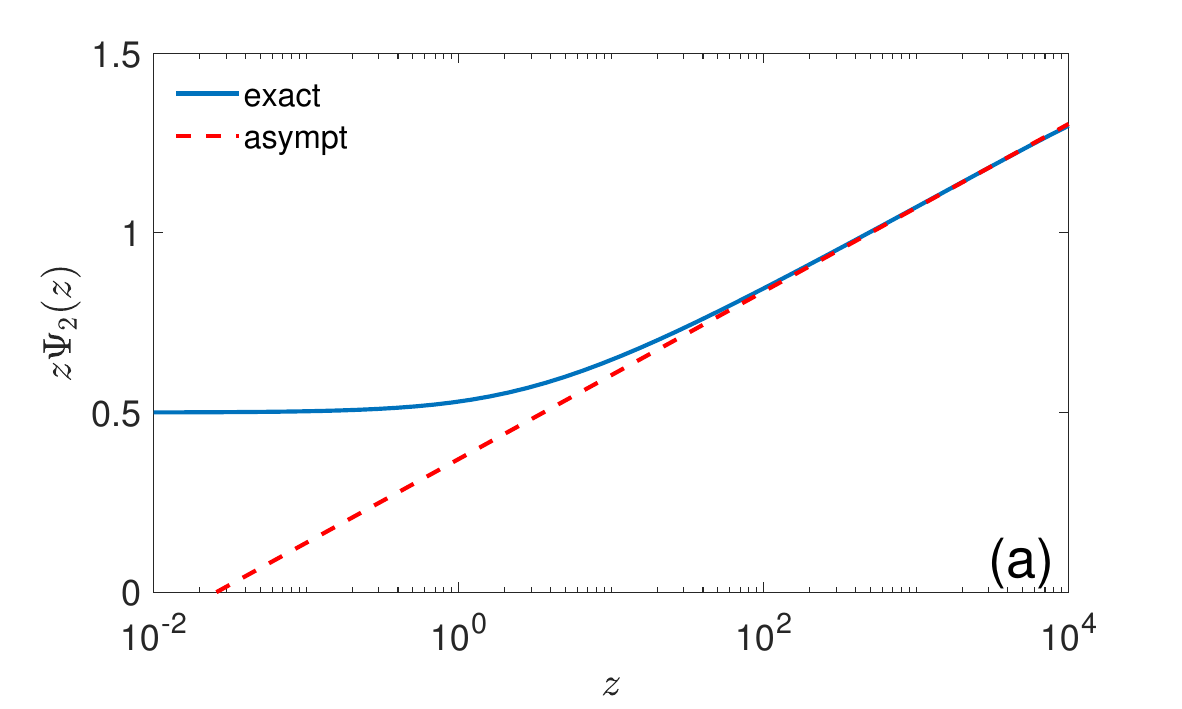} 
\includegraphics[width=0.49\textwidth]{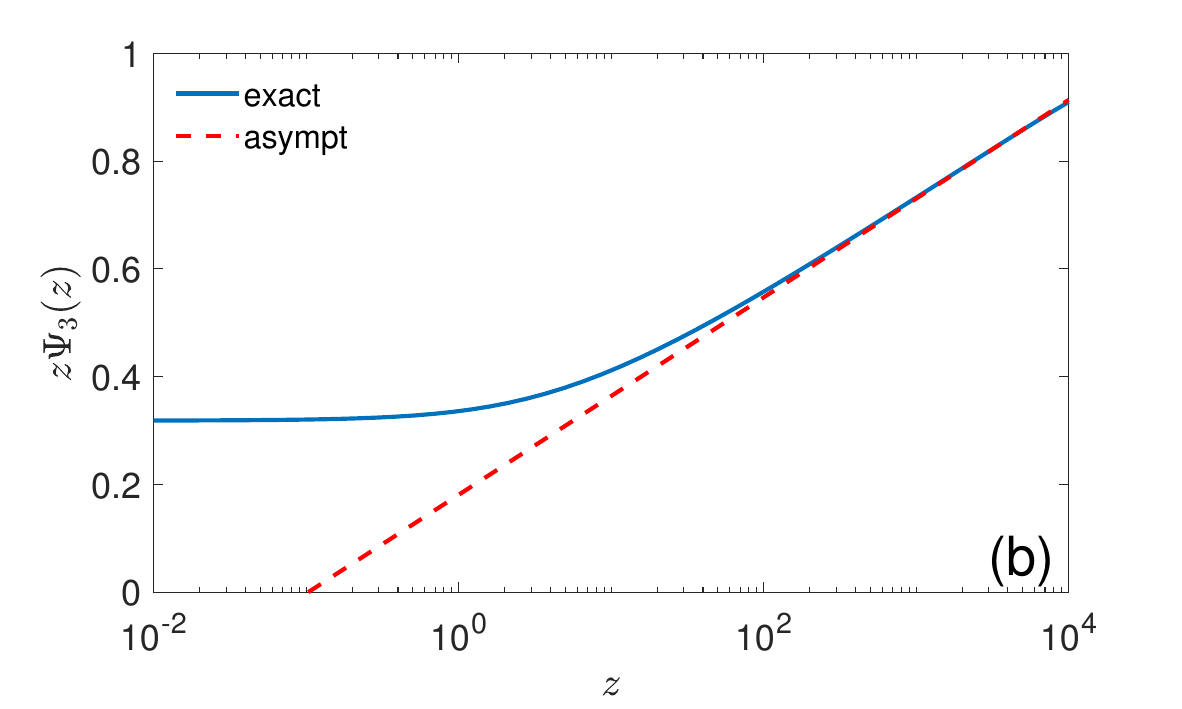} 
\end{center}
\caption{
{\bf (a)} The function $z \Psi_2(z)$ from Eq. (\ref{eq:Tq_Tinf_2d}), shown
by solid line, and its large-$z$ asymptotic behavior
(\ref{eq:Psi_2d_asympt}), shown by dashed line.  {\bf (b)} The
function $z \Psi_3(z)$ from Eq. (\ref{eq:Tq_Tinf_3d}), shown by solid
line, and its large-$z$ asymptotic behavior (\ref{eq:Psi_3d_asympt}),
shown by dashed line. }
\label{fig:psi}
% A_DN_ellipse_int_Psi_fig(2);
% A_DN_ellipse_int_Psi_fig(3);
\end{figure}

(ii) for a partially reactive target ($0 < q < \infty$),
Eq. (\ref{eq:Tq_3d_auxil}) can be rewritten as
\begin{align} \nonumber 
\overline{T_q} & \approx \frac{|\Omega|}{\pi D}\biggl(\frac{1}{q \epsilon^2} + \frac{8}{3\pi \epsilon}
+ \frac{H(\x_\Gamma)}{4} \ln (1/\epsilon) \\  \label{eq:Tq_3d_auxil2}
& - q \pi \sum\limits_{k=1}^\infty \frac{[\hat{V}_k(\infty)]^2}{\hat{\mu}_k^2} 
+ q^2\epsilon \pi \sum\limits_{k=1}^\infty \frac{[\hat{V}_k(\infty)]^2/\hat{\mu}_k}{q\epsilon + \hat{\mu}_k} + O(1) \biggr).
\end{align}
When $q$ is fixed, the last two terms are of the order $O(1)$ and
$O(\epsilon)$, respectively.  In turn, the first three terms provide
the leading and subleading contributions to $\overline{T_q}$.
Moreover, as the second and the third terms came from $\a_\Gamma$,
they are accessible within the constant-flux approximation, which
turns out to be much more accurate for a partially reactive target.
As previously, we obtain the universal dependence of $(\overline{T_q}
- \overline{T_\infty} )/|\Omega|$ on $q\epsilon$, regardless of the
domain:
\begin{equation}   \label{eq:Tq_Tinf_3d}
\overline{T_q} - \overline{T_\infty} \approx \frac{|\Omega|}{\epsilon D} \Psi_3(q\epsilon),
\qquad \Psi_3(z) = \frac{1}{z\pi} + \sum\limits_{k=1}^\infty \frac{[\hat{V}_k(\infty)]^2}{z + \hat{\mu}_k} \,.
\end{equation}
At small $z$, the dominant contribution comes from the first term,
$1/(z\pi)$, whereas the sum gives higher-order corrections in $z =
q\epsilon$:
\begin{equation}
z \Psi_3(z) \approx \frac{1}{\pi} + \sum\limits_{n=1}^\infty C_n z^n, 
\qquad C_n = \sum\limits_{k=1}^\infty \frac{[\hat{V}_k(\infty)]^2}{[\hat{\mu}_k]^n} \,.
\end{equation}
The first two coefficients $C_1$ and $C_2$ are given by
Eqs. (\ref{eq:sum_mubk2_3d}) and (\ref{eq:sum_bk2_3d}), respectively.
Comparing Eq. (\ref{eq:Tq_Tinf_3d}) with Eqs. [2,3] from
\cite{Guerin23}, we get the asymptotic behavior of the function
$\Psi_3(z)$ at large $z$:
\begin{equation}  \label{eq:Psi_3d_asympt}
z \Psi_3(z) \approx \frac{1}{4\pi} \biggl(\ln(2z) + \gamma + 1\biggr)  \qquad (z \to\infty).
\end{equation}
Figure \ref{fig:psi}(b) illustrates excellent accuracy of this
expression.

A common way to account for the effect of partial reactivity
consists in splitting the first-reaction time into two contributions:
the first-passage time to the target (the first arrival) and the
first-reaction time after restarting from the target.  {\it If} the
first arrival point onto the target was {\it uniformly} distributed,
one would simply get
\begin{equation}  \label{eq:Tq_conjectural}
\overline{T}_q \overset{?}{=} \overline{T}_\infty + \frac{1}{|\Gamma|} \int\limits_{\Gamma} d\x \, T_q(\x) 
=  \overline{T}_\infty + \frac{|\Omega|}{qD|\Gamma|} ,
\end{equation}
where the integral over $\Gamma$ was evaluated exactly by integrating
Eq.  (\ref{eq:Tq_problem_1}) over $\x_0\in \Omega$, using the Green's
formula and the boundary condition (\ref{eq:Tq_problem_2}).  One sees
that the hypothetical relation (\ref{eq:Tq_conjectural}) corresponds
to an approximation of the asymptotic relations (\ref{eq:Tq_Tinf_2d},
\ref{eq:Tq_Tinf_3d}) by replacing $\Psi_d(z)$ by $1/(2z)$ ($d = 2$) or
$1/(z\pi)$ ($d = 3$).  In other words, Eq. (\ref{eq:Tq_conjectural})
captures the first term of $\Psi_d(z)$ and ignores the remaining sum.
As a consequence, this sum originates from the fact that the first
arrival point on the target is distributed according to the harmonic
measure density, which is highly non-uniform and even exhibits
singularity at the edges of the target, see Eqs. (\ref{eq:HMD_2d},
\ref{eq:HMD_3d}).

On one hand, the smallness of the coefficients $[\hat{V}_k(\infty)]^2$
(see Table \ref{tab:eta_2D}) suggests that the approximation
(\ref{eq:Tq_conjectural}) is accurate at small or even moderate values
of the parameter $z = q\epsilon$, i.e., it justifies the approximation
(\ref{eq:Tq_conjectural}).  On the other hand, the sum in
Eqs. (\ref{eq:Tq_Tinf_2d}, \ref{eq:Tq_Tinf_3d}) is responsible for the
peculiar large-$z$ asymptotic behavior in
Eqs. (\ref{eq:Psi_2d_asympt}, \ref{eq:Psi_3d_asympt}) that was
discovered by Gu\'erin {\it et al.} \cite{Guerin23}.
While their work does not mention the Steklov-Neumann problem, many
employed tools are similar.  For instance, the integral equation [24]
from \cite{Guerin23} involves the Green's function
$\tilde{G}_0(\x,0|\x_0)$ in the half space and the related kernels.
The asymptotic analysis from \cite{Guerin23} may thus be reformulated
in terms of the Dirichlet-to-Neumann operator and the Steklov-Neumann
eigenfunctions.  To illustrate this point, we mention that the
first-order correction $f_1(x)$ for the three-dimensional setting,
defined via Eq. [88] from \cite{Guerin23}, can be found explicitly by
using the Landen transformation:
\begin{equation} % $K(\sqrt{4xy/(x+y)^2}) = (1+x/y) K(x/y)$
f_1(x) = \frac{2}{\pi^2}\biggl(\frac{4}{3} - E(x)\biggr),  \qquad c_1 = \frac{8}{3\pi^2} \approx 0.2702,
\end{equation}
where the constant $c_1$ (with unknown value) is defined by Eq. [85]
from \cite{Guerin23}, and $E(x)$ is given by Eq. (\ref{eq:EllipticE}).
The function $f_1(x)$, which determines the first-order correction to
the MFRT in the small reactivity regime, is proportional to our
function $\hat{w}_0(\hat{r})$ from Eq. (\ref{eq:hatw0_3d}), as it
should.

To give a broader picture of the small-target limit, we outline a
complementary approach \cite{Bressloff22b} that was developed to study
the narrow capture problem when the target is a small trap inside the
confining domain, which is separated from the reflecting wall $\pa_N$
by a large distance (i.e., the boundary $\pa$ is disconnected, see
Sec. \ref{sec:connectivity}).  Bressloff employed the matched
asymptotic analysis to the encounter-based formulation of
diffusion-controlled reactions and derived an asymptotic expansion of
the joint probability density for particle position and the boundary
local time, without employing spectral expansions like
Eq. (\ref{eq:Gq_spectral}) on the Steklov-Neumann eigenfunctions.  In
this light, his asymptotic analysis brings complementary insights to
our spectral results.

\subsection{Beyond constant reactivity}

A general framework for dealing with surface reactions on a target,
known as the encounter-based approach, was introduced in
\cite{Grebenkov20}.  At each encounter with the target $\Gamma$, the
particle attempts to react but may fail and thus resume its diffusion.
The surface reaction occurs after a sufficient number of encounters,
which is described by the boundary local time $\ell_t$ on $\Gamma$
\cite{Ito,Freidlin,Saisho87,Papanicolaou90}.  The first-reaction time
$\tau$ is thus defined as the first time instance when $\ell_t$
exceeds some threshold $\hat{\ell}$: 
\begin{equation}
\tau = \inf\{ t>0~:~ \ell_t > \hat{\ell}\}.  
\end{equation}
The conventional setting of a target with a constant reactivity $q >
0$, standing in the Robin boundary condition like
Eq. (\ref{eq:Gq_Robin}) or (\ref{eq:Sq_Robin}), corresponds to the
exponentially distributed threshold: $\P\{\hat{\ell} > \ell\} =
e^{-q\ell}$ \cite{Grebenkov20}.  In turn, other distributions of the
random threshold $\hat{\ell}$ yield more sophisticated surface
reactions.  In analogy to Eq. (\ref{eq:tildeHq}), the spectral
expansion of the moment-generating function of the associated
first-reaction time $\tau$ reads \cite{Grebenkov20}
\begin{equation}  \label{eq:tildeH}
\tilde{H}(p|\x_0) = \sum\limits_{k=0}^\infty V_k^{(p,\Gamma)}(\x_0) \Upsilon(\mu_k^{(p,\Gamma)}) \int\limits_{\Gamma} v_k^{(p,\Gamma)} 
\qquad (\x_0 \in \overline{\Omega}),
\end{equation}
where $\Upsilon(\mu) = \E\{ e^{-\mu \hat{\ell}} \}$ is the
moment-generating function of the threshold $\hat{\ell}$.  When
$\hat{\ell}$ obeys the exponential law, its moment-generating function
is $\Upsilon(\mu) = 1/(1 + \mu/q)$, and Eq. (\ref{eq:tildeH}) is
reduced to Eq. (\ref{eq:tildeHq}).  In general, if the mean value of
$\hat{\ell}$ is finite, its moment-generating function behaves as
$\Upsilon(\mu) = 1 - \mu \E\{ \hat{\ell}\} + o(\mu)$ as $\mu
\to 0$.  Repeating the small-$p$ asymptotic analysis from
Sec. \ref{sec:constant}, we deduce the MFRT:
\begin{align} \nonumber
\E_{\x_0}\{ \tau \} & = \frac{|\Omega|}{D} \biggl(\frac{\E\{ \hat{\ell} \}}{|\Gamma|} - W_0^{(\Gamma)}(\x_0) 
 -  \sum\limits_{k=1}^\infty V_k^{(0,\Gamma)}(\x_0) \Upsilon(\mu_k^{(0,\Gamma)}) \frac{b_k^{(\Gamma)}}{\mu_k^{(0,\Gamma)}} \biggr).
\end{align}
Similarly, the volume-averaged MFRT reads
\begin{align} \nonumber
\overline{T} & = \frac{1}{|\Omega|} \int\limits_{\Omega} d\x_0 \, \E_{\x_0}\{ \tau \} 
 = \frac{|\Omega|}{D} \biggl(\frac{\E\{ \hat{\ell} \}}{|\Gamma|} + \a_\Gamma 
  -  \sum\limits_{k=1}^\infty \Upsilon(\mu_k^{(0,\Gamma)}) \frac{[b_k^{(\Gamma)}]^2}{\mu_k^{(0,\Gamma)}} \biggr).
\end{align}
These expressions extend the spectral expansions
(\ref{eq:Tq_spectral}, \ref{eq:Tq_volume}) to more general surface
reactions (note that $\E\{ \hat{\ell} \} = 1/q$ for the exponentially
distributed threshold $\hat{\ell}$).  While the dominant contribution
(the first two terms) does not change (except that $1/q$ is replaced
by $\E\{ \hat{\ell}\}$), the correction term is affected by the chosen
surface reaction via $\Upsilon(\mu)$.  Our asymptotic results for
$\mu_k^{(0,\Gamma)}$ and $b_k^{(\Gamma)}$ can thus help understanding
sophisticated surface reactions on small targets (see also
\cite{Bressloff22b}).  Moreover, if the mean threshold is infinite,
$\E\{ \hat{\ell} \} = +\infty$, the MFRT is also infinite, despite the
fact that diffusion occurs in a bounded domain.

\section{Conclusion}
\label{sec:conclusion}

In this paper, we revisited the mixed Steklov-Neumann spectral problem
in a bounded Euclidean domain $\Omega$, with the Steklov condition on
a subset $\Gamma$ of a smooth boundary $\pa$.  In the first step, we
discussed a general scheme for constructing the eigenvalues
$\mu_k^{(p,\Gamma)}$ and eigenfunctions $V_k^{(p,\Gamma)}$ with $p >
0$ from the restriction of the Green's function
$\tilde{G}_0(\x,p|\x_0)$ to $\Gamma\times \Gamma$.  In the limit $p
\to 0$, this construction involved the pseudo-Green's function
$\G_0(\x|\x_0)$ and the first-order corrections $\a_\Gamma$ and
$W_0^{(\Gamma)}$ to the principal eigenvalue $\mu_0^{(p,\Gamma)}$ and
eigenfunction $V_0^{(p,\Gamma)}$ as $p \to 0$.  These corrections were
shown to emerge in different contexts such as the long-time asymptotic
behavior of the variance of the boundary local time on $\Gamma$, the
asymptotic behavior of the MFRT on $\Gamma$, its constant-flux
approximation, etc.

In the second step, we obtained the asymptotic relations
(\ref{eq:main_asympt}) for the eigenvalues and eigenfunctions in the
small-target limit when the subset $\Gamma$ shrinks.  For this
purpose, we identified two auxiliary Steklov-Neumann problems and also
constructed the explicit kernels of integral operators that determine
the eigenpairs $\hat{\mu}_k$ and $\hat{V}_k$ appearing in
Eq. (\ref{eq:main_asympt}).  We also described efficient matrix
representations to compute these eigenpairs numerically with high
precision.  The asymptotic behavior of the corrections $\a_\Gamma$ and
$w_0^{(\Gamma)}$ and of the coefficients $b_k^{(\Gamma)}$ was derived.
The general results were illustrated for two basic examples: an
arc-shaped subset $\Gamma$ on a circle and a spherical cap on a
sphere.  By solving the original Steklov-Neumann problem numerically,
we revealed high accuracy of the asymptotic relations
(\ref{eq:main_asympt}), even for moderately large subsets $\Gamma$.

In the third step, we presented a direct application of the derived
asymptotic results to the analysis of first-reaction times and related
diffusion-controlled reactions.  Since the moment-generating function
of the first-reaction time on a subset $\Gamma$ admits a spectral
expansion over the Steklov-Neumann eigenpairs, its asymptotic behavior
in the small-target limit can be directly accessed.  In particular, we
retrieved and generalized the asymptotic behavior of the MFRT for both
perfect and partially reactive targets.  We also revealed why the
constant-flux approximation provided the correct leading terms in two
dimensions but yielded wrong prefactors in three dimensions.  We
derived a universal, geometry-independent formula for the difference
between volume-averaged MFRT (for a finite reactivity) and MFPT (for
infinite reactivity) in the small-target limit.  Moreover, we also
discussed an extension to more sophisticated surface reactions beyond
the conventional setting of constant reactivity.  Further
investigations of the narrow escape limit in the context of the mixed
Steklov-Neumann problem can significantly improve our understanding of
diffusion-controlled reactions in complex environments such as
biological tissues or porous media.

\section*{Acknowledgments}
The author thanks professors A. F. M. ter Elst, P. Freitas,
M. Levitin, and I. Polterovich for fruitful discussions, and
A. Chaigneau for preliminary numerical results.  The author
acknowledges the Simons Foundation for supporting his sabbatical
sojourn in 2024 at the CRM (CNRS -- University of Montr\'eal, Canada),
as well as the Alexander von Humboldt Foundation for support within a
Bessel Prize award.

%\section*{Data Availability Statement}
%
%The data that support the findings of this study are available from
%the corresponding author upon reasonable request.

%%%%%%%%%%%%%%%%%%%%%%%%%%%%%%%%%%%%%%%%%%%%%%%%%%%%%%%%%%%%%%%%%%%%%%%%%%%%%%%

\appendix
\section{Some derivations}

\subsection{Identity (\ref{eq:dmu_dp})}
\label{sec:identity}

For the Steklov problem, the identity (\ref{eq:dmu_dp}) was rigorously
established in \cite{Friedlander91} (see also
\cite{Grebenkov19,Chaigneau24}).  Its extension to the Steklov-Neumann
problem seems to rather straightforward but we are not aware of its
proof.  In the following, we provide a formal derivation of this
identity.

Multiplying Eq. (\ref{eq:Helm}) by $V_k^{(p+\delta,\Gamma)}$ with some
$\delta > 0$, multiplying Eq. (\ref{eq:Helm}) with $p' = p+\delta$ by
$V_k^{(p,\Gamma)}$, subtracting them, integrating over $\Omega$ and
using the Green's formula, one gets
\begin{align*}
\delta \int\limits_{\Omega} V_k^{(p,\Gamma)} \, V_k^{(p+\delta,\Gamma)} 
& = D \int\limits_{\Gamma} \bigl(V_k^{(p,\Gamma)} \partial_n V_k^{(p+\delta,\Gamma)} -  
V_k^{(p+\delta,\Gamma)} \partial_n V_k^{(p,\Gamma)} \bigr) \\
& = D \bigl(\mu_k^{(p+\delta,\Gamma)} - \mu_k^{(p,\Gamma)}\bigr) \int\limits_{\Gamma} v_k^{(p,\Gamma)} v_k^{(p+\delta,\Gamma)} .
\end{align*}
Dividing this identity by $\delta$ and taking the limit $\delta
\to 0$ yield
\begin{equation*}
\partial_p \mu_k^{(p,\Gamma)} = \lim\limits_{\delta\to 0} \frac{\mu_k^{(p+\delta,\Gamma)} - \mu_k^{(p,\Gamma)}}{\delta}
= \frac{1}{D} \int\limits_{\Omega} |V_k^{(p,\Gamma)}|^2,
\end{equation*}
where we used the normalization (\ref{eq:L2norm}) of eigenfunctions
$v_k^{(p,\Gamma)}$.  The existence of the limit follows from the
analyticity of the Dirichlet-to-Neumann map, see \cite{Behrndt15} and
the discussion before Eq. (\ref{eq:mu0v0_p0}).  Strictly speaking, one
has to follow the same ($k$-th) analytical branch for $p$ and
$p+\delta$ that may perturb the increasing order of eigenvalues.

\subsection{Spectral expansion}
\label{sec:Gq_spectral}

Even though the spectral expansion (\ref{eq:Gq_spectral}) resembles
that reported in \cite{Grebenkov20}, we sketch here its formal
derivation for mixed Robin-Neumann conditions.  The expression
(\ref{eq:Gq_spectral}) can be deduced by searching
$\tilde{G}_q(\x,p|\x_0)$ as $\tilde{G}_\infty(\x,p|\x_0) +
\tilde{g}_q(\x,p|\x_0)$, with an unknown function
$\tilde{g}_q(\x,p|\x_0)$ that satisfies the homogeneous modified
Helmholtz equation and can thus be expressed by using the completeness
of the eigenfunctions $v_k^{(p,\Gamma)}$ in $L^2(\Gamma)$.  Here we
follow an alternative way by verifying directly that
Eq. (\ref{eq:Gq_spectral}) satisfies the boundary value problem
(\ref{eq:Gq_problem}).

As Eqs. (\ref{eq:Gq_Helm}, \ref{eq:Gq_Neumann}) are satisfied by
construction, it remains to check the Robin boundary condition
(\ref{eq:Gq_Robin}).  Substituting Eq. (\ref{eq:Gq_spectral}) into
Eq. (\ref{eq:Gq_Robin}), one gets for any $\x_0 \in \Omega$ and $\x\in
\Gamma$:
\begin{align}   \label{eq:auxil37}
& D(\partial_n + q) \tilde{G}_q(\x,p|\x_0) 
 = - \tilde{j}_\infty(\x,p|\x_0) + \sum\limits_{k=0}^\infty V_k^{(p,\Gamma)}(\x_0) v_k^{(p,\Gamma)}(\x) ,
\end{align}
where $\tilde{j}_\infty(\x,p|\x_0) = - D \partial_n
\tilde{G}_\infty(\x,p|\x_0)$.  To check that the right-hand side is
zero, we first note that the Steklov eigenfunction
$V_k^{(p,\Gamma)}(\x_0)$ for $\x_0\in \Omega$ can be obtained from its
restriction $v_k^{(p,\Gamma)}$ to $\Gamma$ as
\begin{equation} \label{eq:auxil38}
V_k^{(p,\Gamma)}(\x_0) = \int\limits_{\Gamma} d\x'  \, \tilde{j}_\infty(\x',p|\x_0) \, v_k^{(p,\Gamma)}(\x')
\qquad (\x_0\in\Omega).
\end{equation}
This relation is deduced in a standard way by multiplying
Eq. (\ref{eq:Helm}) by $\tilde{G}_\infty(\x,p|\x_0)$, multiplying
Eq. (\ref{eq:Gq_Helm}) by $V_k^{(p,\Gamma)}(\x)$, subtracting these
equations, integrating over $\x\in\Omega$, applying the Green's
formula, and using the boundary conditions.  Substituting the
representation (\ref{eq:auxil38}) into Eq. (\ref{eq:auxil37}), we have
\begin{align*}
& D(\partial_n + q) \tilde{G}_q(\x,p|\x_0) = - \tilde{j}_\infty(\x,p|\x_0) 
 + \int\limits_{\Gamma} d\x' \, \tilde{j}_\infty(\x',p|\x_0) \sum\limits_{k=0}^\infty  v_k^{(p,\Gamma)}(\x') \,  v_k^{(p,\Gamma)}(\x) ,
\end{align*}
where the order of sum and integral was exchanged.  Second, we employ
the completeness relation
\begin{equation}  \label{eq:vk_completeness}
\sum\limits_{k=0}^\infty  v_k^{(p,\Gamma)}(\x') \,  v_k^{(p,\Gamma)}(\x) = \delta(\x-\x')
\end{equation}
that formally reflects that the basis of the eigenfunctions
$\{v_k^{(p,\Gamma)}\}$ is complete in $L^2(\Gamma)$ (this relation can
also be formally understood as the expansion of the Dirac distribution
on the basis $\{ v_k^{(p,\Gamma)}\}$).  As a consequence, one sees
that the right-hand side is zero, so that the spectral expansion
(\ref{eq:Gq_spectral}) satisfies the Robin boundary condition.

We emphasize that the above derivation is not a mathematical proof; in
particular, we do not discuss the convergence of the above expansions,
as well as the possibility of exchanging the order sum and normal
derivative, integration by parts, etc.  Spectral expansions of Green's
functions over Laplacian eigenfunctions (such as
Eq. (\ref{eq:G0_Laplacian})) are fairly standard.  In turn, we are not
aware of convergence results for Eq. (\ref{eq:Gq_spectral}) and
related expressions used in the manuscript.  For instance, a variant
of Eq. (\ref{eq:Gq_spectral}) for $q = 0$ and Steklov problem (i.e.,
$\Gamma = \pa$) was rigorously established in \cite{Behrndt15}
(Theorem 4.9) in a much more general setting.  We note that similar
expansions involving Steklov eigenfunctions and single/double layer
potentials were rigorously established in \cite{Auchmuty18}.
Moreover, if (i) $\Gamma = \pa$, (ii) the boundary $\pa$ is
real-analytic, and (iii) at least one point $\x$ or $\x_0$ does not
belong to $\pa$, then the Steklov eigenfunctions $V_k^{(p,\pa)}$ are
known to decay exponentially with $k$ \cite{Helffer22}, thus ensuring
the fast convergence of the series.  To our knowledge, similar results
are not yet established for mixed Steklov-Neumann problem.  A rigorous
justification of the spectral expansion (\ref{eq:Gq_spectral})
presents an open mathematical problem.

\section{Disk}
\label{sec:Adisk}

In this Appendix, we recall the exact explicit formulas for the
Green's function $\tilde{G}_0(\x,p|\x_0)$ and the pseudo-Green's
function $\G_0(\x|\x_0)$ for the disk of radius $R$.  We also deduce
the exact expressions for the corrections $\a_\Gamma$ and
$w_0^{(\Gamma)}$.  We finally present an implementation for computing
numerically the eigenpairs $\{\mu_k^{(p,\Gamma)}, V_k^{(p,\Gamma)}\}$.

\subsection{General solution}
\label{sec:disk_general}

The eigenmodes of the conventional Steklov problem, as well as the
Green's function, are fairly well-known for the disk
\cite{Levitin,Grebenkov20c}.  For instance, the restriction of the
Green's function to the boundary $\pa$ reads in polar coordinates as
(see, e.g., \cite{Grebenkov20c})
\begin{align}   \label{eq:G0p_disk}
\tilde{G}_0(\theta,p|\theta_0) & = \frac{1}{\pi D R} \biggl(\frac{1}{2\mu_0^{(p,\pa)}} 
+ \sum\limits_{k=1}^\infty \frac{\cos(k(\theta-\theta_0))}{\mu_k^{(p,\pa)}} \biggr) ,
\end{align}
where 
\begin{equation}
\mu_k^{(p,\pa)} = \sqrt{p/D} \frac{I'_k(R\sqrt{p/D})}{I_k(R\sqrt{p/D})} \,,
\end{equation}
with $I_\nu(z)$ being the modified Bessel function of the first kind,
and prime denoting the derivative with respect to the argument.
Writing $\cos(k(\theta - \theta_0)) = \cos(k\theta)\cos(k\theta_0) +
\sin(k\theta) \sin(k\theta_0)$, one can recognize the spectral
expansion (\ref{eq:G0_spectral}) over the Steklov eigenvalues
$\mu_k^{(p,\pa)}$ and eigenfunctions $v_k^{(p,\pa)}$, which are given
by cosine and sine functions \cite{Grebenkov20c}.  The principal
eigenvalue $\mu_0^{(p,\pa)}$ is simple, whereas the other eigenvalues
are twice degenerate (throughout this Appendix, we ignore the
degeneracy and enumerate by $k$ distinct eigenvalues).

We consider the subset $\Gamma$ to be an arc of angle $2\ve$ on the
boundary of the disk: $\Gamma = \{ (R,\theta)\in\pa ~:~
|\theta|<\ve\}$.  According to Eq. (\ref{eq:Gp_problem}), the
restriction of the Green's function $\tilde{G}_0(\theta,p|\theta_0)$
to $\Gamma \times \Gamma$ defines an integral operator that determines
the eigenvalues and eigenfunctions of the Dirichlet-to-Neumann
operator $\M_p^{(\Gamma)}$ for any $\ve$ and any $p > 0$:
\begin{equation}  \label{eq:spectral_disk_p}
\int\limits_{-\ve}^\ve d\theta \, D\tilde{G}_0(\theta,p|\theta_0) \, v_k^{(p,\Gamma)}(\theta) 
= \frac{v_k^{(p,\Gamma)}(\theta_0)}{R\mu_k^{(p,\Gamma)}}    
\qquad     (|\theta_0| < \ve, ~ k \geq 0).
\end{equation}
Replacing this integral by a Riemann sum yields an approximate matrix
representation of the integral operator, while its numerical
diagonalization approximates eigenvalues and eigenfunctions (see
Appendix \ref{sec:numerics_2d}).

The analysis is a little subtler at $p = 0$.  Since $\mu_k^{(0,\pa)} =
k/R$, one needs to remove the diverging term $1/\mu_0^{(p,\pa)}$ in
Eq. (\ref{eq:G0p_disk}) by subtracting $D/(p |\Omega|)$ that yields
the pseudo-Green's function
\begin{align}
\G_0(\theta|\theta_0) & = \frac{1}{8\pi}
+ \frac{1}{\pi} \sum\limits_{k=1}^\infty \frac{\cos(k(\theta-\theta_0))}{k} 
 = \frac{1}{8\pi} - \frac{1}{2\pi} \ln(2 - 2\cos(\theta-\theta_0))
\end{align}
(note that a more general explicit form of the pseudo-Green's function
inside the unit disk was given in \cite{Kolokolnikov05}, see
Eq. [4.3a]).  In particular, in the limit $\theta \to \theta_0$,
one retrieves the asymptotic behavior (\ref{eq:Gpseudo_2d_asympt}),
with
\begin{equation}
R_0 = \frac{1}{8\pi} \,,
\end{equation}
independently of $\theta_0$.
Substituting this expression into Eqs. (\ref{eq:a_eps},
\ref{eq:omega_eps}), we find
\begin{align} \nonumber 
w_0^{(\Gamma)}(\theta_0) & = - \a_\Gamma + \frac{1}{2\ve} \int\limits_{-\ve}^{\ve} d\theta \, \G_0(\theta|\theta_0)   
 = - \a_\Gamma + \frac{1}{8\pi} + \frac{1}{\pi \ve} \sum\limits_{k=1}^\infty \frac{\sin(k\ve) \cos(k\theta_0)}{k^2} \\ \label{eq:w0_disk}
& = - \a_\Gamma + \frac{1}{8\pi} + \frac{\Li_2(e^{i(\ve + \theta)})   
 - \Li_2(e^{-i(\ve + \theta)}) + \Li_2(e^{i(\ve - \theta)}) - \Li_2(e^{-i(\ve - \theta)})}{4i \pi \ve} \, ,
\end{align}
and
\begin{align}   \label{eq:aGamma_disk}  
\a_\Gamma & = \frac{1}{8\pi} + \frac{1}{\pi \ve^2} \sum\limits_{k=1}^{\infty} \frac{\sin^2(k\ve)}{k^3}  
= \frac{1}{8\pi} + \frac{2\Li_3(1) - \Li_3(e^{2i\ve}) - \Li_3(e^{-2i\ve})}{4\pi \ve^2}  \,,
\end{align}
where
\begin{equation}
\Li_n(z) = \sum\limits_{k=1}^\infty  \frac{z^k}{k^n} 
\end{equation}
is the polylogarithm.
Substitution of these relations in Eq. (\ref{eq:G_def}) yields the
kernel
\begin{align}    \label{eq:Gkernel}
\G(\theta|\theta_0) & = - \frac{1}{2\pi} \ln(2-2\cos(\theta-\theta_0)) \\ \nonumber
& - \frac{1}{\pi} \sum\limits_{k=1}^\infty \biggl(\frac{\sin(k\ve) [\cos(k\theta_0) + \cos(k\theta)]}{\ve k^2} 
- \frac{\sin^2(k\ve)}{\ve^2 k^3} \biggr),
\end{align}
which can also be expressed in terms of polylogarithms.

In summary, the eigenvalue problem (\ref{eq:eigen_G}) reads here as
\begin{equation}  \label{eq:eigenproblem_disk}
\int\limits_{-\ve}^\ve d\theta \, \G(\theta|\theta_0) \, v_k^{(0,\Gamma)}(\theta) 
= \frac{1}{R\mu_k^{(0,\Gamma)}} v_k^{(0,\Gamma)}(\theta_0) 
  \qquad     (|\theta_0| < \ve, ~ k \geq 1).
\end{equation}
This equation is valid for any $0 < \ve \leq \pi$ but requires a
numerical treatment (see Appendix \ref{sec:numerics_2d}).  If the
eigenvalue $\mu_k^{(0,\Gamma)}$ is simple, the symmetry of the
considered Steklov-Neumann problem implies that the associated
eigenfunction $v_k^{(0,\Gamma)}(\theta)$ is either symmetric or
antisymmetric: $v_k^{(0,\Gamma)}(-\theta) = \pm
v_k^{(0,\Gamma)}(\theta)$.  Lower and upper bounds on the eigenvalues
are obtained in Appendix \ref{sec:bounds}.

In the limit $\ve \to 0$, one can use the behavior of the
polylogarithm $\Li_n(z)$ as $z\to 1$ \cite{Abramowitz} to investigate
the asymptotic behavior of Eqs. (\ref{eq:w0_disk},
\ref{eq:aGamma_disk}, \ref{eq:Gkernel}).  In this way, one can retrieve
the asymptotic relations (\ref{eq:aGamma_asympt_2d},
\ref{eq:hatw0_asympt_2d}, \ref{eq:Glim_disk}); moreover, this
alternative derivation illustrates our statement that the curvature of
the boundary does not appear in the leading order.

\subsection{Numerical implementation}
\label{sec:numerics_2d}

Let us briefly discuss the numerical implementation of the considered
spectral problems that was realized in Matlab.

\subsubsection*{Arbitrary arc}

For a numerical implementation of the spectral problem
(\ref{eq:eigenproblem_disk}), the interval $(-1,1)$ is discretized
into $2N$ segments, centered at $y_n = (n-N-1/2)/N$, with $n =
1,2,\ldots,2N$.  Setting $\eta = 1/(2N)$, we compute the matrix of
size $(2N)\times (2N)$:
\begin{align} \nonumber
\GG_{n,n'}^{(\ve)} & = \int\limits_{\ve(y_n - \eta)}^{\ve(y_n + \eta)} d\theta \, \G(\theta|\ve y_{n'}) 
= \ve \int\limits_{y_n - \eta}^{y_n + \eta} dy \, \G(\ve y|\ve y_{n'}) \\  \nonumber
& \approx \frac{2R}{\pi} \sum\limits_{k=1}^{\kmax} \biggl[\frac{\cos(k\ve (y_n - y_{n'})) \sin(k\ve \eta)}{k^2} 
+ \eta \frac{\sin^2(k\ve)}{\ve k^3} \\   \label{eq:Gmatrix_ve}
& - \frac{\sin(k\ve)}{\ve k^2} \biggl(\eta \ve \cos(k\ve y_{n'}) + \cos(k\ve y_n) \frac{\sin(k\ve \eta)}{k}\biggr) \biggr],
\end{align}
where the kernel $\G(\theta|\theta_0)$ was given by
Eq. (\ref{eq:Gkernel}), and $k_{\rm max}$ is the truncation order that
may need to be large if $\ve$ is small.
The integral eigenvalue problem (\ref{eq:eigenproblem_disk}) is then
approximated as
\begin{equation}
\sum\limits_{n=1}^{2N} \GG_{n,n'}^{(\ve)} v_k^{(0,\Gamma)}(\ve y_n) \approx \frac{1}{R \mu_k^{(0,\Gamma)}} v_k^{(0,\Gamma)}(\ve y_{n'})
\qquad (n=1,2,\ldots,2N).
\end{equation}
In other words, one needs to diagonalize the matrix $\GG^{(\ve)}$ in
order to approximate the (inverse) eigenvalues and eigenfunctions of
the Dirichlet-to-Neumann operator $\M_0^{(\Gamma)}$.  We checked
numerically the convergence of several first eigenvalues and
eigenfunctions as $N$ increases.  Moreover, the eigenvalues and
eigenfunctions were compared to those obtained independently by a
finite-element method \cite{Chaigneau24}.

A very similar computation is also applicable to the eigenvalue
problem (\ref{eq:eigen_disk_eps0}) for the kernel $\hat{\G}(x|x_0)$
from Eq. (\ref{eq:Glim_disk}).  However, its simpler form allows for a
more efficient computation described in the next section.

\subsubsection*{Matrix representation of the kernel $\hat{\G}(x|x_0)$}

The eigenvalue problem (\ref{eq:eigen_disk_eps0}) admits a simple
matrix representation that relies on the following expansion in terms
of Chebyshev polynomials of the first kind for $x, x_0
\in [-1,1]$ and $x \ne x_0$ (see, e.g., \cite{Davis07}):
\begin{equation}
\ln (2|x-x_0|) = - \sum\limits_{n=1}^\infty \frac{2}{n} T_n(x) \, T_n(x_0).
\end{equation}  % see A_DN_disk_mixed_G_Cheb0();
The kernel $\hat{\G}(x|x_0)$ can then be represented as
\begin{align} \label{eq:G_Chebyshev2}
\hat{\G}(x|x_0) & = - \frac{1 + 2\ln 2}{2\pi} + \frac{1}{2\pi} \biggl((1+x_0)\ln(1+x_0) + (1-x_0)\ln(1-x_0)\biggr) \\ \nonumber
& + \frac{1}{\pi} \sum\limits_{n=1}^\infty \frac{1}{n} \biggl\{ 2T_n(x) T_n(x_0)
 - T_n(x)\bigl(1+(-1)^n- x(1-(-1)^n)\bigr) \biggr\} .
\end{align} % see A_DN_disk_mixed_G_Cheb1();
Let us search for an eigenfunction $\hat{v}_k(x)$ in the form
\begin{equation}  \label{eq:hatvk_T}
\hat{v}_k(x) = \frac{1}{\sqrt{1-x^2}} \sum\limits_{n'=1}^{\infty} C_{k,n'} T_{n'}(x),
\end{equation}
with unknown coefficients $C_{k,n'}$, where the term $n' = 0$ was
excluded from the sum to ensure that $\hat{v}_k(x)$ is orthogonal to a
constant $T_0(x) = 1$.  Substituting this expression into
Eq. (\ref{eq:eigen_disk_eps0}), multiplying by $T_m(x_0)$ and
integrating over $x_0$ from $-1$ to $1$, we reduce the original
integral problem to the equivalent matrix problem
\begin{equation}
\sum\limits_{n=1}^{\infty} C_{k,n} \GG_{n,m} = \frac{1}{\hat{\mu}_k} C_{k,m}   \qquad (m=1,2,\ldots),
\end{equation}
where
\begin{equation}
\GG_{n,m} = \frac{2}{\pi} \int\limits_{-1}^1 \frac{dx}{\sqrt{1-x^2}} \int\limits_{-1}^1 dx_0 \,  T_n(x)\, \hat{\G}(x|x_0) \, T_m(x_0),
\end{equation}
and we used the orthogonality and normalization of Chebyshev
polynomials.  Substituting Eq. (\ref{eq:G_Chebyshev2}), we can
evaluate the matrix elements explicitly:
\begin{align} \nonumber
\GG_{n,m} & = \frac{1+(-1)^{m+n}}{\pi n} \biggl(\frac{1}{1-(m-n)^2} + \frac{1}{1-(m+n)^2}\biggr) \\  \label{eq:Glim_matrix}
& - \frac{(1+(-1)^n)(1+(-1)^m)}{\pi n (1-n^2)(1-m^2)}  \qquad (m,n=1,2,\ldots)
\end{align}  % A_DN_disk_mixed_G_Cheb2();
(note that the ``seemingly divergent'' terms with $m=1$, $n=1$,
$(m-n)^2 = 1$ or $(n+m)^2 = 1$ are set to zero).  The eigenvalues of
the truncated matrix $\GG$ approximate $1/\hat{\mu}_k$, whereas its
left eigenvectors yield the coefficients $C_{k,n}$ determining the
eigenfunctions $\hat{v}_k(x)$ via Eq. (\ref{eq:hatvk_T}).  We checked
a rapid convergence of the numerical values of $1/\hat{\mu}_k$ as the
truncation size increases.  In fact, this method is much faster and
more accurate that a direct discretization of the integral equation.

\section{Ball}
\label{sec:Aball}

In this Appendix, we recall the exact explicit formulas for the
Green's function $\tilde{G}_0(\x,p|\x_0)$ and the pseudo-Green's
function $\G_0(\x|\x_0)$ for the ball of radius $R$.  We also deduce
the exact expressions for the corrections $\a_\Gamma$ and
$w_0^{(\Gamma)}$.  We finally present an implementation for computing
numerically the eigenpairs $\{\mu_k^{(p,\Gamma)}, V_k^{(p,\Gamma)}\}$.

\subsection{General solution}
\label{sec:ball_general}

The restriction of the Green's function to the boundary $\pa$ reads in
the spherical coordinates as (see, e.g., \cite{Grebenkov20c})
\begin{equation}
\tilde{G}_0(\x,p|\x_0) = \frac{1}{R^2 D} \sum\limits_{n=0}^\infty \sum\limits_{m=-n}^n 
\frac{Y_{mn}(\theta,\varphi) Y_{mn}^*(\theta_0,\varphi_0)}{\mu_n^{(p,\pa)}} \,,
\end{equation}
where 
\begin{equation}
\mu_n^{(p,\pa)} = \sqrt{p/D}\, \frac{i'_n(R\sqrt{p/D})}{i_n(R\sqrt{p/D})} \,,
\end{equation}
$i_n(z)$ is the modified spherical Bessel function of the first kind,
\begin{equation}
Y_{mn}(\theta,\varphi) = \sqrt{\frac{(2n+1)(n-m)!}{4\pi (n+m)!}} P_n^m(\cos\theta) e^{im\varphi} 
\end{equation}
are the normalized spherical harmonics, and $P_n^m(z)$ are the
associated Legendre polynomials.  Note that the eigenvalue
$\mu_n^{(p,\pa)}$ is $(2n+1)$ times degenerate, whereas the
eigenfunctions of the Dirichlet-to-Neumann operator $\M_p^{(\pa)}$,
given by spherical harmonics, do not depend on $p$.  These properties
are specific to the ball and follow from the rotational invariance of
the problem.
According to Eq. (\ref{eq:Gp_problem}), the restriction of
$D\tilde{G}_0(\x,p|\x_0)$ to $\Gamma\times \Gamma$ defines an integral
operator, which determines the eigenvalues and eigenfunctions of the
Dirichlet-to-Neumann operator $\M_p^{(\Gamma)}$.  This construction is
valid for any $0 < \ve \leq \pi$ and $p > 0$, though it requires a
numerical treatment.  In the following, we focus on the limit $p = 0$.

Substituting the expansion 
\begin{equation*}
\frac{1}{\mu_0^{(p,\pa)}} \approx \frac{3}{R p/D} + \frac{R}{5} + O(p)  \qquad (p\to 0)
\end{equation*}
into Eq. (\ref{eq:G0_limit}), we retrieve the restriction of the
pseudo-Green's function to the boundary $\pa$,
\begin{equation}  \label{eq:G0_sphere_whole}
\G_0(\x|\x_0) = \frac{1}{R} \biggl(\frac{1}{20\pi} + \sum\limits_{n=1}^\infty \sum\limits_{m=-n}^n 
\frac{Y_{mn}(\theta,\varphi) Y_{mn}^*(\theta_0,\varphi_0)}{n} \biggr),
\end{equation}
where we used $\mu_n^{(0,\pa)} = n/R$.  A more explicit form of the
pseudo-Green's function inside a ball is given by Eq. [2.26a] in Ref.
\cite{Cheviakov11}.  Its restriction to the boundary $\pa$ reads:
\begin{align} \label{eq:G0_sphere_whole2}
\G_0(\x|\x_0) & = \frac{1}{4\pi R} \biggl\{\frac{\sqrt{2}}{\sqrt{1-\cos\gamma}} - \frac{9}{5} 
 + \ln\biggl(\frac{2}{1-\cos\gamma + \sqrt{2(1-\cos\gamma)}}\biggr)\biggr\},
\end{align}  % checked in [G0,YY,YY0] = A_DN_sphere_pseudoGreen();
where $\gamma$ is the angle between two vectors pointing from the
origin at $\x$ and $\x_0$ on the unit sphere:
\begin{equation} \label{eq:cosgamma}
\cos\gamma = \cos \theta \cos \theta_0 + \sin \theta \sin\theta_0 \cos(\varphi-\varphi_0)
\end{equation}
(note that the expression (\ref{eq:G0_sphere_whole2}) can be
found in the classical textbook \cite{Kellog}, p. 247).
In particular, in the limit $\x \to \x_0$, one retrieves the
asymptotic behavior (\ref{eq:G0_asympt_3d}), with
\begin{equation}  \label{eq:R0_ball}
R_0(\x_0) = \frac{\ln(2) - 9/5 + \ln R}{4\pi R} \,, \qquad H(\x_0) = \frac{1}{R} \,,
\end{equation}
independently of $\x_0$.

We consider the subset $\Gamma$ to be a spherical cap of angle
$\ve$ around the North pole, defined in the spherical coordinates as
$\Gamma = \{ (R,\theta,\varphi)\in \pa ~:~ 0 \leq \theta < \ve\}$. 
Using Eq. (\ref{eq:G0_sphere_whole}), we find then
\begin{align*}
\frac{1}{|\Gamma|} \int\limits_{\Gamma} d\x \, \G_0(\x|\x_0) 
& = \frac{1}{4\pi R} \biggl(\frac{1}{5} 
+ \sum\limits_{n=1}^\infty \frac{\phi_n}{n} \, P_n(\cos\theta_0) \biggr),
\end{align*}
where $|\Gamma| = 2\pi R^2(1 - \cos\ve)$ is the area of the spherical
cap $\Gamma$, $P_n(z)$ are Legendre polynomials, and
\begin{equation}
\phi_n = \frac{P_{n-1}(\cos\ve) - P_{n+1}(\cos\ve)}{1 - \cos\ve} \,.
\end{equation}
As a consequence, we find
\begin{subequations}
\begin{align}  \label{eq:w0_sphere}
w_0^{(\Gamma)}(\x_0) & = -\a_\Gamma + \frac{1}{4\pi R} \biggl(\frac{1}{5} 
+ \sum\limits_{n=1}^\infty \frac{\phi_n}{n}  \, P_n(\cos\theta_0) \biggr) ,\\   \label{eq:aG_sphere}
\a_\Gamma & = \frac{1}{4\pi R} \biggl(\frac15 + \sum\limits_{n=1}^\infty \frac{\phi_n^2}{n(2n+1)} \biggr).
\end{align}
\end{subequations}
Substituting Eq. (\ref{eq:R0_ball}) into
Eq. (\ref{eq:aGamma_asympt_3d}) gives the asymptotic behavior of
$\a_\Gamma$ as $\ve\to 0$,
\begin{equation}
\a_\Gamma \approx \frac{1}{4\pi R} \biggl(\frac{32}{3\pi \ve} + \ln(1/\ve) - \frac{31}{20} + \ln 2 + O(\ve)\biggr),
\end{equation}
that was earlier derived by a different method in \cite{Grebenkov17}.
Combining the above results, we get the kernel
\begin{align} \nonumber
\G(\x|\x_0) & = \frac{1}{4\pi R} \biggl\{ 4\pi \sum\limits_{n=1}^\infty \sum\limits_{m=-n}^n 
\frac{Y_{mn}(\theta,\varphi) Y_{mn}^*(\theta_0,\varphi_0)}{n}  \\  \label{eq:Gp0_ball}
&- \sum\limits_{n=1}^\infty \biggl(\frac{\phi_n}{n}  \, \bigl[P_n(\cos\theta_0) + P_n(\cos\theta)\bigr]
- \frac{\phi_n^2}{n(2n+1)}\biggr) \biggr\} ,
\end{align}
restricted to $\Gamma \times \Gamma$, that determines the eigenvalues
and eigenfunctions of the Dirichlet-to-Neumann operator
$\M_0^{(\Gamma)}$ for the mixed Steklov-Neumann problem.

The axial symmetry of the problem allows one to decompose the
kernel as
\begin{equation}  \label{eq:G0_sphere_Fourier}
\G(\x|\x_0) = \frac{1}{2\pi R} \sum\limits_{m=-\infty}^\infty e^{im(\varphi-\varphi_0)} \G^{(m)}(\theta|\theta_0),
\end{equation}
with the one-dimensional kernels
\begin{subequations}
\begin{align} \nonumber
\G^{(0)}(\theta|\theta_0) & = \sum\limits_{n=1}^\infty \biggl\{ \frac{n+1/2}{n} P_n(\cos\theta) P_n(\cos\theta_0) \\ \label{eq:G00_def}
&- \frac{\phi_n}{n}  \, \bigl[P_n(\cos\theta_0) + P_n(\cos\theta)\bigr] + \frac{\phi_n^2}{n(2n+1)} \biggr\} , \\  \label{eq:Gm_def}
\G^{(m)}(\theta|\theta_0) & = \sum\limits_{n=m}^\infty 
\frac{(n-m)!}{(n+m)!} \, \frac{n+1/2}{n} \, P_n^m(\cos\theta) P_n^{m}(\cos\theta_0) \qquad (m \geq 1), \\
\G^{(m)}(\theta|\theta_0) & = \G^{(-m)}(\theta|\theta_0) \qquad (m \leq -1). 
\end{align}
\end{subequations}
As a consequence, the eigenfunctions of the kernel $\G(\x|\x_0)$ are
of the form $e^{im\varphi} v_{m,n}^{(0,\Gamma)}(\theta)$, where the
second factor is the $n$-th eigenfunction of the kernel
$\G^{(m)}(\theta|\theta_0)$:
\begin{equation}  \label{eq:Gm_eigen_sphere}
\int\limits_0^\ve d\theta \, \sin\theta \, \G^{(m)}(\theta|\theta_0) \, v_{m,n}^{(0,\Gamma)}(\theta) = \frac{1}{R\mu_{m,n}^{(0,\Gamma)}}
v_{m,n}^{(0,\Gamma)}(\theta_0)  \qquad (0 < \theta_0 < \ve)
\end{equation} 
(the additional index $m$ is used to distinguish different kernels
$\G^{(m)}(\theta|\theta_0)$ and their eigenmodes).  For convenience,
we used here a complex-valued form; however, the factors $e^{\pm
im\varphi}$ can be replaced by a linear combination of sine and cosine
functions that we employed in Sec. \ref{sec:generic}.

One sees that the original problem on the two-dimensional spherical
cap $\Gamma$ is decomposed into separate spectral problems for the
kernels $\G^{(m)}(\theta|\theta_0)$.  When the target covers the whole
sphere (i.e., $\Gamma = \pa$), it is easy to check that
$P_n^m(\cos\theta)$ is an eigenfunction of the kernel
$\G^{(m)}(\theta|\theta_0)$, whereas the associated eigenvalue is
$R/n$, as it should be (given that $n/R$ is an eigenvalue of the
Dirichlet-to-Neumann operator $\M_0^{(\pa)}$ on the sphere).

In the limit $\ve \to 0$, one can investigate the asymptotic
behavior of Eqs. (\ref{eq:aG_sphere}, \ref{eq:w0_sphere},
\ref{eq:G00_def}).  For instance, one can check that the leading term
of each kernel $\G^{(m)}(\theta|\theta_0)$ is $\ve^{-1}\,
\hat{\G}^{(m)}(\theta/\ve|\theta_0/\ve)$, given by
Eqs. (\ref{eq:Ghatm_3d}, \ref{eq:Ghat_3d}), as expected.  As these
cumbersome computations do not bring new insights, they are not
presented.

\subsection{Numerical implementation}
\label{sec:numerics_3d}

By substituting $x = \cos\theta$, $x_0 = \cos\theta_0$ and $a =
\cos\ve$, it is convenient to rewrite the spectral problem
(\ref{eq:Gm_eigen_sphere}) as
\begin{equation}
\int\limits_a^1 dx \, \G^{(m)'}(x|x_0) \, v_{m,n}^{(0,\Gamma)}(x) = \frac{1}{\mu_{m,n}^{(0,\Gamma)}} \, v_{m,n}^{(0,\Gamma)}(x_0)
\qquad (a < x_0 < 1),
\end{equation}
where $\G^{(m)'}(\cos\theta|\cos\theta_0) =
\G^{(m)}(\theta|\theta_0)$.  The interval $(a,1)$ is discretized into
$N$ segments, centered at $x_n = a + (1-a)(n-1/2)/N$, with $n =
1,2,\ldots,N$.  Setting $\eta = (1-a)/(2N)$ and using the identity
\begin{equation}
(2k+1)P_k(x) = \frac{d}{dx} (P_{k+1}(x) - P_{k-1}(x)) ,
\end{equation}
we evaluate the matrix elements for $m = 0$ as
\begin{align} \label{eq:G_matrix_sphere}
\GG_{n,n'}^{(\ve)} & = \int\limits_{x_n-\eta}^{x_n+\eta} dx \, \G^{(0)'}(x|x_{n'}) 
 \approx \frac{R}{2} \sum\limits_{k=1}^{k_{\rm max}} \frac{1}{k} 
\biggl\{\biggl(P_k(x_{n'}) - \frac{\phi_k}{2k+1} \biggr) \\  \nonumber
& \times \biggl([P_{k+1}(x_n+\eta)-P_{k-1}(x_n+\eta)] - [P_{k+1}(x_n-\eta)-P_{k-1}(x_n-\eta)]\biggr) \\  \nonumber
&  - 2\eta \biggl(\phi_k \, P_k(x_{n'}) - \frac{\phi_k^2}{2k+1} \biggr)\biggr\} ,
\end{align}
where the infinite sum was truncated up to some $k_{\rm max}$. 
% See \verb|[G, Ed,V] = A_DN_sphere_mixed_G();|
A numerical diagonalization of the matrix $\GG^{(\ve)}$ of size
$N\times N$ allows one to approximate the axially symmetric
eigenfunctions $v_{0,n}^{(0,\Gamma)}(x)$ and the associated
eigenvalues $\mu_{0,n}^{(0,\Gamma)}$.  A similar computation can be
done for kernels with $m = 1,2,\ldots$ to approximate
$v_{m,n}^{(0,\Gamma)}(x)$ and $\mu_{m,n}^{(0,\Gamma)}$.

In the same vein, one can discretize the integral in
Eq. (\ref{eq:Geigen_3d}) and then diagonalize the resulting matrix to
approximate the eigenvalues $\hat{\mu}_{m,n}$ and eigenfunctions
$\hat{v}_{m,n}$.  However, as the semi-analytical method described in
Appendix \ref{sec:spheroid} is much faster and more accurate, we do
not present the details of the direct diagonalization.

\section{Variance of the boundary local time}
\label{sec:variance}

In this Appendix, we sketch the main steps to get the variance of the
boundary local time by following Ref. \cite{Grebenkov19c} (see also
\cite{Grebenkov07a}).

Let $\X_t$ denote the reflected Brownian motion in a bounded domain
$\Omega\subset \R^d$ with a smooth boundary $\pa$ (see
\cite{Ito,Freidlin,Saisho87,Papanicolaou90} for a mathematical
formulation and \cite{Grebenkov20} for physical insights).  The
boundary local time $\ell_t$ on a subset $\Gamma$ of the boundary
$\pa$ can be introduced as the rescaled residence time in a thin
boundary layer $\Gamma_a = \{\x\in\Omega ~:~ |\x - \Gamma| < a\}$ of
thickness $a$ near $\Gamma$:
\begin{equation}
\ell_t = \lim\limits_{a\to 0} \frac{D}{a} \int\limits_0^t dt' \, \Theta(a - |\X_{t'} - \Gamma|),
\end{equation}
where $|\X_{t'} - \Gamma|$ is the Euclidean distance between the
position $\X_{t'}$ at time $t'$ and the subset $\Gamma$, and
$\Theta(z)$ is the Heaviside step function: $\Theta(z) = 1$ for $z>0$
and $0$ otherwise.  In other words, the above integral determines how
long the diffusing particle stayed near the subset $\Gamma$ up to time
$t$.  As the layer thickness $a$ goes to $0$, the residence time also
vanishes but its rescaling by $a$ yields a nontrivial limit -- the
boundary local time $\ell_t$ (despite its name, $\ell_t$ has units of
length, given that the diffusivity $D$ has units of squared length
over time).

In \cite{Grebenkov19c}, the distribution of the boundary local time on
the boundary $\pa$ was obtained in terms of the Steklov eigenpairs.  A
straightforward extension of its derivation to the case of a subset
$\Gamma$ consists in using the eigenpairs of the Steklov-Neumann
problem.  In particular, Eq. [37] from \cite{Grebenkov19c} for the
$n$-th order moment of $\ell_t$, $L_n(t|\x_0) = \E_{\x_0}\{
\ell_t^n\}$, is generalized in our setting as (see also
\cite{Grebenkov20})
\begin{equation}
\tilde{L}_n(p|\x_0) = \int\limits_0^\infty dt \, e^{-pt} \, L_n(t|\x_0) 
= \sum\limits_{k = 0}^\infty \frac{V_k^{(p,\Gamma)}(\x_0)}{[\mu_k^{(p,\Gamma)}]^n} \int\limits_{\Gamma} v_k^{(p,\Gamma)},
\end{equation}
where $\x_0 \in \overline{\Omega}$ is the starting point of the
particle.  According to the properties of the Laplace transform
\cite{Watson}, the growth of the moments $L_n(t|\x_0)$ at long times
can be determined from the asymptotic behavior of
$\tilde{L}_n(p|\x_0)$ as $p\to 0$.  Using Eqs. (\ref{eq:mu0v0_p0}), we
get
\begin{subequations}
\begin{align}
\tilde{L}_1(p|\x_0) & \approx \frac{D|\Gamma|}{|\Omega| p^2} 
\biggl(1 + p\frac{|\Omega|}{D} \bigl(\a_\Gamma + W_0^{(\Gamma)}(\x_0)\bigr) + O(p^2)\biggr)  \qquad (p\to 0), \\
\tilde{L}_2(p|\x_0) & \approx \frac{2 D^2|\Gamma|^2}{|\Omega|^2 p^3} 
\biggl(1 + p\frac{|\Omega|}{D} \bigl(2\a_\Gamma + W_0^{(\Gamma)}(\x_0)\bigr) + O(p^2)\biggr)  \qquad (p\to 0), 
\end{align}
\end{subequations}
from which a formal Laplace transform inversion yields
\begin{subequations}
\begin{align}
L_1(t|\x_0) & \approx \frac{D|\Gamma|}{|\Omega|} \biggl(t + \frac{|\Omega|}{D} 
\bigl(\a_\Gamma + W_0^{(\Gamma)}(\x_0)\bigr) + \ldots\biggr)  \qquad (t\to \infty), \\
L_2(t|\x_0) & \approx \frac{D^2|\Gamma|^2}{|\Omega|^2} 
\biggl(t^2 + 2t\frac{|\Omega|}{D} \bigl(2\a_\Gamma + W_0^{(\Gamma)}(\x_0)\bigr) + \ldots\biggr)  \qquad (t\to \infty).
\end{align}
\end{subequations}
These relations imply the long-time asymptotic behavior
(\ref{eq:variance}) of the variance: 
\begin{equation}
\mathrm{Var}_{\x_0}\{ \ell_t\} = L_2(t|\x_0) - L_1^2(t|\x_0).  
\end{equation}
Expectedly, the leading-order term does not depend on the starting
point $\x_0$, which appears in the subleading correction (not shown).

\section{Exterior Steklov problem for an ellipse}
\label{sec:ellipse}

The auxiliary spectral problem (\ref{eq:Steklov_interval}) in the
upper half-plane $\H_2 = \R \times \R_+$ is actually the specific case
of the Steklov problem in the exterior of an ellipse, $\Omega =
\{(x,y)\in \R^2 ~:~ (x/a)^2 + (y/b)^2 > 1\}$, with semi-axes $a > b$
(Fig. \ref{fig:scheme_ellipse}e):
\begin{subequations}  \label{eq:Steklov_ellipse}
\begin{align}  \label{eq:Steklov_ellipse_Laplace}
\Delta V_k & = 0 \quad \textrm{in}~\Omega, \\
\partial_n V_k & = \mu_k V_k  \quad \textrm{on}~ \pa, \\ 
|\x| \, |\nabla V_k| & \to 0 \quad (|\x|\to \infty) 
\end{align}
\end{subequations}
(see \cite{Christiansen23,Grebenkov25} for more details).  In this
Appendix, we briefly discuss this more general problem by reproducing
the analysis of the exterior Steklov problem for oblate spheroids
\cite{Grebenkov24}.

To solve the Laplace equation (\ref{eq:Steklov_ellipse_Laplace}), it
is convenient to use the elliptic coordinates $(\alpha,\theta)$,
\begin{equation}
x = a_E \cosh \alpha \cos \theta, \quad y = a_E \sinh\alpha \sin \theta ,
\end{equation}
with $0 \leq \alpha < +\infty$, $-\pi < \theta \leq \pi$, and $a_E =
\sqrt{a^2-b^2}$ being one-half of the focal distance.  In these coordinates,
the exterior of an ellipse reads $\Omega = \{ \alpha > \alpha_0,~ -\pi
< \theta \leq \pi\}$, with $\tanh \alpha_0 = b/a$.  The Steklov
boundary condition is
\begin{equation}  \label{eq:Steklov_ellipse_BC}
\partial_n V_{k}\biggr|_{\pa} = - \frac{1}{h_\alpha} \partial_\alpha V_{k}\biggr|_{\alpha = \alpha_0}
= \mu_{k} V_{k} \biggr|_{\alpha = \alpha_0} ,
\end{equation}
where $h_\alpha = a_E \sqrt{\cosh^2\alpha - \cos^2\theta}$ is the
scale factor accounting for the curvature.
Since the Laplace operator reads
\begin{equation}
\Delta = \frac{1}{a_E^2 (\cosh^2\alpha - \cos^2\theta)} \bigl(\partial_\alpha^2 + \partial_\theta^2),
\end{equation}
a Steklov eigenfunction can be searched in either of two forms:
\begin{subequations}
\begin{align}  \label{eq:V2k_ellipse}
V_{2k}(\alpha,\theta) & = \sum\limits_{n=0}^\infty c_{2k,n}  \cos(n\theta) e^{-n(\alpha - \alpha_0)} , \\
V_{2k+1}(\alpha,\theta) & = \sum\limits_{n=1}^\infty c_{2k+1,n} \sin(n\theta) e^{-n(\alpha - \alpha_0)} ,
\end{align}
\end{subequations}
with unknown coefficients $c_{k,n}$.  Even and odd indices are used to
explicitly distinguish eigenfunctions that are symmetric and
antisymmetric with respect to the horizontal axis:
\begin{align*}
& V_{2k}(\alpha,-\theta) = V_{2k}(\alpha,\theta) \quad \Leftrightarrow \quad V_{2k}(x,-y) = V_{2k}(x,y) ,\\
& V_{2k+1}(\alpha,-\theta) = -V_{2k+1}(\alpha,\theta) \quad \Leftrightarrow \quad  
 V_{2k+1}(x,-y) = -V_{2k+1}(x,y) 
\end{align*}
(with some abuse of notations, we wrote the symmetries in both
elliptic and Cartesian coordinates).  These symmetries imply that any
eigenfunction $V_{2k}(x,y)$ is actually the solution of the mixed
Steklov-Neumann problem in the exterior of the half-ellipse in the
upper half-plane (Fig. \ref{fig:scheme_ellipse}d):
\begin{subequations}  \label{eq:StekN_int}
\begin{align}
\Delta V_{2k} & = 0 \quad (\textrm {in}~\Omega \cap \H_2), \\
\partial_n V_{2k} & = \mu_{2k} V_{2k}  \quad (\textrm{on}~\Gamma \cap \H_2), \\
-\partial_y V_{2k} & = 0 \quad (y = 0,~ |x| \geq a)
\end{align}
\end{subequations}
(in turn, antisymmetric eigenfunctions $V_{2k+1}(x,y)$ solve the mixed
Steklov-Dirichlet problem).  In particular, one retrieves the
auxiliary spectral problem (\ref{eq:Steklov_interval}) by setting $a =
1$ and $b = 0$.  In the following, we focus on the symmetric
eigenfunctions $V_{2k}$.

Since the Steklov condition (\ref{eq:Steklov_ellipse_BC}) should be
satisfied for any $\theta$, it is convenient to multiply it by
$h_{\alpha_0}(\theta) \cos (m\theta)$ and integrate over $\theta$ from
$0$ to $\pi$ to get
\begin{equation}
\frac{\pi}{2} m\, c_{2k,m} = \mu_{2k} \sum\limits_{n=0}^\infty c_{2k,n} 
\int\limits_0^{\pi} d\theta \, h_{\alpha_0}(\theta)  \cos(n\theta)  \cos(m\theta)  \qquad (m = 1,2,\ldots).
\end{equation}
Denoting 
\begin{align}    \label{eq:Mhat_ellipse}
\AA_{n,m}(\alpha_0) & = \frac{2}{\pi} \int\limits_0^{\pi} 
d\theta \cos(m\theta) \cos(n\theta) \sqrt{\cosh^2\alpha_0 - \cos^2\theta} \,,
\end{align}
we rewrite the above equation as
\begin{equation}  \label{eq:auxil12}
\frac{m}{a_E \mu_{2k}} c_{2k,m} = c_{2k,0} \AA_{m,0}(\alpha_0) + \sum\limits_{n=1}^\infty c_{2k,n} \AA_{n,m}(\alpha_0).
\end{equation}
In turn, multiplying Eq. (\ref{eq:Steklov_ellipse_BC}) by
$h_{\alpha_0}(\theta)$ and integrating over $\theta$ from $0$ to $\pi$
yield
\begin{equation} \label{eq:auxil13}
c_{2k,0} = - \frac{1}{\AA_{0,0}(\alpha_0)} \sum\limits_{n=1}^\infty c_{2k,n} \AA_{n,0}(\alpha_0). 
\end{equation}
Substituting this expression into Eq. (\ref{eq:auxil12}), we finally
get the matrix spectral problem
\begin{equation}  \label{eq:eigen_ellipse}
\sum\limits_{n=1}^\infty c_{2k,n} \MM_{n,m}(\alpha_0) = \frac{1}{a_E \mu_{2k}} c_{2k,m}  \qquad (m=1,2,\ldots),
\end{equation}
where
\begin{equation}  \label{eq:M_ellipse}
\MM_{n,m}(\alpha_0) = \frac{1}{m} \biggl[\AA_{n,m}(\alpha_0) - \frac{\AA_{n,0}(\alpha_0) \AA_{0,m}(\alpha_0)}{\AA_{0,0}(\alpha_0)}\biggr].
\end{equation} 
A numerical diagonalization of the truncated matrix $\MM(\alpha_0)$
allows one to approximate the eigenvalues $\mu_{2k}$ and to construct
the symmetric eigenfunctions $V_{2k}$ of the exterior Steklov problem.

Note that the antisymmetric eigenmodes satisfy a similar eigenvalue
problem,
\begin{equation}  \label{eq:eigen_ellipseA}
\sum\limits_{n=1}^\infty c_{2k+1,n} \MM'_{n,m}(\alpha_0) = \frac{1}{a_E \mu_{2k+1}} c_{2k+1,m}  \qquad (m=1,2,\ldots),
\end{equation}
and are obtained by diagonalizing the matrix
\begin{align}  \label{eq:Mhat_ellipse_sin}
\MM'_{n,m}(\alpha_0) & = \frac{2}{\pi m} \int\limits_0^{\pi} 
d\theta \sin(m\theta) \sin(n\theta) \sqrt{\cosh^2\alpha_0 - \cos^2\theta} \,.
\end{align}
% See [mu,V,M,Mp,Vinf, vk_norm2] = A_DN_ellipse(a,b,fsym, N);

Finally, we compute the normalization of the Steklov eigenfunctions:
\begin{align} 
\int\limits_{\pa} d\x \, |V_{2k}|^2 & = \int\limits_{-\pi}^{\pi} d\theta \, h_{\alpha_0}(\theta) \, |V_{2k}(\alpha_0,\theta)|^2 
 = a_E \pi \sum\limits_{m=0}^{\infty} \sum\limits_{n=0}^{\infty}  c_{2k,n} c_{2k,m} A_{n,m}(\alpha_0) ,
\end{align}
where we substituted Eq. (\ref{eq:V2k_ellipse}) and used
Eq. (\ref{eq:Mhat_ellipse}).  Using Eqs. (\ref{eq:auxil12},
\ref{eq:auxil13}) to evaluate the sums over $n$, we get
\begin{align} \nonumber
\int\limits_{\pa} d\x\, |V_{2k}|^2 & = a_E \pi \biggl[c_{2k,0} \biggl(c_{2k,0} A_{0,0}(\alpha_0) +
\underbrace{\sum\limits_{n=1}^{\infty} c_{2k,n} A_{n,0}(\alpha_0)}_{= -c_{2k,0} A_{0,0}(\alpha_0)} \biggr) \\  
& + \sum\limits_{m=1}^{\infty} c_{2k,m}
\underbrace{\sum\limits_{n=0}^{\infty} c_{2k,n} A_{n,m}(\alpha_0)}_{= m c_{2k,m}/(a_E \mu_{2k})} \biggr] 
= \frac{\pi}{\mu_{2k}} \sum\limits_{m=1}^{\infty} m |c_{2k,m}|^2 .
\end{align}
A similar relation holds for $V_{2k+1}$:
\begin{equation}
\int\limits_{\pa} d\x\, |V_{2k+1}|^2 = \frac{\pi}{\mu_{2k+1}} \sum\limits_{m=1}^{\infty} m |c_{2k+1,m}|^2 .
\end{equation}
These relations can be used to ensure the $L^2(\pa)$ normalization of
the Steklov eigenfunctions by rescaling the coefficients $c_{k,m}$.

Setting $a = 1$ and $b = 0$ (i.e., $\alpha_0 =0$), we deal with the
exterior of the interval $(-1,1)$.  In this limit, the elements of the
matrices $\AA(0)$, $\MM(0)$, and $\MM'(0)$ can be found explicitly:
\begin{subequations}
\begin{align} \label{eq:Mhat_ellipse0}
\AA_{n,m}(0) & = \frac{1+(-1)^{m+n}}{\pi}  \biggl(\frac{1}{1-(m-n)^2} + \frac{1}{1-(m+n)^2}\biggr), \\ \nonumber
\MM'_{n,m}(0) & = \frac{1+(-1)^{m+n}}{\pi m} \biggl(\frac{1}{1-(m-n)^2} - \frac{1}{1-(m+n)^2}\biggr)
\end{align}
\end{subequations}
(note that the elements with $m = \pm n \pm 1$ are zero).
Substituting $\AA_{n,m}(0)$ into Eq. (\ref{eq:M_ellipse}), one gets
the matrix $\MM(0)$ that determines the symmetric eigenmodes.
Comparing its expression with Eq. (\ref{eq:Glim_matrix}), we realize
that $\MM(0) = \GG^\dagger$, i.e., the symmetric eigenfunctions
$v_{2k}$ and their eigenvalues $\mu_{2k}$ of the Steklov-Neumann
problem for the exterior of the interval $(-1,1)$ are therefore the
eigenmodes of the kernel $\hat{\G}(x|x_0)$ from
Eq. (\ref{eq:Glim_disk}).
%%%  !!!  Function [mu,V,c0] = A_DN_ellipse_vk_fig(mu,V,c0);  illustrates the agreement of eigenmodes !!!

We also note that the interior Steklov problem for the ellipse can be
solved in the same way.  The Steklov eigenfunctions are searched in
symmetric and antisymmetric forms
\begin{subequations}
\begin{align}  \label{eq:V2k_ellipse_int}
V_{2k}(\alpha,\theta) & = \sum\limits_{n=0}^\infty c_{2k,n}  \cos(n\theta) \frac{\cosh(n\alpha)}{\cosh(n\alpha_0)} , \\
V_{2k+1}(\alpha,\theta) & = \sum\limits_{n=1}^\infty c_{2k+1,n} \sin(n\theta) \frac{\sinh(n\alpha)}{\sinh(n\alpha_0)} .
\end{align}
\end{subequations}
Repeating the above computation, one gets the matrix eigenvalue
problem (\ref{eq:eigen_ellipse}) for symmetric eigenmodes, in which
the matrix elements $\MM_{n,m}(\alpha_0)$ are multiplied by
$\ctanh(m\alpha_0)$, and the matrix eigenvalue problem
(\ref{eq:eigen_ellipseA}) for antisymmetric eigenmodes, in which the
matrix elements $\MM'_{n,m}(\alpha_0)$ are multiplied by
$\tanh(m\alpha_0)$.
% See [mu, mus,Vs, mua,Va, M,Mp] = A_DN_ellipse_interior(a,b, N);;  % confirmed by FEM

\section{Exterior Steklov problem for an oblate spheroid}
\label{sec:spheroid}

In analogy to Appendix \ref{sec:ellipse}, one can get a
semi-analytical solution of the auxiliary Steklov-Neumann problem
(\ref{eq:Steklov_disk}).  In \cite{Grebenkov24}, we studied a more
general Steklov problem for the exterior of an oblate spheroid (see
also \cite{Grebenkov25b}).  However, the published results are not
directly applicable because the Dirichlet boundary condition at
infinity was employed so that all eigenvalues were strictly positive,
whereas the principal eigenfunction was not a constant.  Indeed, in
three dimensions, the choice of the boundary condition at infinity
(Dirichlet or Neumann) affects the resulting Dirichlet-to-Neumann
operators $\M_0^D$ and $\M_0^N$ and their spectral properties (see
\cite{Arendt15} for a rigorous definition of both operators and their
comparison).  In order to get the correct asymptotic behavior in the
small-target limit, one needs to impose Neumann boundary condition
(\ref{eq:Vk3d_inf}), which ensures that the principal eigenvalue is
zero, while the associated eigenfunction is constant, as in the
original Steklov-Neumann problem.
In this Appendix, we sketch the main steps and resulting formulas in
this setting, whereas notations, explanations, and technical details
are skipped and can be found in \cite{Grebenkov24}.

We consider the Steklov problem in the exterior of an oblate spheroid
with semi-axes $b > a$, $\Omega = \{ (x,y,z)\in\R^3 ~:~ (x/b)^2 +
(y/b)^2 + (z/a)^2 > 1\}$, i.e., we search for the eigenpairs $\{\mu_k,
V_k\}$ satisfying
\begin{subequations}  \label{eq:Steklov_spheroid}
\begin{align}  
\Delta V_k & = 0 \quad \textrm{in}~\Omega, \\
\partial_n V_k & = \mu_k V_k  \quad \textrm{on}~ \pa, \\   \label{eq:Steklov_spheroid_inf}
|\x|^2 \, |\nabla V_k| & \to 0 \quad (|\x|\to \infty) .
\end{align}
\end{subequations}
The starting point is the expansion of an eigenfunction
$v_{m,n}(\theta,\varphi)$ in the oblate spheroidal coordinates onto
the basis of (modified) spherical harmonics
$\bar{Y}_{m',n'}(\theta,\varphi)$ (in which $P_n^m(\cos\theta)$ is
replaced by $P_n^m(\sin\theta)$, see \cite{Grebenkov24})
\begin{equation}  \label{eq:vmn_expansion}
v_{m,n}(\theta,\varphi) = \sum\limits_{m',n'} [\VV_{m,n}]_{m',n'} \bar{Y}_{m',n'}(\theta,\varphi),
\end{equation}
where the unknown coefficients $[\VV_{m,n}]_{m',n'}$ are determined
from the boundary condition, and we keep using the double indices.
Replicating the derivation of Ref. \cite{Grebenkov24}, one obtains the
system in Eq. [B13] of linear algebraic equations:
\begin{equation} \label{eq:oblate_matrix_auxil}
\mu_{m,n} \sum\limits_{n'=0}^\infty [\VV_{m,n}]_{m,n'} \bar{G}_{m,n';m,n''} = [\VV_{m,n}]_{m,n''} c_{m,n''} 
\qquad (n'' = 0,1,2,\cdots).
\end{equation}
Here the matrix elements $\bar{G}_{m,n;m',n'}$ are given in Eq. [B14],
and the coefficients $c_{m,n''}$ are given by Eq. [43], except for
$c_{0,0}$, which is now equal to $0$ to represent the Neumann boundary
condition (\ref{eq:Steklov_spheroid_inf}) at infinity (we use square
brackets when referring to equations from Ref. \cite{Grebenkov24}).
As a consequence, the construction remains unchanged for any $m > 0$
as the associated eigenfunctions were periodically oscillating and
thus already orthogonal to a constant.  In turn, the analysis for
axially symmetric eigenfunctions with $m = 0$ requires modifications.
For any $n'' = 1,2,\cdots$, the system (\ref{eq:oblate_matrix_auxil})
can be rewritten as
\begin{equation} 
\frac{1}{c_{0,n''}} \biggl([\VV_{0,n}]_{0,0} \bar{G}_{0,0;0,n''} +
\sum\limits_{n'=1}^\infty [\VV_{0,n}]_{0,n'} \bar{G}_{0,n';0,n''}\biggr) = \frac{[\VV_{0,n}]_{0,n''}}{\mu_{0,n}}  \,.
\end{equation}
In turn, as $c_{0,0} = 0$, the equation for $n'' = 0$ yields
\begin{equation} \label{eq:V0n_00}
[\VV_{0,n}]_{0,0} = - \frac{1}{\bar{G}_{0,0;0,0}} \sum\limits_{n'=1}^\infty [\VV_{0,n}]_{0,n'} \bar{G}_{0,n';0,0} , 
\end{equation}
allowing one to exclude $[\VV_{0,n}]_{0,0}$ from the above equations and
to get a closed system
\begin{equation} 
\sum\limits_{n'=1}^\infty [\VV_{0,n}]_{0,n'} \GG_{n',n''} = \frac{1}{\mu_{0,n}} [\VV_{0,n}]_{0,n''} \qquad (n'' = 1,2,\cdots),
\end{equation}
where
\begin{equation}
\GG_{n',n''} = \biggl[\bar{G}_{0,n';0,n''} - \frac{\bar{G}_{0,n';0,0} \, \bar{G}_{0,0;0,n''}}{\bar{G}_{0,0;0,0}}\biggr]
\frac{1}{c_{0,n''}} \,.
\end{equation}
One sees that $\{\VV_{0,n}\}$, enumerated by the index $n =
1,2,\cdots$, are the {\it left} eigenvectors of the matrix $\GG$ that
correspond to the eigenvalues $1/\mu_{0,n}$.  Truncation and numerical
diagonalization of this matrix allow one to approximate $\mu_{0,n}$ and
$[\VV]_{0,n}$, from which the eigenfunctions $v_{0,n}$ are obtained via
Eq. (\ref{eq:vmn_expansion}).  Note that the coefficient
$[\VV_{0,n}]_{0,0}$ is evaluated via Eq. (\ref{eq:V0n_00}).  

The exterior of the disk corresponds to the limit $a = 0$.  Since the
disk has two ``faces'', it is convenient to describe any point on the
top face at distance $r$ from the center by the variable $\chi =
\sqrt{1-r^2}$, whereas any point on the bottom face by $\chi = -
\sqrt{1-r^2}$.  This convention naturally follows from the oblate
spheroidal coordinates (see \cite{Grebenkov24} for details).
The eigenvalues $\mu_{0,n}$ of the axially symmetric eigenfunctions of
$\M_0^N$ are then obtained by diagonalizing the matrix
\begin{align}  \label{eq:GG_disk_upper}
\GG_{n',n''} & = \frac{\sqrt{(n'+1/2)(n''+1/2)}}{c_{0,n''}} 
 \biggl[G_{n',n''}^0(0) - \frac{G_{n',0}^0(0) G_{0,n''}^0(0)}{G_{0,0}^0(0)}\biggr],
\end{align}
with $n',n'' = 1,2,3,\cdots$, where $c_{0,n} =
2(\Gamma(n/2+1)/\Gamma(n/2+1/2))^2$ was given in Eq. [51], and
\begin{equation}
G_{n',n''}^0(0) = \int\limits_{-1}^1 dx \, |x| \, P_{n'}(x) P_{n''}(x),
\end{equation}
where $P_n(x)$ are the Legendre polynomials.  The elements of this
matrix can be rapidly obtained via recurrence relations derived in
\cite{Grebenkov24}.  In turn, the {\it left} eigenvectors of this
matrix, $\VV_{0,n}$, determine the axially symmetric eigenfunctions
$v_{0,n}$ and thus $V_{0,n}$; for instance,
\begin{equation}
v_{0,n}(\chi) = \sum\limits_{n'=0}^\infty [\VV_{0,n}]_{0,n'} \sqrt{n'+1/2}\, P_{n'}(\chi)  \qquad (-1 < \chi < 1).
\end{equation}
Note that these eigenfunctions need to be normalized, see details in
\cite{Grebenkov24}.

As discussed in \cite{Grebenkov24}, the Steklov eigenfunctions
$V_{0,n}$ are symmetric with respect to the horizontal plane for even
$n$, and antisymmetric for odd $n$.  As a consequence, the symmetric
eigenfunctions provide solutions of the mixed Steklov-Neumann problem
(\ref{eq:Steklov_disk}) in the upper half-space.  In other words, the
eigenvalues $\mu_{0,2n}$ and eigenfunctions $v_{0,2n}(\chi)$,
restricted to positive $\chi$ (i.e., to the ``top face'' of the disk),
coincide with the eigenpairs $\{ \hat{\mu}_k, \hat{v}_k(\hat{r})\}$ of
the kernel $\hat{\G}^{(0)}(\hat{r}|\hat{r}_0)$ from
Eq. (\ref{eq:Ghat_3d}), by setting $\chi = \sqrt{1-\hat{r}^2}$.  This
is confirmed by a numerical diagonalization of the matrix $\GG$ from
Eq. (\ref{eq:GG_disk_upper}).

\section{First-order corrections in the small-target limit}
\label{sec:small}

In this Appendix, we discuss the small-target limit of the
corrections $\a_\Gamma$ and $w_0^{(\Gamma)}$ and the coefficients
$b_k^{(\Gamma)}$ for a bounded domain with a smooth boundary.

\subsection{Two dimensions} 

For a connected subset $\Gamma$ of perimeter $|\Gamma| = 2\epsilon$,
centered at a boundary point $\x_\Gamma\in\pa$, substitution of the
expansion (\ref{eq:Gpseudo_2d_asympt}) of the pseudo-Green's function
$\G_0(\x|\x_0)$ into Eq. (\ref{eq:a_eps}) yields in the limit
$\epsilon\to 0$:
\begin{align}  \label{eq:aGamma_2d}
\a_\Gamma & \approx \frac{1}{(2\epsilon)^2} \int\limits_{-\epsilon}^\epsilon ds \int\limits_{-\epsilon}^\epsilon ds_0 \, 
\biggl[- \frac{1}{\pi} \ln |s-s_0| + R_0(\x_\Gamma)\biggr] %   
 = - \frac{\ln|\Gamma|}{\pi} + \frac{3}{2\pi} + R_0(\x_\Gamma) + o(1),
\end{align}
where $s$ and $s_0$ are the curvilinear coordinates of the points $\x$
and $\x_0$ on the subset $\Gamma$.  As the boundary $\pa$ is smooth,
the small subset $\Gamma$ can be approximated as a flat interval
$(-\epsilon,\epsilon)$ (the curvature effect emerging in the
next-order term in $\epsilon$).  Similarly, we compute
\begin{align*}  
w_0^{(\Gamma)}(\x_0) & \approx -\a_\Gamma + \frac{1}{2\epsilon} \int\limits_{-\epsilon}^\epsilon ds
\biggl[- \frac{1}{\pi} \ln |x-x_0| + R_0(\x_0)\biggr] \\
& = - \a_\Gamma + R_0(\x_\Gamma) 
 - \frac{(1+s_0/\epsilon)\ln(\epsilon+s_0) + (1-s_0/\epsilon)\ln(\epsilon-s_0) - 2}{2\pi} \,,
\end{align*}
which yields Eq. (\ref{eq:hatw0_asympt_2d}) after substitution of
$\a_\Gamma$ from Eq. (\ref{eq:aGamma_2d}).
As a consequence, the coefficients $b_k^{(\Gamma)}$ from
Eq. (\ref{eq:bk_w0}) admit the following scaling in the leading order:
\begin{align*}
b_k^{(\Gamma)} & \approx - \frac{\hat{\mu}_k}{\epsilon} \int\limits_{-\epsilon}^{\epsilon} ds \, 
\hat{w}_0(s/\epsilon) \frac{1}{\sqrt{\epsilon}} \hat{v}_k(s/\epsilon) ,
\end{align*}
yielding
\begin{equation}   \label{eq:bk_2d_bis}
b_k^{(\Gamma)} \approx \epsilon^{-1/2} \, \hat{b}_k,  \qquad
\hat{b}_k = - \hat{\mu}_k \int\limits_{-1}^1 dy \, \hat{w}_0(y) \hat{v}_k(y).
\end{equation}

In order to avoid the numerical integration in
Eq. (\ref{eq:bk_2d_bis}), we derive another representation of the
coefficients $\hat{b}_k$.  We start by an intuitive ``hand-waving''
argument that relies on Eq. (\ref{eq:Vk0_int}), which is valid for a
bounded domain $\Omega$.  Let $\Omega_L$ to be the half-disk of radius
$L$ in the upper half-plane, with $\Gamma = (-1,1)$ lying on the
horizontal diameter.  When $L\to \infty$, the domain $\Omega_L$
approaches $\H_2$, so that an eigenfunction $V_k^{(0,\Gamma)}$
approaches $\hat{V}_k$, which remains bounded.  As a consequence,
Eq. (\ref{eq:Vk0_int}) yields in the limit $L\to\infty$:
\begin{equation}  \label{eq:Vinf_bk}
\hat{b}_k = \hat{V}_k(\infty) .
\end{equation}

A more formal derivation of this relation employs the pseudo-Green's
function $\G_0(\x|\x_0)$ for the upper half-plane, 
\begin{equation}  \label{eq:kernel_interval0}
\G_0(\x|\x_0) = -\frac{1}{2\pi} \biggl(\ln|\x-\x_0| + \ln|\x-\x'_0|\biggr),
\end{equation}
where $\x'_0 = (x_0,-y_0)$ is is the mirror reflection of the point
$\x_0 = (x_0,y_0)$.  This function satisfies
\begin{subequations}
\begin{align}  \label{eq:G0_H2}
-\Delta \G_0(\x|\x_0) & = \delta(\x-\x_0) \quad (\x\in \H_2), \\
\partial_n \G_0(\x|\x_0) & = 0 \quad (\x \in \partial \H_2), \\
\G_0(\x|\x_0) & \sim - \frac{1}{\pi} \ln |\x| + o(1) \quad (|\x|\to\infty).
\end{align}
\end{subequations}
Multiplying Eq. (\ref{eq:Steklov_interval_1}) by $\G_0(\x|\x_0)$,
multiplying Eq. (\ref{eq:G0_H2}) by $\hat{V}_k(\x)$, subtracting them,
integrating over $\x = (x,y)\in \H_2$ and using the Green's formula,
we get for any $\x_0 = (x_0,y_0)\in\H_2$:
\begin{equation}
\hat{V}_k(x_0,y_0) = \hat{V}_k(\infty) + \hat{\mu}_k \int\limits_{-1}^1 dx \, \G_0(x,0|x_0,y_0)\, \hat{v}_k(x) .
\end{equation}
Substituting
\begin{equation}
\frac12 \int\limits_{-1}^1 dx \, \G_0(x,0|x_0,0) = \hat{w}_0(x_0) + a_0  \qquad (-1 < x_0 < 1)
\end{equation} 
(with $a_0 = (3-2\ln 2)/(2\pi)$) into Eq. (\ref{eq:bk_2d_bis}) yields
\begin{align*}
\hat{b}_k 
& = - \hat{\mu}_k \int\limits_{-1}^1 dx \, \hat{v}_k(x) \biggl[-a_0 + \frac{1}{2} \int\limits_{-1}^1 dx_0 \, \G_0(x,0|x_0,0)\biggr] \\
& = - \frac{\hat{\mu}_k}{2} \int\limits_{-1}^1 dx_0 \underbrace{\int\limits_{-1}^1 dx \, \hat{v}_k(x) \G_0(x,0|x_0,0)}_{ 
= (\hat{v}_k(x_0) - \hat{V}_k(\infty))/\hat{\mu}_k} 
= \hat{V}_k(\infty) \,,
\end{align*}
where we used the orthogonality of $\hat{v}_k$ to a constant.  
In order to compute numerically $\hat{\mu}_k$ and $\hat{V}_k(\infty)$,
it is sufficient to diagonalize a truncated matrix $\MM(0)$ with
explicitly known coefficients, as described in Appendix
\ref{sec:ellipse}.  The first ten values of $\hat{V}_k(\infty)$ are
reported in Table \ref{tab:eta_2D}.
%  see \verb|[bk, vk, y, muint] = A_DN_ellipse_int_bk;|]}

In addition, we compute two sums used in Sec. \ref{sec:application}.
First, we establish the following identity
\begin{align} \nonumber
\sum\limits_{k=1}^\infty \frac{[b_k^{(\Gamma)}]^2}{[\mu_k^{(0,\Gamma)}]^2} & = \int\limits_\Gamma d\x_1\, w_0^{(\Gamma)}(\x_1)
\int\limits_\Gamma d\x_2\, w_0^{(\Gamma)}(\x_2) 
\underbrace{\sum\limits_{k=1}^\infty v_k^{(0,\Gamma)}(\x_1) v_k^{(0,\Gamma)}(\x_2)}_{=\delta(\x_1-\x_2) - 1/|\Gamma|}  \\
\label{eq:bk_sum}
& = \int\limits_\Gamma d\x [w_0^{(\Gamma)}(\x)]^2 = \| w_0^{(\Gamma)}\|^2_{L^2(\Gamma)} ,
\end{align}
where we used Eq. (\ref{eq:bk_w0}), the orthogonality
(\ref{eq:w0_int}) of $w_0^{(\Gamma)}$ to a constant, and the
completeness relation (\ref{eq:vk_completeness}) at $p = 0$.
In the limit $\epsilon\to 0$, Eq. (\ref{eq:bk_sum}) yields
\begin{equation}  \label{eq:sum_bk2_2d}
C_2 = \sum\limits_{k=1}^\infty \frac{[\hat{V}_k(\infty)]^2}{\hat{\mu}_k^2} 
= \int\limits_{-1}^1 dx \, [\hat{w}_0(x)]^2 = \frac{21 -2\pi^2}{18\pi^2} \approx 0.007 ,
\end{equation}
where we used Eq. (\ref{eq:hatw0_2d}) for $\hat{w}_0(x)$.
Second, the divergence theorem implies 
\begin{equation}  \label{eq:HMD_2d}
\hat{V}_k(\infty) = \int\limits_{-1}^1 dx \, \hat{v}_k(x) \, h(x|\infty)  \qquad
\textrm{where} \quad h(x|\infty) = \frac{1}{\pi \sqrt{1-x^2}}
\end{equation}
is the harmonic measure density on the interval $(-1,1)$, seen from
infinity.  Combining this representation with
Eqs. (\ref{eq:bk_2d_bis}, \ref{eq:Vinf_bk}), we get
\begin{align*} 
\sum\limits_{k=1}^\infty \frac{[\hat{V}_k(\infty)]^2}{\hat{\mu}_k} & = - \sum\limits_{k=1}^\infty \int\limits_{-1}^1 dx \, \hat{v}_k(x)
h(x|\infty) \int\limits_{-1}^1 dx' \, \hat{w}_0(x') \, \hat{v}_k(x') \\
& = - \int\limits_{-1}^1 dx \, h(x|\infty) \int\limits_{-1}^1 dx' \, \hat{w}_0(x') 
\underbrace{\sum\limits_{k=1}^\infty \hat{v}_k(x) \,\hat{v}_k(x')}_{=\delta(x-x') - 1/2} .
\end{align*}
As a consequence, we obtain
\begin{equation}   \label{eq:sum_mu_bk2_2d}
C_1 = \sum\limits_{k=1}^\infty \frac{[\hat{V}_k(\infty)]^2}{\hat{\mu}_k} 
= - \int\limits_{-1}^1 dx \, \frac{\hat{w}_0(x)}{\pi \sqrt{1-x^2}} = \frac{3-4\ln 2}{2\pi} 
\approx 0.0362.
\end{equation}

\subsection{Three dimensions}

We consider the subset $\Gamma$ to be a small disk of radius
$\epsilon$, centered at a point $\x_\Gamma\in
\pa$.  Substitution of the expansion (\ref{eq:G0_asympt_3d}) of the
pseudo-Green's function $\G_0(\x|\x_0)$ into Eq. (\ref{eq:a_eps})
yields as $\epsilon\to 0$:
\begin{align} \label{eq:aGamma_auxil1}
\a_\Gamma & \approx R_0(\x_\Gamma) + \underbrace{
\frac{1}{|\Gamma|^2} \int\limits_{\Gamma} d\x_0 \int\limits_{\Gamma} \frac{d\x}{2\pi |\x-\x_0|}}_{=\a_\Gamma^{(1)}}  
 - \underbrace{\frac{H(\x_\Gamma)}{4\pi |\Gamma|^2} \int\limits_{\Gamma} 
d\x_0 \int\limits_{\Gamma} d\x \,\ln |\x-\x_0| }_{=\a_\Gamma^{(2)}}  + o(1),
\end{align}
where we used the smoothness of $R_0(\x_0)$ and $H(\x_0)$ to replace
them by $R_0(\x_\Gamma)$ and $H(\x_\Gamma)$, respectively.

To compute the first integral, denoted as $\a_\Gamma^{(1)}$, we use
the spherical coordinates $(r,\pi/2,\varphi)$, in which $\theta =
\pi/2$ corresponds to the disk $\Gamma$ on the horizontal plane, and
the Legendre expansion:
\begin{equation}
\frac{1}{|\x-\x_0|} = \sum\limits_{n=0}^\infty P_n(\cos(\varphi-\varphi_0)) \frac{r_<^n}{r_>^{n+1}} \,,
\end{equation}
where $r_< = \min\{r,r_0\}$ and $r_> = \max\{r,r_0\}$.  As a
consequence, we get
\begin{align*}
\a_\Gamma^{(1)} & = \frac{1}{2\pi (\pi \epsilon^2)^2} 
\int\limits_0^\epsilon dr_0 \, r_0  \int\limits_0^{2\pi} d\varphi_0 \int\limits_0^\epsilon dr \, r  \int\limits_0^{2\pi} d\varphi 
\sum\limits_{n=0}^\infty P_n(\cos(\varphi-\varphi_0)) \frac{r_<^n}{r_>^{n+1}} \\
&  = \frac{2}{3\pi \epsilon} \sum\limits_{n=0}^\infty \frac{P_{2n}^2(0)}{n+1} \,,
\end{align*}
where we used the addition theorem to evaluate 
\begin{equation*}
\int\limits_0^{2\pi} d\varphi \, P_{2n}(\cos \varphi) = 2\pi [P_{2n}(0)]^2 = 2\pi \biggl(\frac{(2n-1)!!}{(2n)!!}\biggr)^2 \,,
\end{equation*} 
whereas odd-order contributions vanished.  Finally, we employ the
expansion of the complete elliptic integral of the first kind
\cite{Abramowitz},
\begin{equation}
K(z) = \frac{\pi}{2} \sum\limits_{n=0}^\infty [P_{2n}(0)]^2 \, z^{2n}    \qquad (|z| < 1),
\end{equation}
to get
\begin{equation} \label{eq:aGamma_3d}
\a_\Gamma^{(1)} = \frac{2}{3\pi \epsilon} \, \frac{2}{\pi} \underbrace{\int\limits_0^1 dz \, K(\sqrt{z})}_{=2}  
= \frac{8}{3\pi^2 \epsilon} \,.  
\end{equation}
In the same way, we also have
\begin{align}  \nonumber
\int\limits_0^{2\pi} \frac{d\varphi}{|\x-\x_0|} & = \sum\limits_{n=0}^\infty \frac{r_<^n}{r_>^{n+1}} 
\underbrace{\int\limits_0^{2\pi} d\varphi P_n(\cos(\varphi-\varphi_0))}_{=2\pi [P_n(0)]^2}  
= \frac{2\pi}{r_>} \sum\limits_{n=0}^\infty [P_{2n}(0)]^2 (r_</r_>)^{2n} \\  \label{eq:dist_average}
& = \frac{4}{r_>} K\bigl(r_< /r_>\bigr) .
\end{align}

For the second integral in Eq. (\ref{eq:aGamma_auxil1}), we write
$|\x-\x_0| = \sqrt{r^2 + r_0^2 - 2rr_0 \cos(\varphi-\varphi_0)}$ and
use the relation
\begin{equation} \label{eq:integral_log}
\int\limits_0^{2\pi} d\varphi \, \ln(A - B \cos\varphi) = 2\pi \ln\biggl(\frac{A+\sqrt{A^2-B^2}}{2}\biggr) 
\end{equation}
to evaluate first the integral over $\varphi$:
\begin{equation}  \label{eq:logdist_average}
\int\limits_0^{2\pi} d\varphi \, \ln|\x-\x_0| = \pi \ln \biggl(\frac{r^2 + r_0^2 + |r^2 - r_0^2|}{2}\biggr).
\end{equation}
As a consequence, the integral of this expression over $r$ yields
\begin{align} 
& \frac{1}{|\Gamma|} \int\limits_\Gamma d\x \, \ln|\x-\x_0| =
\frac{1}{\epsilon^2}  \int\limits_0^{\epsilon} dr \, r \ln \biggl(\frac{r^2 + r_0^2 + |r^2 - r_0^2|}{2}\biggr) 
= \ln \epsilon - \frac{1-(r_0/\epsilon)^2}{2} \,,
\end{align}
from which
\begin{equation}
\a_\Gamma^{(2)} = \frac{H(\x_\Gamma)}{4\pi} \biggl(\ln \epsilon - \frac14\biggr).
\end{equation}
Combining these results, we obtain the asymptotic relation
(\ref{eq:aGamma_asympt_3d}) for $\a_\Gamma$.  For instance, for a
sphere of radius $R$, one can substitute $H(\x_\Gamma)$ and
$R_0(\x_\Gamma)$ from Eq. (\ref{eq:R0_ball}) to get
\begin{equation}  \label{eq:aGamma_asympt_sphere}
\a_\Gamma \approx \frac{1}{4\pi R} \biggl(\frac{32}{3\pi \ve} + \ln (1/\ve) + \ln 2 - \frac{31}{20}\biggr) + o(1)
\end{equation}
(with $\ve = \epsilon/R$), which agrees with the asymptotic analysis
of the series (\ref{eq:aG_sphere}) reported in \cite{Grebenkov17}.

Similarly, we can compute the leading-order term in the correction
$w_0^{(\Gamma)}(\x_0)$ by substituting Eq. (\ref{eq:G0_asympt_3d}) to
Eq. (\ref{eq:omega_eps}):
\begin{align*}
w_0^{(\Gamma)}(\x_0) & \approx  
\frac{1}{2\pi^2 \epsilon^2} \int\limits_0^{\epsilon} dr \, r \int\limits_0^{2\pi} d\varphi
\sum\limits_{n=0}^\infty P_n(\cos(\varphi-\varphi_0)) \frac{r_<^n}{r_>^{n+1}}  \\
& -\a_\Gamma  + R_0(\x_0) - \frac{H(\x_0)}{4\pi |\Gamma|} \int\limits_{\Gamma} d\x \, \ln |\x-\x_0| \\
& = -\a_\Gamma + \frac{r_0}{\pi \epsilon^2} \sum\limits_{n=0}^\infty [P_{2n}(0)]^2 \biggl(\frac{1}{2n+2} 
+ \frac{1 - (r_0/\epsilon)^{2n-1}}{2n-1} \biggr) \\
& + R_0(\x_0) - \frac{H(\x_0)}{4\pi} \biggl(\ln \epsilon - \frac{1-(r_0/\epsilon)^2}{2}\biggr).
\end{align*}
Using the expansion \cite{Abramowitz}
\begin{equation}
E(z) = - \frac{\pi}{2} \sum\limits_{n=0}^\infty \frac{[P_{2n}(0)]^2}{2n-1} \, z^{2n}    \qquad (|z| < 1),
\end{equation}
we evaluate the sums and obtain as $\epsilon\to 0$:
\begin{equation}
w_0^{(\Gamma)}(\x_0) \approx \frac{2}{\pi^2 \epsilon} \biggl( E(r_0/\epsilon) - \frac43\biggr)
+ \frac{H(\x_\Gamma) (1 - 2(r_0/\epsilon)^2)}{16\pi} \,,
\end{equation}
where we replaced $H(\x_0)$ by $H(\x_\Gamma)$ (the error being
$o(1)$).  In the leading order, one can neglect the second term that
yields Eq. (\ref{eq:w0_asympt_3d}).

Substituting Eqs. (\ref{eq:main_asympt}, \ref{eq:w0_asympt_3d}) into
Eq. (\ref{eq:bk_w0}), we deduce the asymptotic behavior of the
coefficients $b_k^{(\Gamma)}$:
\begin{equation} \label{eq:bk_3d_def2}
b_k^{(\Gamma)} \approx \epsilon^{-1} \hat{b}_k,  \qquad
\hat{b}_k = - \frac{\hat{\mu}_k}{2\pi} \int\limits_{\Gamma_1} d\hat{\x} \, \hat{w}_0(|\hat{\x}|) \, \hat{v}_k(\hat{\x}) ,
\end{equation}
where $\Gamma_1$ is the unit disk on the horizontal plane.  As the
integral of the axially symmetric function $\hat{w}_0(|\hat{\x}|)$
with any periodically oscillating eigenfunction vanishes, only axially
symmetric eigenfunctions $\hat{v}_{0,n_k}$ can provide contributions
$O(\epsilon^{-1})$ to $b_k^{(\Gamma)}$ in the leading order.

As in the planar case, one can employ Eq. (\ref{eq:Vk0_int}) to deduce
the following relation
\begin{equation}  \label{eq:Vinf_bk_3d}
\hat{b}_k = \hat{V}_k(\infty) .
\end{equation}
We also provide a more formal derivation of this relation by using the
pseudo-Green's function $\G_0(\x|\x_0)$ for the upper half-space,
\begin{equation}  \label{eq:kernel_disk0}
\G_0(\x|\x_0) = \frac{1}{4\pi} \biggl(\frac{1}{|\x-\x_0|} + \frac{1}{|\x-\x'_0|}\biggr),
\end{equation}
where $\x'_0 = (x_0,y_0,-z_0)$ is the mirror reflection of the
point $\x_0 = (x_0,y_0,z_0)$.  This function satisfies
\begin{subequations}
\begin{align}  \label{eq:G0_H3}
-\Delta \G_0(\x|\x_0) & = \delta(\x-\x_0) \quad (\x\in \H_3), \\
\partial_n \G_0(\x|\x_0) & = 0 \quad (\x \in \partial \H_3), \\
\G_0(\x|\x_0) & \sim \frac{1}{2\pi |\x|} + o(1) \quad (|\x|\to\infty).
\end{align}
\end{subequations}
Multiplying Eq. (\ref{eq:Steklov_disk_1}) by $\G_0(\x|\x_0)$,
multiplying Eq. (\ref{eq:G0_H3}) by $\hat{V}_k(\x)$, subtracting them,
integrating over $\x\in \H_3$ and using the Green's formula,
we get for any $\x_0 \in \H_3$
\begin{equation}
\hat{V}_k(\x_0) = \hat{V}_k(\infty) + \hat{\mu}_k \int\limits_{\Gamma_1} d\x \, \G_0(\x|\x_0)\, \hat{v}_k(\x) .
\end{equation}
Substituting
\begin{equation}
\int\limits_{\Gamma_1} d\x_0 \, \G_0(\x|\x_0) = \frac{\hat{w}_0(|\x|)}{2} + \frac{8}{3\pi}  \qquad (\x \in \Gamma_1)
\end{equation}
into Eq. (\ref{eq:bk_3d_def2}), we have
\begin{align*}
\hat{b}_k 
& = - \frac{\hat{\mu}_k}{\pi} \int\limits_{\Gamma_1} d\x \, \hat{v}_k(\x) \biggl[-\frac{8}{3\pi} 
+ \int\limits_{\Gamma_1} d\x_0 \, \G_0(\x|\x_0)\biggr] \\
& = - \frac{\hat{\mu}_k}{\pi} \int\limits_{\Gamma_1} d\x_0 \underbrace{\int\limits_{\Gamma_1} d\x \, \hat{v}_k(\x) \G_0(\x|\x_0)}_{ 
= (\hat{v}_k(\x_0) - \hat{V}_k(\infty))/\hat{\mu}_k} 
= \hat{V}_k(\infty) \,,
\end{align*}
where we used the orthogonality of $\hat{v}_k$ to a constant.

To complete this section, we also evaluate two sums used in
Sec. \ref{sec:application}.  According to Eqs. (\ref{eq:bk_sum},
\ref{eq:bk_3d_def2}), we have
\begin{align}  \label{eq:sum_bk2_3d}
C_2 = \sum\limits_{k=1}^\infty \frac{\hat{b}_k^2}{\hat{\mu}_k^2} 
& = \frac{1}{2\pi} \int\limits_0^1 d\hat{r} \, \hat{r} \, \hat{w}_0^2(\hat{r}) 
 = \frac{8}{\pi^3} \int\limits_0^1 d\hat{r} \, \hat{r} \bigl(E(\hat{r})-4/3\bigr)^2 \approx 0.0031 ,
\end{align}
where this integral involving the explicit function
$\hat{w}_0(\hat{r})$ from Eq. (\ref{eq:hatw0_3d}) was computed
numerically.
%%% CONFIRMED, SEE \verb|A_DN_disk_ext_bk;|

To proceed, we recall that a harmonic function $\hat{V}_k(\x_0)$ can
be extended via the harmonic measure density $h(\x|\x_0)$ on the unit
disk $\Gamma_1$ in $\H_3$: 
\begin{equation}
\hat{V}_k(\x_0) = \int\limits_{\Gamma_1} d\x \, \hat{v}_k(\x) \, h(\x|\x_0)   \qquad (\x_0 \in \H_3).
\end{equation}
In the limit $|\x_0| \to \infty$, the density $h(\x|\x_0)$ admits a
simple form, 
\begin{equation}  \label{eq:HMD_3d}
h(\x|\infty) = \frac{1}{2\pi \sqrt{1-|\x|^2}} \qquad (\x\in\Gamma_1),
\end{equation}
that follows from the Weber's solution for an electrified disk
\cite{Sneddon}.  Combining this expression with
Eqs. (\ref{eq:bk_3d_def2}, \ref{eq:Vinf_bk_3d}), we obtain
\begin{align*}
\sum\limits_{k=1}^\infty \frac{[\hat{V}_k(\infty)]^2}{\hat{\mu}_k} &
= - \sum\limits_{k=1}^\infty \int\limits_{\Gamma_1} \frac{d\x}{2\pi} \,
\hat{w}_0(|\x|) \, \hat{v}_k(\x) \int\limits_{\Gamma_1} d\x' \, h(\x'|\infty) \, \hat{v}_k(\x') \\
& = - \frac{1}{2\pi} \int\limits_{\Gamma_1} d\x \, \hat{w}_0(|\x|)  \int\limits_{\Gamma_1} d\x' \, h(\x'|\infty) 
\underbrace{\sum\limits_{k=1}^\infty \hat{v}_k(\x)\, \hat{v}_k(\x')}_{=\delta(\x-\x') - 1/\pi} \\
& = - \frac{1}{2\pi} \int\limits_{\Gamma_1} d\x \, \hat{w}_0(|\x|)  h(\x|\infty) ,
\end{align*}
where we used the orthogonality of $\hat{v}_k$ to a constant.  We
conclude that
\begin{align} \label{eq:sum_mubk2_3d}
C_1 = \sum\limits_{k=1}^\infty  \frac{[\hat{V}_k(\infty)]^2}{\hat{\mu}_k} & = 
 - \frac{2}{\pi^2} \int\limits_0^1 d\hat{r} \, \hat{r} \frac{E(\hat{r}) - 4/3}{\sqrt{1 - \hat{r}^2}}   
= \frac{8}{3\pi^2} - \frac14 \approx 0.0202 .
\end{align}
%%% CONFIRMED, SEE  \verb|A_DN_disk_ext_bk;|  %%%

\section{Lower and upper bounds for eigenvalues}
\label{sec:bounds}

When $\Omega$ is the disk of radius $R$ and $\Gamma$ is the arc of
angle $2\ve$, one can easily get lower and upper bounds for the
eigenvalues $\mu_k^{(p,\Gamma)}$ via variational arguments:
\begin{equation}  \label{eq:bounds}
\mu_k^{(p,2\ve,N)} \leq \mu_k^{(p,\Gamma)} \leq \mu_k^{(p,2\ve,D)} ,
\end{equation}
where $\mu_k^{(p,\alpha,N)}$ and $\mu_k^{(p,\alpha,D)}$ are the
eigenvalues of two auxiliary mixed Steklov problems in the sector of
angle $\alpha$ and radius $R$, $\Omega_\alpha = \{ (r,\theta) ~:~ 0 <
r < R,~ 0 < \theta < \alpha\}$, when the Steklov condition is imposed
on the arc ($r = R$), while either Neumann or Dirichlet condition is
imposed on two radial segments ($\theta =0$ and $\theta = \alpha$).
As this technique is fairly standard (see \cite{Levitin} for details),
we just sketch the arguments for the principal eigenvalue.  For the
original Steklov-Neumann problem in the disk $\Omega$, one has to
minimize the Rayleigh quotient:
\begin{equation}
\mu_0^{(p,\Gamma)} = \inf\limits_{U\in H^1(\Omega)} \left\{ \frac{\|\nabla U\|^2_{L^2(\Omega)} + \frac{p}{D} \|U\|^2_{L^2(\Omega)}}
{\| U|_\Gamma \|^2_{L^2(\Gamma)}} \right\} ,
\end{equation}
where the infimum is taken over all suitable functions (in the space
$H^1(\Omega)$).  As the sector $\Omega_{2\ve}$ is a subset of the disk
$\Omega$, the principal eigenvalue $\mu_0^{(p,2\ve,N)}$ in the sector
is determined by the same Rayleigh quotient, in which the two norms in
the numerator are evaluated over a smaller domain, implying the lower
bound in Eq. (\ref{eq:bounds}).  In turn, the principal eigenvalue
$\mu_0^{(p,2\ve,D)}$ is determined by imposing the Dirichlet condition
on two radial segments that reduces the space of functions and thus
increases the infimum, implying the upper bound in
Eq. (\ref{eq:bounds}).

The eigenmodes of the mixed Steklov-Neumann and Steklov-Dirichlet
problems in the sector are known explicitly due to the separation of
variables:
\begin{subequations}
\begin{align}
V_k^{(p,\alpha,N)} & \propto I_{\nu_k^N}(r\sqrt{p/D}) \cos(\nu_k^N \theta), \qquad \nu_k^N = \frac{\pi k}{\alpha} \,, \\
V_k^{(p,\alpha,D)} & \propto I_{\nu_k^D}(r\sqrt{p/D}) \sin(\nu_k^D \theta), \qquad \nu_k^D = \frac{\pi (k+1)}{\alpha} \,, 
\end{align}
\end{subequations}
with $k = 0,1,2,\ldots$, so that
\begin{equation}
\mu_k^{(p,\alpha,N/D)} = \sqrt{p/D} \frac{I'_{\nu_k^{N/D}}(R\sqrt{p/D})}{I_{\nu_k^{N/D}}(R\sqrt{p/D})} \,.
\end{equation}
One sees that these eigenvalues are of the same form, except for
$\nu_k^{N/D}$, which are shifted between the Neumann and Dirichlet
cases.  We conclude that
\begin{equation}  \label{eq:bounds_disk}
z \frac{I'_{\frac{\pi k}{2\ve}}(z)}{I_{\frac{\pi k}{2\ve}}(z)} \leq R \mu_k^{(p,\Gamma)} \leq 
z \frac{I'_{\frac{\pi (k+1)}{2\ve}}(z)}{I_{\frac{\pi (k+1)}{2\ve}}(z)} \qquad (z = R\sqrt{p/D}),
\end{equation}
which are valid for any $k$, $\ve$ and $p$.  In the limit $p\to 0$,
one gets much simpler bounds
\begin{equation}  \label{eq:bounds_disk_p0}
\frac{\pi k}{2\ve} \leq \mu_k^{(0,\Gamma)} \leq \frac{\pi (k+1)}{2\ve}  \,.
\end{equation}
These bounds agree with the asymptotic behavior (\ref{eq:main_asympt})
as $\ve\to 0$, together with the asymptotic behavior
(\ref{eq:etak_asympt_2d}) for the eigenvalues $\hat{\mu}_k$ as $k\to
\infty$.  Moreover, one can check that each eigenvalue $\hat{\mu}_k$
listed in Table \ref{tab:eta_2D}, lies indeed between the bounds $\pi
k/2$ and $\pi (k+1)/2$.

In the same vein, one can derive bounds for the case of a spherical
cap of angle $\ve$ on the sphere of radius $R$ by considering two
auxiliary mixed Steklov problems in the spherical sector of angle
$\ve$.  We focus on the axially symmetric eigenfunctions in the
sector, which have the form $i_{\nu_k^{N/D}}(r\sqrt{p/D})
P_{\nu_k^{N/D}}(\cos\theta)$, where the $\nu_k^{N/D}$ are fixed by
imposing boundary conditions on the conical boundary:
$P_{\nu_k^D}(\cos \ve) = 0$ for the Dirichlet case, and
$P'_{\nu_k^N}(\cos \ve) = 0$ for the Neumann case.  The associated
eigenvalues are
\begin{equation}
\mu_k^{(p,\ve,N/D)} = \sqrt{p/D}\, \frac{i'_{\nu_k^{N/D}}(R\sqrt{p/D})}{i_{\nu_k^{N/D}}(R\sqrt{p/D})} \,, 
\end{equation}
which are reduced to $\mu_k^{(0,\ve,N/D)} = \nu_k^{N/D}/R$ at $p = 0$.
While the values of $\nu_k^{N/D}$ need to be found numerically, their
asymptotic behavior for large $k$ can be obtained by using the
large-$\nu$ asymptotic relation \cite{Abramowitz}
\begin{align} 
P_\nu(\cos\ve) & \approx \frac{\Gamma(\nu+1)}{\Gamma(\nu+3/2)} \biggl(\frac{\pi \sin\ve}{2}\biggr)^{-1/2}  
 \cos\bigl((\nu+1/2)\ve - \pi/4\bigr) + O(\nu^{-1}),
\end{align}
so that
\begin{equation}
\nu_k^D \approx \frac{\pi (k+3/4)}{\ve} - \frac12  \qquad (k\gg 1).
\end{equation}
Similarly, the asymptotic analysis of $P'_\nu(\cos\ve)$ yields 
\begin{equation}
\nu_k^N \approx \frac{\pi (k+1/4)}{\ve} - \frac12  \qquad (k\gg 1)
\end{equation}
(see more discussions in \cite{Grebenkov19f}).  As a consequence, we
get two bounds for the eigenvalues associated to axially symmetric
eigenfunctions, which are valid for any $0< \ve \leq \pi$:
\begin{equation}  \label{eq:bounds_sphere_p0}
\frac{\pi (k+1/4)}{\ve} - \frac12 \leq R \mu_{0,k}^{(0,\Gamma)} \leq \frac{\pi (k+3/4)}{\ve} - \frac12  \qquad (k\gg 1).
\end{equation}
Combining these bounds with the asymptotic relation
(\ref{eq:main_asympt}) as $\ve\to 0$, one gets 
\begin{equation}
\hat{\mu}_{0,k} \simeq \pi k  \qquad (k \gg 1).
\end{equation}

%%%%%%%%%%%%%%%%%%%%%%%%%%%%%%%%%%%%%%%%%%%%%%%%%%%%%%%%%%%%%%%%%%%%%%%%%%%%%%%%%%%%%%
\bibliographystyle{siamplain}

\begin{thebibliography}{132}


\bibitem{North66}			A. M. North,  
					{\it Diffusion-controlled reactions},
					Q. Rev. Chem. Soc., 20 (1966), 421--440.

\bibitem{Wilemski73}			G. Wilemski and M. Fixman,
					{\it General theory of diffusion-controlled reactions}, 
					J. Chem. Phys., 58 (1973), 4009--4019.

\bibitem{Calef83}			D. F. Calef and J. M. Deutch, 
					{\it Diffusion-Controlled Reactions},
					Ann. Rev. Phys. Chem., 34 (1983), 493--524.

\bibitem{Berg85}			O. G. Berg and P. H. von Hippel,  
					{\it Diffusion-Controlled Macromolecular Interactions}, 
					Ann. Rev. Biophys. Biophys. Chem., 14 (1985), 131--160.

\bibitem{Rice85}			S. Rice,
					{\it Diffusion-Limited Reactions}
					(Elsevier: Amsterdam, The Netherlands, 1985).

\bibitem{Grebenkov23c}			D. S. Grebenkov,
					{\it Diffusion-Controlled Reactions: An Overview},
					Molecules, 28 (2023), 7570.





\bibitem{Redner}			S. Redner, 
					{\it A Guide to First Passage Processes} 
					(Cambridge: Cambridge University press, 2001).

\bibitem{Schuss}			Z. Schuss, 
					{\it Brownian Dynamics at Boundaries and Interfaces in Physics, Chemistry and Biology}
					(Springer, New York, 2013).

\bibitem{Metzler}         		R. Metzler, G. Oshanin, and S. Redner (Eds.) 
                            		{\it First-Passage Phenomena and Their Applications}
                            		(Singapore: World Scientific, 2014).

\bibitem{Benichou14}			O. B\'enichou and R. Voituriez,
					{\it From first-passage times of random walks in confinement to geometry-controlled kinetics},
					Phys. Rep., 539 (2014), 225--284.

\bibitem{Holcman14}			D. Holcman and Z. Schuss,
					{\it The Narrow Escape Problem},
					SIAM Rev., 56 (2014), 213--257.

\bibitem{Masoliver}			J. Masoliver,
					{\it Random Processes: First-passage And Escape}
					(World Scientific Publishing, 2018).

\bibitem{Lindenberg}			K. Lindenberg, R. Metzler and G. Oshanin (Eds.)
					{\it Chemical Kinetics: Beyond the Textbook}
					(World Scientific Publishers Europe: London, 2019).


\bibitem{Grebenkov}			D. S. Grebenkov, R. Metzler, and G. Oshanin (Eds), 
					{\it Target Search Problems} 
					(Springer: Cham, Switzerland, 2024). 


\bibitem{Dagdug}			L. Dagdug, J. Pe{\~n}a and I. Pompa-Garc{\'\i}a,
					{\it Diffusion Under Confinement. A Journey Through Counterintuition}
					(Springer, 2024).




\bibitem{Grebenkov19}			D. S. Grebenkov,
					{\it Spectral theory of imperfect diffusion-controlled reactions on heterogeneous catalytic surfaces},
					J. Chem. Phys., 151 (2019), 104108.

\bibitem{Grebenkov19c}			D. S. Grebenkov,
					{\it Probability distribution of the boundary local time of reflected Brownian motion in Euclidean domains},
					Phys. Rev. E, 100 (2019), 062110.

\bibitem{Grebenkov20}			D. S. Grebenkov,
					{\it Paradigm shift in diffusion-mediated surface phenomena},
					Phys. Rev. Lett., 125 (2020), 078102.


\bibitem{Grebenkov20b}			D. S. Grebenkov,
					{\it Joint distribution of multiple boundary local times and related first-passage time problems},
					J. Stat. Mech. (2020), 103205.

\bibitem{Grebenkov20c}			D. S. Grebenkov, 
					{\it Surface Hopping Propagator: An Alternative Approach to Diffusion-Influenced Reactions}, 
					Phys. Rev. E, 102 (2020), 032125.

\bibitem{Grebenkov21a}			D. S. Grebenkov, 
					{\it Statistics of boundary encounters by a particle diffusing outside a compact planar domain}, 
					J. Phys. A: Math. Theor., 54 (2021), 015003.

\bibitem{Grebenkov22a}			D. S. Grebenkov, 
					{\it An encounter-based approach for restricted diffusion with a gradient drift}, 
					J. Phys. A: Math. Theor., 55 (2022), 045203. 

\bibitem{Grebenkov22b}			D. S. Grebenkov, 
					{\it Depletion of Resources by a Population of Diffusing Species}, 
					Phys. Rev. E, 105 (2022), 054402.

\bibitem{Grebenkov22d}			D. S. Grebenkov, 
					{\it Statistics of diffusive encounters with a small target: Three complementary approaches}, 
					J. Stat. Mech. (2022), 083205. 


\bibitem{Bressloff22d}			P. C. Bressloff,
					{\it Diffusion-mediated surface reactions and stochastic resetting},
					J. Phys. A: Math. Theor., 55 (2022), 275002.

\bibitem{Benkhadaj22}			Z. Benkhadaj and D. S. Grebenkov, 
					{\it Encounter-based approach to diffusion with resetting}, 
					Phys. Rev. E, 106 (2022), 044121.

\bibitem{Bressloff22b}			P. C. Bressloff,
					{\it Narrow capture problem: an encounter-based approach to partially reactive targets},
					Phys. Rev. E, 105 (2022), 034141.


\bibitem{Bressloff22a}			P. C. Bressloff,
					{\it Diffusion-mediated absorption by partially-reactive targets: 
					Brownian functionals and generalized propagators},
					J. Phys. A: Math. Theor., 55 (2022), 205001.

\bibitem{Bressloff22c}			P. C. Bressloff, 
					{\it A probabilistic model of diffusion through a semipermeable barrier},
					Proc. Roy. Soc. A, 478 (2022), 20220615.

\bibitem{Grebenkov23a}			D. S. Grebenkov, 
					{\it Diffusion-controlled reactions with non-Markovian binding/unbinding kinetics}, 
					J. Chem. Phys., 158 (2023), 214111. 

\bibitem{Grebenkov23b}			D. S. Grebenkov, 
					{\it Encounter-based approach to the escape problem}, 
					Phys. Rev. E, 107 (2023), 044105.

\bibitem{Bressloff23a}			P. C. Bressloff,
					{\it Renewal equation for single-particle diffusion through a semipermeable interface},
					Phys. Rev. E, 107 (2023), 014110.

\bibitem{Bressloff23b}			P. C. Bressloff,
					{\it Renewal equations for single-particle diffusion in multilayered media},
					SIAM J. Appl. Math., 83 (2023), 1518--1545.





\bibitem{Steklov1902}			W. Steklov (V. A. Steklov), 
					{\it Sur les probl\`emes fondamentaux de la physique math\'ematique},
					Ann. Sci. Ec. Norm. Sup\'er., 19 (1902), 455--490.

\bibitem{Kuznetsov14}			N. Kuznetsov, T. Kulczycki, M. Kwa\'snicki, A. Nazarov, S. Poborchi, I. Polterovich, and B. Siudeja, 
					{\it The legacy of Vladimir Andreevich Steklov}, 
					Not. Am. Math. Soc., 61 (2014), 9--22.




\bibitem{Girouard17}			A. Girouard and I. Polterovich,
					{\it Spectral geometry of the Steklov problem}, 
					J. Spectr. Th., 7 (2017), 321--359.

\bibitem{Colbois24}			B. Colbois, A. Girouard, C. Gordon, and D. Sher,
					{\it Some recent developments on the Steklov eigenvalue problem},
					Rev. Mat. Complut., 37 (2024), 1--161.


\bibitem{Levitin}			M. Levitin, D. Mangoubi, and I. Polterovich,
					{\it Topics in Spectral Geometry}
					(Graduate Studies in Mathematics, vol. 237; American Mathematical Society, 2023).




\bibitem{Auchmuty04}			G. Auchmuty,
					{\it Steklov eigenproblems and the representation of solutions of 
					elliptic boundary value problems},
					Numer. Funct. Anal. Optim., 25 (2005), 321--348.

\bibitem{Auchmuty13}			G. Auchmuty and Q. Han,
					{\it Spectral representations of solutions of linear elliptic equations on exterior regions},
					J. Math. Anal. Appl. 398 (2013), 1--10.

\bibitem{Auchmuty14}			G. Auchmuty and Q. Han,
					{\it Representations of Solutions of Laplacian Boundary Value Problems on Exterior Regions},
					Appl. Math. Optim., 69 (2014), 21--45.  

\bibitem{Auchmuty15}			G. Auchmuty and M. Cho, 
					{\it Boundary integrals and approximations of harmonic functions}, 
					Numer. Funct. Anal. Optim., 36 (2015), 687--703.

\bibitem{Auchmuty18}			G. Auchmuty,
					{\it Steklov Representations of Green's Functions for Laplacian Boundary Value Problems},
					Appl. Math. Optim., 77, 687-703 (2018), 173--195.




\bibitem{Smith96}			B. F. Smith, 
					{\it Domain Decomposition Methods for Partial Differential Equations},
					in ``Parallel Numerical Algorithms'', Eds. D. E. Keyes, A. Sameh, V. Venkatakrishnan 
					(Springer, 1996), pp. 225-243. 

\bibitem{Levitin08}			M. Levitin and M. Marletta,
					{\it A simple method of calculating eigenvalues and resonances in domains with infinite regular ends},
					Proc. Royal Soc. Edinburgh A, 138 (2008), 1043--1065.

\bibitem{Delitsyn12}			A. L. Delitsyn, B.-T. Nguyen, and D. S. Grebenkov,
					{\it Trapped modes in finite quantum waveguides},
					Eur. Phys. J B, 85 (2012), 176.

\bibitem{Delitsyn18}			A. L. Delitsyn and D. S. Grebenkov,
					{\it Mode matching methods in spectral and scattering problems},
					Quart J. Mech. Appl. Math., 71 (2018), 537--580.


\bibitem{Cheney99}			M. Cheney, D. Isaacson, and J. C. Newell,
					{\it Electrical Impedance Tomography},
					SIAM Rev. 41 (1999), 85--101.

\bibitem{Calderon80}			A. P. Calder\'on, 
					{\it On an inverse boundary value problem}, 
					Seminar on Numerical Analysis and its Applications to Continuum Physics, 
					Soc. Brasileira de Matem\'atica, R\'io de Janeiro, (1980), 65-73;
					Reprinted in Comput. Appl. Math. {\bf 25}, 2-3, 133-138 (2006).

\bibitem{Borcea02}			L. Borcea,
					{\it Electrical impedance tomography},
					Inv. Prob., 18 (2002), R99.

\bibitem{Sylvester87}			J. Sylvester and G. Uhlmann,
					{\it A Global Uniqueness Theorem for an Inverse Boundary Value Problem},
					Ann. Math., 125 (1987), 153--169.

\bibitem{Curtis91}			E. Curtis and J. Morrow,
					{\it The Dirichlet to Neumann map for a resistor network},
					SIAM J. Appl. Math., 51 (1991), 1011--1029.















\bibitem{Henrici70}			P. Henrici, B. A. Troesch, L. Wuytack,
					{\it Sloshing frequencies for a half-space with circular or strip-like aperture},
					Z. Angew. Math. Phys., 21 (1970), 285--318.

\bibitem{Troesch72}			B. A. Troesch and H. R. Troesch,
					{\it A remark on the sloshing frequencies for a half-space},
					J. Appl. Math. Phys., 23 (1972), 703--711.

\bibitem{Miles72}			J. W. Miles,
					{\it On the eigenvalue problem for fluid sloshing in a half-space},
					Z. Angew. Math. Phys., 23 (1972), 861--868.

\bibitem{Fox83}				D. W. Fox and J. R. Kuttler,
					{\it Sloshing frequencies},
					Z. Angew. Math. Phys., 34 (1983), 668-696.

\bibitem{Kozlov04}			V. Kozlov and N. Kuznetsov,
					{\it The ice-fishing problem: the fundamental sloshing frequency versus geometry of holes},
					Math. Methods Appl. Sci., 27 (2004), 289-312.

\bibitem{Levitin22}			M. Levitin, L. Parnovski, I. Polterovich, and D. A. Sher,
					{\it Sloshing, Steklov and corners: Asymptotics of Steklov eigenvalues for curvilinear polygons},
					Proc. London Math. Soc., 125 (2022), 359--487.



\bibitem{Banuelos10}			R. Ba\~nuelos, T. Kulczycki, I. Polterovich and B. Siudeja,
					{\it Eigenvalue inequalities for mixed Steklov problems},
					Amer. Math. Soc. Transl., 231 (2010), 19--34.

\bibitem{Mayrand20}			J. Mayrand, C. Sen\'ecal, and S. St-Amant,
					{\it Asymptotics of sloshing eigenvalues for a triangular prism},
					Math. Proc. Camb. Phil. Soc., 173 (2022), 539--571. 


\bibitem{Ammari20b}			H. Ammari, K. Imeri, and N. Nigam,
					{\it Optimization of Steklov-Neumann eigenvalues},
					J. Comput. Phys., 406 (2020), 109211.






\bibitem{Ward93}			M. J. Ward and J. B. Keller,
					{\it Strong Localized Perturbations of Eigenvalue Problems}, 
					SIAM J. Appl. Math., 53 (1993), 770--798.

\bibitem{Grigoriev02}			I. V. Grigoriev, Y. A. Makhnovskii, A. M. Berezhkovskii, and V. Y. Zitserman, 
					{\it Kinetics of escape through a small hole},
					J. Chem. Phys., 116 (2002), 9574--9577.

\bibitem{Holcman04}			D. Holcman and Z. Schuss,
					{\it Escape Through a Small Opening: Receptor Trafficking in a Synaptic Membrane},
					J. Stat. Phys., 117 (2004), 975-1014.

\bibitem{Kolokolnikov05}		T. Kolokolnikov, M. S. Titcombe, and M. J. Ward,
					{\it Optimizing the Fundamental Neumann Eigenvalue for the Laplacian in a Domain with Small Traps},
					Eur. J. Appl. Math., 16 (2005), 161--200.

\bibitem{Singer06a}			A. Singer, Z. Schuss, D. Holcman, and R. S. Eisenberg,
					{\it Narrow Escape, Part I},
					J. Stat. Phys., 122 (2006), 437--463.


\bibitem{Singer06b}			A. Singer, Z. Schuss, and D. Holcman,
					{\it Narrow Escape, Part II. The circular disk},
					J. Stat. Phys., 122 (2006), 465--489.

\bibitem{Singer06c}			A. Singer, Z. Schuss, and D. Holcman,
					{\it Narrow Escape, Part III Riemann surfaces and non-smooth domains},
					J. Stat. Phys., 122 (2006), 491--509.

\bibitem{Schuss07}			Z. Schuss, A. Singer, and D. Holcman,
					{\it The narrow escape problem for diffusion in cellular microdomains},
					Proc. Nat. Acad. Sci. USA, 104 (2007), 16098--16103.

\bibitem{Benichou08}			O. B\'enichou and R. Voituriez,
					{\it Narrow-Escape Time Problem: Time Needed for a Particle to Exit a Confining Domain through a Small Window},
					Phys. Rev. Lett., 100 (2008), 168105.


\bibitem{Singer08}			A. Singer, Z. Schuss, and D. Holcman, 
					{\it Narrow escape and leakage of Brownian particles}, 
					Phys. Rev. E, 78 (2008), 051111.



\bibitem{Reingruber09}			J. Reingruber, E. Abad, and D. Holcman,
					{\it Narrow escape time to a structured target located on the boundary of a microdomain},
					J. Chem. Phys., 130 (2009), 094909.


\bibitem{Pillay10}			S. Pillay, M. J. Ward, A. Peirce, and T. Kolokolnikov,
					{\it An Asymptotic Analysis of the Mean First Passage Time 
					for Narrow Escape Problems: Part I: Two-Dimensional Domains},
					SIAM Multi. Model. Simul., 8 (2010), 803--835.

\bibitem{Cheviakov10}			A. F. Cheviakov, M. J. Ward, and R. Straube,
					{\it An Asymptotic Analysis of the Mean First Passage Time for Narrow Escape Problems: Part II: The Sphere,}
					SIAM Multi. Model. Simul., 8 (2010), 836--870.

\bibitem{Cheviakov11}			A. F. Cheviakov and M. J. Ward,
					{\it Optimizing the principal eigenvalue of the Laplacian in a sphere with interior traps},
					Math. Computer Model., 53 (2011), 1394--1409. 

\bibitem{Cheviakov12}			A. F. Cheviakov, A. S. Reimer, and M. J. Ward,
					{\it Mathematical modeling and numerical computation of narrow escape problems},
					Phys. Rev. E, 85 (2012), 021131.

\bibitem{Caginalp12}			C. Caginalp and X. Chen, 
					{\it Analytical and Numerical Results for an Escape Problem},
					Arch. Rational. Mech. Anal., 203 (2012), 329-342.

\bibitem{Rojo12}			F. Rojo, H. S. Wio, and C. E. Budde,
					{\it Narrow-escape-time problem: The imperfect trapping case},
					Phys. Rev. E, 86 (2012), 031105.

\bibitem{Rupprecht15}			J.-F. Rupprecht, O. B\'enichou, D. S. Grebenkov, and R. Voituriez,
					{\it Exit time distribution in spherically symmetric two-dimensional domains},
					J. Stat. Phys., 158 (2015), 192--230.


\bibitem{Gomez15}			D. Gomez and A. F. Cheviakov,
					{\it Asymptotic analysis of narrow escape problems in nonspherical three-dimensional domains},
					Phys. Rev. E, 91 (2015), 012137.


\bibitem{Grebenkov16}			D. S. Grebenkov, 
					{\it Universal formula for the mean first passage time in planar domains}, 
					Phys. Rev. Lett., 117 (2016), 260201.

\bibitem{Marshall16}			J. S. Marshall,
					{\it Analytical Solutions for an Escape Problem in a Disc with an Arbitrary 
					Distribution of Exit Holes Along Its Boundary},
					J. Stat. Phys., 165 (2016), 920--952.

\bibitem{Grebenkov17}			D. S. Grebenkov and G. Oshanin, 
					{\it Diffusive escape through a narrow opening: new insights into a classic problem}, 
					Phys. Chem. Chem. Phys., 19 (2017), 2723--2739.

\bibitem{Grebenkov19d}			D. S. Grebenkov, R. Metzler, and G. Oshanin,
					{\it Full distribution of first exit times in the narrow escape problem},
					New J. Phys., 21 (2019), 122001.


\bibitem{Kaye20}			J. Kaye and L. Greengard,
					{\it A fast solver for the narrow capture and narrow escape problems in the sphere},
					J. Comput. Phys. X, 5 (2020), 100047.

\bibitem{Guerin23}			T. Gu\'erin, M. Dolgushev, O. B\'enichou, and R. Voituriez,
					{\it Imperfect narrow escape problem},
					Phys. Rev. E, 107 (2023), 034134.


\bibitem{Friedlander91}			L. Friedlander,
					{\it Some inequalities between Dirichlet and Neumann eigenvalues},
					Arch. Rational. Mech. Anal., 116 (1991), 153--160.

\bibitem{Chaigneau24}			A. Chaigneau and D. S. Grebenkov,
					{\it A numerical study of the generalized Steklov problem in planar domains},
					J. Phys. A: Math. Theor., 57 (2024), 445201.



\bibitem{McLean}			W. C. H. McLean, 
					{\it Strongly elliptic systems and boundary integral equations}
					(Cambridge University Press, 2000).



\bibitem{Behrndt15}			J. Behrndt and A. F. M ter Elst,
					{\it Dirichlet-to-Neumann maps on bounded Lipschitz domains},
					J. Diff. Eq., 259 (2015), 5903--5926.

\bibitem{Kato}				T. Kato, 
					{\it Perturbation Theory for Linear Operators} 
					(2nd Ed., Springer-Verlag, Berlin, 1980).

\bibitem{Bucur17}			D. Bucur, P. Freitas, and J. Kennedy,
					{\it The Robin problem},
					In ``Shape optimization and spectral theory'' (De Gruyter Open, Warsaw, 2017), pp 78-119.








\bibitem{Condamin07}			S. Condamin, O. B\'enichou, and M. Moreau,
					{\it Random walks and Brownian motion: A method of computation for first-passage
					times and related quantities in confined geometries},
					Phys. Rev. E, 75 (2007), 021111.

\bibitem{Giuggioli20}			L. Giuggioli,
					{\it Exact Spatiotemporal Dynamics of Confined Lattice Random Walks in Arbitrary Dimensions: 
					A Century after Smoluchowski and Polya},
					Phys. Rev. X, 10 (2020), 021045.





\bibitem{Grebenkov17g}			D. S. Grebenkov, R. Metzler, and G. Oshanin, 
					{\it Effects of the target aspect ratio and intrinsic reactivity onto diffusive search in bounded domains}, 
					New J. Phys., 19 (2017), 103025.

\bibitem{Grebenkov18a}			D. S. Grebenkov, R. Metzler, and G. Oshanin, 
					{\it Towards a full quantitative description of single-molecule reaction kinetics in biological cells}, 
					Phys. Chem. Chem. Phys., 20 (2018), 16393--16401.

\bibitem{Grebenkov21g}			D. S. Grebenkov, R. Metzler, and G. Oshanin, 
					{\it Distribution of first-reaction times with target regions on boundaries of shell-like domains}, 
					New J. Phys., 23 (2021), 123049.







\bibitem{Dittmar04}			B. Dittmar,
					{\it Sums of reciprocal Stekloff eigenvalues}, 
					Math. Nachr., 268 (2004), 44--49.

\bibitem{Hitotumatu54}			S. Hitotumatu, 
					{\it On the Neumann function of a sphere},
					Comment. Math. Univ. St. Paul., 3 (1954), 1--5.

\bibitem{McCann01}			R. C. McCann, R. D. Hazlett, and D. K. Babu, 
					{\it Highly accurate approximations of Green's and Neumann functions on rectangular domains},
					Proc. R. Soc. Lond. A, 457 (2001), 767--772.


\bibitem{Friedman89}			A. Friedman and M. Vogelius, 
					{\it Identification of small inhomogeneities of extreme conductivity by boundary 
					measurements: a theorem on continuous dependence}, 
					Arch. Rat. Mech. Anal., 105 (1989), 299--326.


\bibitem{Grebenkov24}			D. S. Grebenkov, 
					{\it Spectral properties of the Dirichlet-to-Neumann operator for spheroids}, 
					Phys. Rev. E, 109 (2024), 055306. 












 

\bibitem{Bonnetier22}			E. Bonnetier, C. Dapogny, and M. S. Vogelius, 
					{\it Small perturbations in the type of boundary conditions for an elliptic operator}, 
					J. Math. Pures Appl., 167 (2022), 101--174.





\bibitem{Widom64}			H. Widom,
					{\it Asymptotic behavior of the eigenvalues of certain integral equations. II},
					Arch. Rat. Mech. Anal., 17 (1964), 215--229.

\bibitem{Simic06}			S. Simi\'c,
					{\it An estimation of the singular values of integral operator with logarithmic kernel},
					Facta Univer. Ser. Math. Inform., 21 (2006), 49--55.

\bibitem{Chen09}			Z. Chen, G. Nelakanti, Y. Xu, and Y. Zhang,
					{\it A Fast Collocation Method for Eigen-Problems of Weakly Singular Integral Operators},
					J. Sci. Comput., 41 (2009), 256--272.

\bibitem{Mandal21}			M. Mandal, K. Kant, and G. Nelakanti,
					{\it Eigenvalue problem of a weakly singular compact integral operator by discrete Legendre projection methods},
					J. Appl. Anal. Comput., 11 (2021), 2090--2101.

\bibitem{Polosin22}			A. A. Polosin, 
					{\it On the Asymptotic Behavior of Eigenvalues and Eigenfunctions of an Integral 
					Convolution Operator with a Logarithmic Kernel on a Finite Interval},
					Diff. Eq., 58 (2022), 1242--1257. 
%https://doi.org/10.1134/S0012266122090099







\bibitem{Popov92}			I. Yu. Popov, 
					{\it Extension theory and localization of resonances for domains of trap type},
					Math. USSR. Sb., 71 (1992), 209--234.

\bibitem{Silbergleit03}			A. Silbergleit, I. Mandel, and I. Nemenman, 
					{\it Potential and field singularity at a surface point charge},
					J. Math. Phys., 44 (2003), 4460--4466.









\bibitem{Colbois21}			B. Colbois and S. Verma,
					{\it Sharp Steklov Upper Bound for Submanifolds of Revolution},
					J. Geom. Anal., 31 (2021), 11214--11225.






\bibitem{Collins49}			F. C. Collins and G. E. Kimball,
					{\it Diffusion-controlled reaction rates},
					J. Coll. Sci., 4 (1949), 425--437.

\bibitem{Sapoval94}			B. Sapoval,
					{\it General Formulation of Laplacian Transfer Across Irregular Surfaces},
					Phys. Rev. Lett., 73 (1994), 3314--3316.

\bibitem{Grebenkov06}			D. S. Grebenkov, 
					{\it Partially Reflected Brownian Motion: A Stochastic Approach to Transport Phenomena}, 
					in ``Focus on Probability Theory'', Ed. L. R. Velle, pp. 135-169 (Nova Science Publishers, 2006). 

\bibitem{Singer08b}			A. Singer, Z. Schuss, A. Osipov, and D. Holcman,
					{\it Partially reflected diffusion},
					SIAM J. Appl. Math., 68 (2008), 844--868.

\bibitem{Grebenkov19b}			D. S. Grebenkov,
					``Imperfect Diffusion-Controlled Reactions'', 
					in {\it Chemical Kinetics: Beyond the Textbook}, Eds. K. Lindenberg, R. Metzler, and G. Oshanin 
					(World Scientific, 2019; available online as ArXiv: 1806.11471).

\bibitem{Piazza22}			F. Piazza,
					{\it The physics of boundary conditions in reaction-diffusion problems},
					J. Chem. Phys., 157 (2022), 234110.




\bibitem{Rayleigh}			J. W. S. Baron Rayleigh, 
					{\it The Theory of Sound}, Vol. 2, 2nd Ed.
					(Dover, New York, 1945).






\bibitem{Ito}				K. Ito and H. P. McKean,
					{\it Diffusion Processes and Their Sample Paths}
					(Springer-Verlag, Berlin, 1965).

\bibitem{Freidlin}			M. Freidlin,
					{\it Functional Integrationand Partial Differential Equations} 
					(Annals of Mathematics Studies, Princeton
					University Press, Princeton, New Jersey, 1985).

\bibitem{Saisho87}			Y. Saisho,
					{\it Stochastic Differential Equations for Multi-Dimentional Domain with Reflecting Boundary},
					Probab. Theory Rel. Fields, 74 (1987), 455--477.

\bibitem{Papanicolaou90}		V. G. Papanicolaou,
					{\it The probabilistic solution of the third boundary value problem for second order elliptic equations},
					Probab. Th. Rel. Fields, 87 (1990), 27--77.





\bibitem{Helffer22}			B. Helffer and A. Kachmar,
					{\it Semi-classical edge states for the Robin Laplacian},
					Mathematika, 68 (2022), 454--485.


\bibitem{Abramowitz}      		 M. Abramowitz and I. A. Stegun,
			   		{\it Handbook of Mathematical Functions} 
			   		(Dover Publisher, New York, 1965).


\bibitem{Sneddon}			I. N. Sneddon, 
					{\it Mixed Boundary Value Problems in Potential Theory}
					(Wiley, New York, NY, 1966).





\bibitem{Grebenkov07a}			D. S. Grebenkov,
					{\it Residence times and other functionals of reflected Brownian motion},
					Phys. Rev. E, 76 (2007), 041139. 


\bibitem{Watson}			E. J. Watson,
					{\it Laplace transformations and applications} 
					(Van Nostrand, Wokingham, England, 1981).



\bibitem{Davis07}			A. M. J. Davis and S. G. L. Smith,
					{\it Perturbation of eigenvalues due to gaps in two-dimensional boundaries},
					Proc. R. Soc. A, 463 (2007), 759--786.



\bibitem{Kellog}			O. D. Kellog, 
					{\it Foundations of Potential Theory} 
					(Dover Publications, New York, 1954).





\bibitem{Grebenkov19f}			D. S. Grebenkov,
					{\it Reversible reactions controlled by surface diffusion on a sphere},
					J. Chem. Phys., 151 (2019), 154103.



\bibitem{Christiansen23}		T. J. Christiansen and K. Datchev, 
					{\it Low energy scattering asymptotics for planar obstacles}, 
					Pure Appl. Anal., 5 (2023), 767--794.

\bibitem{Grebenkov25}			D. S. Grebenkov and A. Chaigneau,
					{\it The Steklov problem for exterior domains: asymptotic behavior and applications}
					J. Math. Phys., 66 (2025), 061502.


\bibitem{Arendt15}			W. Arendt and A. F. M. ter Elst,
					{\it The Dirichlet-to-Neumann Operator on Exterior Domains},
					Potential Anal., 43 (2015), 313--340.

\bibitem{Grebenkov25b}			D. S. Grebenkov, 
					{\it Imperfect diffusion-controlled reactions on a torus and on a pair of balls}, 
					J. Chem. Phys., 163 (2025), 034106.




\end{thebibliography}

\end{document}